\documentclass[twocolumn,epjc3]{svjour3}
\journalname{Eur.\ Phys.\ J.\ A}
\usepackage{amsmath,amssymb,bm,fix-cm,graphicx,hyperref}
\usepackage{ulem}
\usepackage{xcolor}
\usepackage{breakcites}

\newcommand{\secref}[1]{Sec.~\ref{#1}}

\begin{document}

\title{Three-body forces and Efimov physics in nuclei and atoms}

\author{Shimpei Endo\thanksref{eSE,uec}\and
Evgeny Epelbaum\thanksref{eEE,bochum}\and
Pascal Naidon\thanksref{ePN,riken}\and
Yusuke Nishida\thanksref{eYN,titech}\and
Kimiko Sekiguchi\thanksref{eKS,titech,riken}\and
Yoshiro Takahashi\thanksref{eYT,kyoto}}

\institute{Department of Engineering Science, The University of Electro-Communications, Tokyo 182-8585, Japan\label{uec}\and
Institut f\"ur Theoretische Physik II, Ruhr-Universit\"at Bochum, D-44780 Bochum, Germany\label{bochum}\and
RIKEN Nishina Center for Accelerator-Based Science, Wako, Saitama 351-0198, Japan\label{riken}\and
Department of Physics, Institute of Science Tokyo, Ookayama, Meguro, Tokyo 152-8551, Japan\label{titech}\and
Department of Physics, Graduate School of Science, Kyoto University, Kyoto
606-8502, Japan
\label{kyoto}}


\thankstext{eSE}{e-mail: shimpei.endo@uec.ac.jp}
\thankstext{eEE}{e-mail: evgeny.epelbaum@rub.de}
\thankstext{ePN}{e-mail: pascal@riken.jp}
\thankstext{eYN}{e-mail: nishida@phys.titech.ac.jp}
\thankstext{eKS}{e-mail: kimiko@phys.titech.ac.jp}
\thankstext{eYT}{e-mail: takahashi.yoshiro.7v@kyoto-u.ac.jp}

\date{\today}

\maketitle

\begin{abstract}
This review article presents historical developments and recent advances in our understanding on the three-body forces and Efimov physics, from an interdisciplinary viewpoint encompassing nuclear physics and cold atoms. Theoretical attempts to elucidate the three-body force with the chiral effective field theory are explained, followed by an overview of experiments aimed at observing signatures of the nuclear three-body force. Some recent experimental and theoretical works in the field of cold atoms devoted to measuring and engineering three-body forces among atoms are also presented. As a phenomenon arising from the three-body effect, Efimov physics in both cold atoms and nuclear systems is reviewed. 
\end{abstract}

\section{\label{sec:intro}Introduction}

All visible matter in the universe is organized into re\-so\-lution-dependent
hierarchical structures. 
A precise description of the interactions at each hierarchical level
starting from elementary quarks to composite systems like had\-rons, nuclei,
atoms and molecules may help to improve our understanding of the structure
and dynamics of strongly interacting matter. Generally, effective
interactions at all hierarchical levels are dominated by pairwise forces
acting between two constituent particles. 
Meanwhile, recent advances in computational, theoretical 
and experimental techniques
allow one to go beyond the two-body-force level and to
quantitatively probe three-body
interactions, which are known to figure importantly in atomic and
nuclear systems~\cite{gibson1988three,Hammer:2012id}.

\begin{figure}
\centering
  \includegraphics[width=0.2\textwidth]{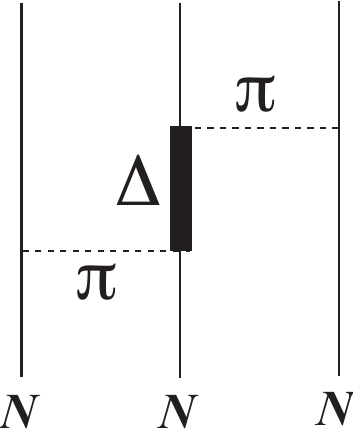}
\caption{\label{fig:FM-3NF}
Feynmann diagram of Fujita-Miyazawa type three-nucleon force.}
\end{figure}

Historically, the most well known three-body force model 
was proposed 
by Fujita and Miyazawa in 1957~\cite{10.1143/PTP.17.360} to describe 
nuclear systems using protons and neutrons, jointly called nucleons, as constituent particles.
The Fujita-Miyazawa type three-nucleon force 
($3N$F), visualized by the Feynmann diagram 
in Fig.~\ref{fig:FM-3NF}, is driven by virtual
pion-nucleon scattering with an intermediate excitation of the nucleon
into the $\Delta (1232)$ resonance. 
Three-nucleon forces are thus intimately related 
to the inner structures of the nucleons and their short-distance virtual excitations.

Following the idea of Fujita and Miyazawa, one may expect three-body forces to play an important role 
in quantum systems other than atomic nuclei. With their high controllability, cold atoms have recently emerged as an
excellent platform to explore quantum systems of interests. It is possible to realize cold-atom systems where 
the three-body forces among atoms appear from similar exchanges of virtual excitations, thereby {\it quantum-simulating} the Fujita-Miyazawa mechanism. It is even possible to engineer the strength of 
the three-body force and thereby realize a system where the three-body force has significant effects on few- and many-body properties, in stark contrast to conventional systems where the two-body force dominates over the three-body force.

The emergence of a three-body force from exchange processes can lead to an even more exotic phenomenon, the Efimov effect. When three particles scatter with each other via resonant two-body interactions, a long-range three-body force emerges via multiple scatterings. The three-body force is attractive and forms a series of weakly bound three-body states, known as the Efimov states. The Efimov states have not only been observed in cold-atom experiments, but they appear universally in various physical systems. Thanks to its universal properties, Efimov physics provides a unified description of various classes of three-body phenomena in cold atoms and nuclear physics.

In this review article, we discuss three-body forces 
in nuclear systems and in cold atomic gases
as a key aspect for bridging different hierarchies. We present an overview of the three-body force in nuclear systems in \secref{sec:nuc3BF}. After briefly presenting historical developments of the three-body forces in \secref{sec:nuc3BF_hist}, we show in \secref{sec:nuc3BF_EFT} the effective field theory (EFT) description of the three-body force in nuclei, followed by a review of the experimental studies in \secref{sec:nuc3BF_exp}. In \secref{sec:atom3BF}, we present recent studies in the field of cold atoms to realize and simulate the three-body forces. In \secref{sec:Efimov}, we show the recent developments of Efimov physics in cold atoms and nuclear systems.

\section{\label{sec:nuc3BF}Three-body forces in nuclei}

\subsection{\label{sec:nuc3BF_hist}Three-nucleon force and its importance in nuclear phenomena}

%
Since Yukawa's meson theory proposed in 1935~\cite{Yukawa193548}, 
the nuclear force has been modeled in terms of meson exchange
interactions between nucleons. Beside the dominant two-nucleon forces, 
the three nucleon forces ($3N$Fs) have attracted an increasing attention in the last three decades. 
The development of high-precision two-nucleon potentials in the 90s
of the last century, coupled with advances in {\it ab initio} few-body
calculations based on these interactions, have confirmed the important
role played by $3N$Fs in various nuclear phenomena.  
Following the seminal work by Fujita and Miyazawa, a number of the
$3N$F models utilizing the longest-range $2\pi$-exchange mechanism
have been developed such as, e.g., the Tucson-Melbourne 99~\cite{tm99}, 
the Urbana IX~\cite{PhysRevC.56.1720} $3N$F models.
A new impetus to study $3N$Fs has come from chiral perturbation theory, 
the low-energy effective field theory of 
QCD~\cite{Weinberg:1990rz,kolk1994,Epelbaum:2002vt}.

Historically, the first indication for a missing $3N$F came from
the three-nucleon bound states $^3\rm H$ and $^3 \rm He$
\cite{PhysRevC.33.1740,FewBodySys.1.3}.
The binding energies of these nuclei were found to be not reproduced by 
an exact solution of the three-nucleon Faddeev equation using 
high-precision phe\-no\-me\-no\-logical $NN$ 
forces including the Argonne $V_{18}$ (AV18) \cite{AV18}, 
CD Bonn~\cite{cdb} as well as Nijmegen I and II~\cite{nijm} potentials.  
The underbinding of $^3$H and $^3$He could be 
explained by adding a $2\pi$-exchange-type 
$3N$F~\cite{PhysRevC.33.1740,FewBodySys.1.3,PhysRevC.65.054003}. 
The importance of $3N$Fs has also been noted in other instances. 
In particular, 
microscopic calculations of light and medium-mass nuclei carried out 
using {\it ab initio} methods such as, e.g., quantum Monte Carlo~\cite{PhysRevC.66.044310,Piarulli:2017dwd}, no-core shell
model~\cite{PhysRevC.68.034305}, coupled cluster theory
\cite{Hagen:2012sh}, self-consistent Green's function method
\cite{Cipollone:2014hfa} and nuclear lattice simulations
\cite{Lahde:2019npb} highlight the important role of $3N$Fs 
in explaining the corresponding binding energies. 
Furthermore, short-range repulsive $3N$Fs are considered key elements 
in describing the nuclear equation of state and two-solar-mass neutron star properties
\cite{PhysRevC.58.1804,Gandolfi:2013baa,PhysRevLett.116.062501}.
In the past two decades, low-energy nucleon-deuteron scattering, binding energies 
of light and medium-mass nuclei as well as the equation of state of nuclear matter 
have also been extensively 
studied in the framework of the chiral effective field theory~\cite{epelbaum2009,Kalantar-Nayestanaki_2012,Hammer:2012id,PhysRep21Hebeler}.
In all these investigations, it became evident that $3N$Fs 
are key elements to understand various nuclear phenomena.

Discussions of three-body forces in nuclei currently 
extend to strange baryonic systems, e.g.~$NN\Lambda$ interactions,
especially for the neutron star properties~\cite{Haidenbauer:2016vfq}, 
which are needed to establish a universal understanding of the forces
acting in nuclear phenomena.
Also, it is notable that discussions of $3N$Fs 
stimulate the investigation of three-body forces in different hierarchies.
As described in \secref{sec:atom3BF},
study of three-body forces in the cold-atom systems
are in progress not only from the theoretical but also from the
experimental point of view.


\subsection{\label{sec:nuc3BF_EFT}Three-nucleon forces in chiral effective field theory}

\subsubsection{Chiral perturbation theory}

The interactions and dynamics of pions can be described using the most
general effective Lagrangian $\mathcal{L}_{\pi}$ that features the approximate chiral
symmetry of QCD. It includes all possible terms allowed by symmetry, multiplied with coefficients that are
commonly referred to as low-energy constants (LECs). These LECs are not
fixed by symmetry and carry information about short-range
QCD dynamics. It is convenient to classify terms in the
effective Lagrangian according to the number of derivatives and/or
insertions of the pion mass $M_\pi$: $\mathcal{L}_{\pi} =
\mathcal{L}_{\pi}^{(2)} +  \mathcal{L}_{\pi}^{(4)} + \ldots$.
Multi-pion scattering amplitudes
for the physically intersting case of $p \sim M_\pi \neq 0$ can be
calculated from the effective Lagrangian using chiral perturbation
theory (ChPT) \cite{Weinberg:1978kz,Gasser:1983yg}, i.e.~via a perturbative expansion
in $Q \in \{ p/\Lambda_b, \,
M_\pi/\Lambda_b \}$, where $\Lambda_b$ is the breakdown scale of ChPT,
which may be estimated by the masses of lowest-lying resonances in
the $\pi\pi$ system.  

The perturbative approach outlined above can be straightforwardly
generalized to processes involving a single non-Goldstone-boson
particle such as, e.g., the nucleon. The most general pion-nucleon
effective Lagrangian $\mathcal{L}_{\pi N} = \mathcal{L}_{\pi N}^{(1)} +
\mathcal{L}_{\pi N}^{(2)} + \mathcal{L}_{\pi N}^{(3)} + \ldots$ can be constructed using the methods described in
Refs.~\cite{Weinberg:1968de,Coleman:1969sm,Callan:1969sn}. The
explicit form of $\mathcal{L}_{\pi N}^{(1)}$ and $\mathcal{L}_{\pi
  N}^{(2)}$ can be found, e.g., in Ref.~\cite{Bernard:1995dp}, while
the Lagrangians $\mathcal{L}_{\pi
  N}^{(3)}$ and $\mathcal{L}_{\pi
  N}^{(4)}$ are given in Refs.~\cite{Fettes:1998ud} and \cite{Fettes:2000gb}, respectively. 
Compared to the Goldstone-boson sector, special care is required for
processes involving a nucleon to ensure that the hard scale set by the
nucleon mass $m_N$ does not spoil the chiral power counting when
computing loop diagrams. The simplest way to achieve this is to
perform a non-relativistic expansion of the effective Lagrangian
$\mathcal{L}_{\pi N}$ \cite{Jenkins:1990jv,Bernard:1992qa}. This ensures that $m_N$ appears in
$\mathcal{L}_{\pi N}$ only in the form of $1/m_N^n$-corrections with
$n>0$, and the corresponding framework is referred to as the
heavy-baryon ChPT.  It is also possible
to perform calculations in a manifestly Lorentz-invariant way by
choosing the appropriate renormalization conditions \cite{Becher:1999he,Gegelia:1999gf,Fuchs:2003qc}. 

\subsubsection{Chiral EFT for nuclear systems}
\label{sec:ChEFT}

Clearly, a direct application of ChPT to systems involving two and more nucleons
is not possible due to the non-perturbative nature of the nuclear
interactions, as reflected in the existence of shallow bound states
such as $^2$H, $^3$H, $^3$He, $^4$He, etc. These bound states manifest
themselves as subthreshold poles of the corresponding scattering
amplitudes and signal the breakdown of the perturbative
expansion.

In the early 1990s, Weinberg came up with an approach that is based
on the effective Lagrangian and allows one to analyze low-energy few- and
many-nuc\-leon systems in a model-independent and systematic
fashion \cite{Weinberg:1990rz,Weinberg:1991um}, which is nowadays commonly referred to as chiral effective field theory (ChEFT).
He 
has
attributed the failure of perturbation theory
in the 
few-nucleon sector
to the appearance of enhanced
few-nucleon-reducible ladder-type diagrams and argued that they need
to be resummed non-per\-tur\-ba\-ti\-ve\-ly. He also noticed that a
resummation of ladder-type diagrams is performed automatically by
solving the corresponding Lippmann-Schwinger-type integral equa\-ti\-ons
for the scattering amplitude. Thus, Weinberg's ChEFT approach to
few-nucleon systems technically reduces to the conventional A-body problem, 
\begin{equation}
\bigg[ \bigg(\sum_{i=1}^A \frac{- \vec \nabla_i^2}{2 m_N} + \mathcal{O}
(m_N^{-3} )\bigg) + V_{NN} + V_{3N} + \ldots \bigg] \big| \Psi
\big\rangle = E \big| \Psi \big\rangle ,
\end{equation}  
but it opens the possibility to derive nuclear interactions $V_{NN}$,
$V_{3N}$, $V_{4N}$, $\ldots$, via a {\it systematically improvable} 
ChPT expansion in harmony with the symmetries of QCD. Here, nuclear
potentials
are defined by means of all
possible few-nucleon-irreducible contributions to the scattering
amplitude, which are not affected by the above-mentioned enhancement
and can be derived in the framework of ChPT using a variety of
methods.
Following
Weinberg's original work \cite{Weinberg:1990rz,Weinberg:1991um},
time-ordered perturbation theory was employed in Refs.~\cite{Ordonez:1995rz,Pastore:2008ui,Pastore:2009is,Pastore:2011ip,Baroni:2015uza}
to derive nuclear forces and electroweak currents. Another
approach, the so-called method of unitary transformation (MUT), ma\-kes
use of a unitary transformation of the pion-nucleon Hamiltonian to
decouple the purely nucleonic subspace of the Fock space from the
rest. The MUT was applied to derive few-nucleon forces as well as
electroweak and scalar nuclear currents in   
Refs.~\cite{Epelbaum:1998ka,Epelbaum:1999dj,Epelbaum:2002gb,Epelbaum:2005fd,Epelbaum:2005bjv,Epelbaum:2007us,Bernard:2007sp,Bernard:2011zr,Krebs:2012yv,Krebs:2013kha,Kolling:2009iq,Kolling:2011mt,Krebs:2016rqz,Krebs:2019aka,Krebs:2020plh}.
Yet
another me\-thod to derive the $NN$ two- and three-pion exchange
potentials from matching to the scattering amplitude was applied in  
Refs.~\cite{Kaiser:1997mw,Kaiser:1998wa,Kaiser:1999ff,Kaiser:1999jg,Kaiser:2001dm,Kaiser:2001pc,Kaiser:2001at,Entem:2015xwa}. For a detailed discussion of these techniques and applications the reader
is referred to the review articles
\cite{Epelbaum:2005pn,epelbaum2009,Machleidt:2011zz,Epelbaum:2019kcf,Krebs:2020pii,deVries:2020iea}.

Non-perturbative resummations of reducible di\-ag\-rams in ChEFT pose complications as compared with ChPT in the Goldstone-boson or
single-nucleon sectors. Consider the longest-range $NN$ force due to the one-pion exchange 
\begin{equation}
  \label{OPEP}
V_{NN}^{1 \pi} (\vec q \, )= - \frac{g_A^2}{4 F_\pi^2} \frac{\vec \sigma_1 \cdot
  \vec q \vec \sigma_2 \cdot \vec q}{q^2 + M_\pi^2}  \mbox{\boldmath $\tau$}_1 \cdot \mbox{\boldmath $\tau$}_2\,,
\end{equation}
where $g_A$ and $F_\pi$ are the nucleon axial vector coupling and the
pion decay constant, respectively. Further, $\vec \sigma_i$
($\mbox{\boldmath $\tau$}_i$) denote the spin (isospin) Pauli matrices of
nucleon $i$, while $\vec q = \vec p \, '
- \vec p$, with $\vec p$ ($\vec p \, '$) being the nucleon
center-of-mass (CM) momentum in the initial (final) state, is the nucleon
momentum transfer. The one-pion exchange potential (OPEP) contributes to the leading-order (i.e., order-$Q^0$)
nuclear force and thus needs to be resummed non-perturbatively.
However, the $1/r^3$ singularity in the tensor part of the OPEP 
implies the appearance of ultraviolet divergences in all
  spin-triplet channels upon performing iterations of the Lippmann-Schwinger (LS)
  equation. That is, removing ultraviolet divergences from the iterative solution of the
  LS equation requires the introduction of an infinite number of counter terms from
  the Lagrangians $\mathcal{L}_{NN}^{(0)}$,  $\mathcal{L}_{NN}^{(2)}$,
  $\mathcal{L}_{NN}^{(4)}$, $\ldots$. This is in strong contrast to
  ChPT, where a finite number of counter terms are required to remove
  ultraviolet divergences at any given order. 

Clearly, the singular
short-distance nature of $V_{NN}^{1 \pi}(\vec q \, )$ is unphysical and
represents an artifact of using the low-momentum approximation in
Eq.~(\ref{OPEP}) beyond its validity range (i.e., at short distances
or large values of the momentum transfer). Renormalization of the
Schr\"odinger equation in EFTs with singular interactions like
$V_{NN}^{1 \pi}(\vec q \, )$ can be achieved in the way compatible
with the weak-interaction limit by introducing a
finite cutoff $\Lambda$ of the order of the pertinent hard
scale, $\Lambda \sim \Lambda_b$ \cite{Lepage:1997cs}. At each order of
ChEFT and for every value of $\Lambda$, renormalization is carried out
implicitly by expressing bare LECs from $\mathcal{L}_{NN}$,
$\mathcal{L}_{NNN}$, $\ldots$, in terms of measurable quantities. In
practice, this is achieved by tuning few-nucleon contact interactions
to experimental data. The residual dependence of the renormalized
results on the cutoff $\Lambda$ serves as  a measure of the neglected
contributions of terms beyond the EFT truncation level. It is expected
to decrease with the chiral order and provides an {\it a
posteriori} consistency check. In Refs.~\cite{Gasparyan:2021edy,Gasparyan:2023rtj}, the finite-$\Lambda$
formulation of ChEFT was formally proven to be renormalizable (in the
EFT sense) up to next-to-leading order $Q^2$ (NLO). Here and in what
follows, we restrict ourselves to the finite-cutoff formulation of
ChEFT as described in detail in Ref.~\cite{Epelbaum:2019kcf}. 
A pedagogical introduction into the considered framework can be found
in Sec.~6.3 of Ref.~\cite{Gross2022hyw}, while a discretized
(lattice) formulation is described in Ref.~\cite{Lahde:2019npb}. For a collection of different views on renormalization and power counting issues in nuclear effective field theories see Ref.~\cite{Tews:2022yfb} and references therein.

\subsubsection{Current status of the $NN$ potentials}
\label{sec:NN}

Presently, the $NN$ force has been worked out completely up through
fifth order $Q^5$ (i.e., N$^4$LO) including isospin-breaking corrections due to
$m_u \neq m_d$ and QED effects. It involves the one- two-
and three-pion exchange potentials, supplemented by short-range NN
interactions from $\mathcal{L}_{NN}^{(0)}$,  $\mathcal{L}_{NN}^{(2)}$,
and $\mathcal{L}_{NN}^{(4)}$, which depend on $2$, $7$ and
  $15$\footnote{The above numbers refer to isospin-invariant contact interactions.} LECs,
  respectively, that need to be tuned to $NN$ data. Notice that $3$ out
  of $15$ contact interactions at N$^4$LO contribute only to the off-shell
  part of the $NN$ potential and thus cannot be determined from
  two-nucleon scattering data. In the NN potentials of Refs.~\cite{Reinert:2017usi,Reinert:2020mcu}, these off-shell LECs are set to zero. To analyze $A \geq 3$ nuclear systems at N$^3$LO, these LECs will have to be determined from experimental data beyond the NN system, such as, e.g., three-nucleon scattering data.

\begin{figure}[tbp]
\centering
  \includegraphics[width=0.49\textwidth]{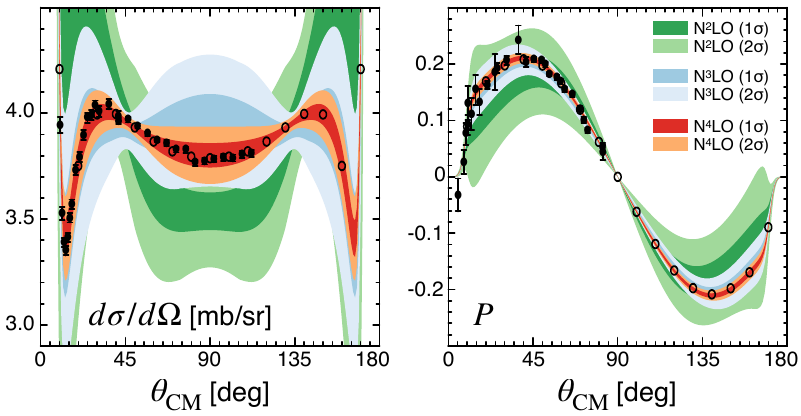}
\caption{\label{fig:pp}
 Proton-proton differential cross section at $E_{\rm lab} = 144.1$~MeV
 (left panel) and the analyzing power $P$ at $E_{\rm lab} = 142$~MeV
 (right panel) at N$^2$LO (green bands), N$^3$LO (blue bands) and
 N$^4$LO$^+$ (red bands) in ChEFT. Dark- and light-shaded bands show
 $1\sigma$ and $2\sigma$ confidence levels at the corresponding order, respectively, estimated
 using the Bayesian model $\bar {\rm C}^{650}_{0.5-10}$ from
 Ref.~\cite{Epelbaum:2019zqc}. Experimental data are shown by filled symbols and taken
 from Refs.~\cite{Cox:1968jxz,Jarvis:1971fla,TAYLOR1960320}. Open circles are the results of the Nijmegen
 partial wave analysis \cite{Stoks:1993tb}. See Ref.~\cite{Epelbaum:2019kcf} for more details.}
\end{figure}

  While
  semi-quantitative ChEFT potentials have a long history that goes
  back to the beginning of 2000s, see Refs.~\cite{Epelbaum:2004fk,Entem:2003ft} for the
  first-generation order-$Q^4$ (i.e., N$^3$LO) $NN$ potentials constructed
  using non-local regulators, the field has received new impetus in
  the last decade by developing a local regularization scheme for pion-exchange
  contributions that preserves the long-range behavior of the nuclear
  force, which allowed to significantly improve the predictive power of ChEFT \cite{Epelbaum:2014efa,Epelbaum:2014sza}.  
  The state-of-the-art $NN$ N$^4$LO$^+$ potential of Ref.~\cite{Reinert:2017usi}
  employs a local momentum-space regulator for the OPEP and
  two-pion exchange potential (TPEP) via
  \begin{eqnarray}
    \label{reg_opep}
V_{NN}^{1\pi} (\vec q \, ) &\to & 
                                  V_{NN}^{1\pi} (\vec q \, ) \,  e^{-
                                  \frac{q^2 + M_\pi^2}{\Lambda^2}} +
                                  \ldots\,,\\
      \label{reg_tpep}
 V_{NN}^{2\pi} (q \, ) &\to &  \frac{2}{\pi} \int_{2
                                   M_\pi}^\infty d\mu \mu
                                   \frac{\rho_{2 \pi}
                                   (\mu)}{q^2 + \mu^2} \,  e^{-
                                  \frac{q^2 + \mu^2}{2\Lambda^2}} + \ldots\,,
    \end{eqnarray}
where
 ``$\ldots$''
denote subtractions in the form of $NN$ contact
interactions, chosen in such a way that the corresponding $r$-space
potentials vanish at the origin. In the second equation, we have used
a dispersive representation of the TPEP \cite{Kaiser:1997mw}, where the spectral function 
$\rho_{2 \pi} (\mu )$ is obtained from an analytic continuation of the
unregularized potential $V_{NN}^{2 \pi} (q)$ to imaginary values of
$q$ via $\rho_{2 \pi} (\mu) :=
\Im \big[ V_{NN}^{2 \pi} (0^+ - i \mu)\big]$. Equations (\ref{reg_opep}) and
(\ref{reg_tpep}) show that all finite-$\Lambda$ artifacts stemming from
the $1/\Lambda$-expansion of the regularized OPEP and TPEP have the
form of short-range contact interactions. 
The latter are regularized in Ref.~\cite{Reinert:2017usi} using a Gaussian
non-local regulator $\exp[-(p^2 + p'^2)/\Lambda^2]$.  The resulting
semi-local momentum-space regularized (SMS) $NN$ potentials are available
at orders LO through N$^4$LO and cutoff values of $\Lambda =
400$, $450$, $500$ and $550$ MeV.
The N$^4$LO$^+$ potentials lead to a
statistically perfect description of the Granada-2013 database of
mutually consistent neutron-proton and proton-proton scattering data
below pion production threshold \cite{NavarroPerez:2013usk} and
show, as expected, a very weak residual $\Lambda$-dependence.    
In Ref.~\cite{Reinert:2020mcu}, the SMS $NN$ potential of
Ref.~\cite{Reinert:2017usi} was updated to include
all relevant isospin-breaking contributions up through N$^4$LO,  and it was
used for  a
precision determination of the $\pi N$ coupling
constants from $NN$ scattering and to perform a full fledged partial wave
analysis of $NN$ data  (including a
selection of mutually compatible data), see
Ref.~\cite{Epelbaum:2022cyo} for details and Fig.~\ref{fig:pp} for 
representative examples

These new developments allowed one to test the
predictive power of ChEFT by  quantitatively
addressing the impact of the TPEP. In Refs.~\cite{Epelbaum:2014efa,Epelbaum:2014sza,Reinert:2017usi}, a clear evidence of
the parameter-free contributions of the TPEP was observed at orders $Q^3$
(N$^2$LO) and $Q^5$ (N$^4$LO), where no additional $NN$ contact
interactions appear. A significantly smaller number of
adjustable parameters in the SMS potential of
Refs.~\cite{Epelbaum:2014sza,Reinert:2017usi} as compared
to the phenomenological high-precision potential models provides yet another
indication of the importance of the TPEP and demonstrates the predictive power of ChEFT. 
For related earlier studies along this line see Refs.~\cite{Rentmeester:1999vw,Birse:2003nz}. 

Other notable recent additions to the ChEFT $NN$ potentials involve the nonlocal
N$^4$LO$^+$ potentials by the Idaho group \cite{Entem:2017gor} as well
as the (nearly) local N$^3$LO potentials of
Refs.~\cite{Piarulli:2014bda,Somasundaram2023sup,Saha:2022oep}. 

\subsubsection{Chiral expansion of the $3N$F}
\label{sec:3NFTheory}

We now turn to the main subject of this article and review the
applications of ChEFT to the $3N$F. It is instructive to first discuss 
the most general structure of a $3N$F. In the static limit of infinitely
heavy nucleons, the potentials mediated by the exchange of one or
multiple pions take a local form, i.e.~they depend only on the
momentum transfers $\vec q_i$ and not on the individual momenta $\vec
p_i$, $\vec p_i^{\, \prime}$  of the nucleons. 
Assuming parity, time-reversal invariance and isospin symmetry, 
the most general local $3N$F can be written as \cite{Phillips:2013rsa,Epelbaum:2014sea}
\begin{equation}
\label{20structures}  
V_{3N} = \sum_{\alpha = 1}^{20} \hat O_\alpha \, f_\alpha (q_1,  q_2,  q_3)
\text{ + permutations}\,,
\end{equation}
where $q_i \equiv | \vec q_i \,|$, $\hat O_\alpha$,  are rotationally
and isospin-inva\-ri\-ant Hermitian operators
constructed out of $\vec \sigma_i$, $\mbox{\boldmath $\tau$}_i$ and
$\vec q_i$, while $f_\alpha$ are the corresponding scalar functions.
Upon performing the permutations, the  $20$ operators $\hat O_\alpha$ give rise to
$80$ different spin-isospin-momentum structures.  
When relaxing the locality constraint, the structure of the $3N$F
becomes more involved, comprising $320$ spin-isospin-momentum
operators \cite{Topolnicki:2017rnt}. This enormous complexity of
three-nucleon interactions, along with the significant computational
effort needed to solve the three-body Faddeev equations, make the
development of high-precision $3N$F models a challenging task that
requires a guidance from theory to constrain the structure and
identify the dominant contributions. ChEFT is  well su\-i\-ted
to tackle the $3N$F challenge by {\it predicting}
its long-distance behavior in a parameter-free and model independent
way and offering a systematic scheme for classifying short-range
$3N$ interactions according to their importance. Based on the effective
Lagrangian for pions, nucleons and external sources, ChEFT also naturally
allows one to maintain off-shell consistency between $NN$ potentials, $3N$Fs and the
corresponding current operators as discussed in Sec.~\ref{sec:ChEFT}.   

\begin{figure}[tbp]
\centering
  \includegraphics[width=0.49\textwidth]{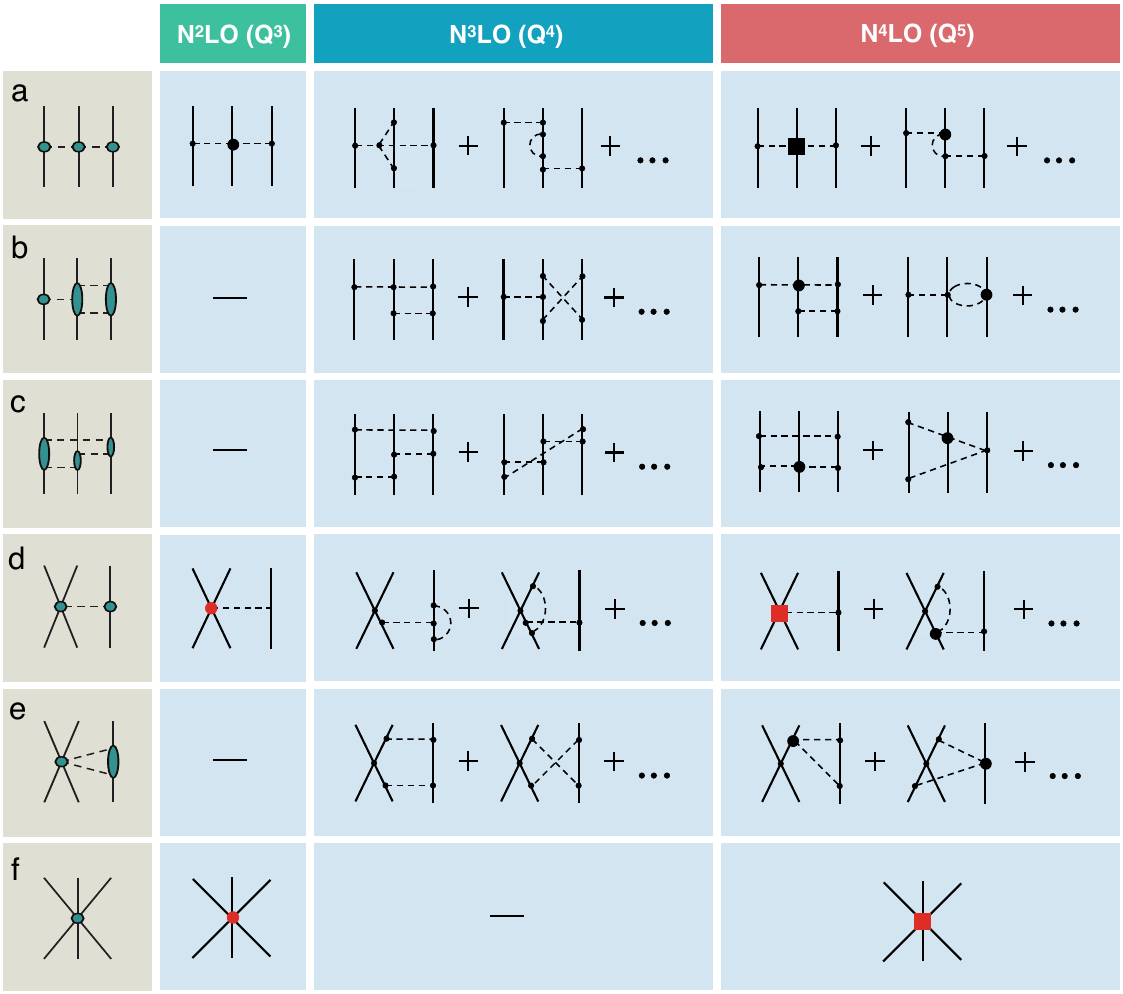}
\caption{\label{fig:3NFchiral}
  Chiral expansion of the $3N$F.
Solid and dashed lines refer to nucleons
and pions, respectively. 
  Diagrams in the leftmost column visualize
the two-pion exchange (a), two-pion-one-pion exchange (b), ring (c),
one-pion-contact (d), two-pion-contact (e) and purely contact (f)
$3N$F topologies. Cyan circles and ellipses in these diagrams represent 
scattering amplitudes of the corresponding sub-processes.  
The second, third and fourth columns show examples of diagrams, which
contribute to the individual $3N$F topologies at increasing orders of the ChEFT expansion. 
Here, solid dots, filled circles and filled squares
denote vertices from the effective Lagrangian of increasing chiral
dimension $\Delta$ as explained in the text. Vertices shown in red involve LECs that need to be determined
from three- or more-nucleon data. }
\end{figure}

Up-to-and-including N$^4$LO, the $3N$F is given by 
tree-level and one-loop diagrams, which can be grouped into six
distinct topologies depicted in the left column of Fig.~\ref{fig:3NFchiral}. 
In this figure, solid dots, filled circles and filled squares denote
vertices from the effective Lagrangians of the dimension $\Delta = 0$,
$\Delta = 1$ and $\Delta = 2$, respectively, defined as $\Delta = d +
\frac{1}{2} n-2$ with $d$ being the number of derivatives or
$M_\pi$-insertions and $n$ the number of nucleon fields
\cite{Weinberg:1990rz}.

The leading $3N$F contributions arise at N$^2$LO from tree-level
diagrams contributing to the topologies (a), (d) and (f), which are
made out of the $\Delta=0$-vertices and a single insertion of a
$\Delta = 1$ interaction \cite{Weinberg:1992yk,kolk1994}. The
short-range topologies (d) 
and (f) depend on the LECs $c_D$ and $c_E$, respectively, which cannot
be fixed in the $NN$ system.  

The first corrections to the leading $3N$F are generated by one-loop
graphs made out of the lowest-order vertices with $\Delta = 0$, see
the third column in Fig.~\ref{fig:3NFchiral} for representative
examples. Here, the shown diagrams represent
sets of irreducible time-ordered-like graphs, whose
precise meaning (and the corresponding algebraic expressions) depend
on the employed choice for the off-shell part of the $NN$ and $3N$ potentials.
For example, the $3N$F from the first of
the two three-pion exchange 
diagrams of type (c) depends on the (ambiguous) choice made for the
$1/m_N^2$-correction to the OPE $NN$ potential and the $1/m_N$-corrections to the
tree-level two-pion exchange $3N$F of type (a), which appear at the same
order.

The N$^3$LO contribution
to the longest-range topology (a) was derived in Ref.~\cite{Ishikawa:2007zz} based on
the order-$Q^3$ $\pi N$ amplitude. These results were confirmed using
the MUT in Ref.~\cite{Bernard:2007sp}, where also the expressions for the topologies
(b) and (c) were derived. The shorter-range $3N$Fs of type (d) and (e),
along with the $1/m_N$-corrections to the topologies (c) and (d),
are worked out in Ref.~\cite{Bernard:2011zr}. All results mentioned
above are obtained using dimensional regularization (DimReg) to evaluate 
divergent loop integrals and are parameter-free.\footnote{As already
  mentioned in Sec.~\ref{sec:NN}, the N$^3$LO $NN$ potential in the CM system
  depends on $3$ off-shell LECs. Two further off-shell LECs contribute
  to the $NN$ potential away from the CM system
  \cite{Girlanda:2020pqn}. These LECs, being redundant in the NN
  system, will generally affect $3N$ observables calculated at N$^3$LO \cite{Girlanda:2023znc}.
  Their contributions to $3N$ observables can,
  however, be absorbed into redefinitions of LECs accompanying
  short-range $3N$F at N$^4$LO and, therefore, can be ignored
  beyond the N$^3$LO level.} Finally, it is worth mentioning that one can
also draw tree-level $3N$ diagrams constructed from the leading
$\Delta=0$-vertices and a single insertion of a $\Delta=2$-interaction
from $\mathcal{L}_{\pi N}^{(3)}$ that could potentially contribute to
the $3N$F at N$^3$LO. However, all irreducible contributions generated
by such diagrams either contribute to renormalization of the $\pi N$
coupling constant or vanish. 

Subleading corrections to the $3N$F at N$^4$LO are visualized in the last
column of Fig.~\ref{fig:3NFchiral} and comprise one-loop diagrams made
out of the leading $\Delta=0$-vertices and a single insertion of a
$\Delta=1$-interaction from $\mathcal{L}_{\pi N}^{(2)}$, as well as
tree-level graphs from the lowest-order interactions and a single insertion of a
$\Delta=3$-vertex. The longest-range two-pion exchange $3N$F and 
the intermediate-range N$^4$LO contributions of types (b), (c) are derived
using the MUT in Refs.~\cite{Krebs:2012yv} and \cite{Krebs:2013kha},
respectively.  The corresponding potentials do not involve any unknown
LECs. Short-range $3N$F terms of type (f) are considered in Ref.~\cite{Girlanda:2011fh}
and depend on $13$ unknown LECs. Finally, one-loop contributions to the
topologies (d) and (e) at N$^4$LO are still to be worked out (and 
involve further unknown LECs).  

In addition to the isospin-invariant $3N$F contributions discussed above
and depicted in Fig.~\ref{fig:3NFchiral}, one also has to account for 
isospin-breaking corrections stemming from different masses of the up
and down quarks and QED effects. The expressions for isospin-violating
$3N$F up through N$^4$LO are worked out using the
MUT in Ref.~\cite{Epelbaum:2004xf}, see also Ref.~\cite{Friar:2004ca} for a related
work. Charge-dependent $3N$Fs involving virtual photon ex\-change are
considered in Refs.~\cite{Yang:1979zz,Yang:1983pd} and found to be rather weak. 
Parity- and time-reversal-violating $3N$Fs have also been studied,
see the review article \cite{deVries:2020iea} and references therein. 

To demonstrate the predictive power of ChEFT we consider below the
chiral expansion of the longest-range two-pion exchange $3N$F topology
(a) as a representative example. We restrict ourselves to
isospin-invariant contributions in the static limit, whose most general structure
is given by 
\begin{eqnarray}
\label{2pi_general}
V^{(a)}_{3N} &=& \frac{\vec \sigma_1 \cdot \vec q_1\,  \vec \sigma_3 \cdot
  \vec q_3}{(q_1^2 + M_\pi^2 ) \, (q_3^2 + M_\pi^2 )}  \nonumber  \\
&\times & \Big[ 
  \mbox{\boldmath $\tau$}_1 \cdot \mbox{\boldmath $\tau$}_3  \, {\cal A}(q_2) + \mbox{\boldmath $\tau$}_1 \times  \mbox{\boldmath $\tau$}_3 \cdot \mbox{\boldmath $\tau$}_2  \,   \vec q_1 \times  \vec q_3   \cdot \vec
          \sigma_2 \, {\cal B}(q_2) \Big]   \nonumber \\
&+ & 
\text{ short-range terms}\; + \; \text{permutations}\,,            
\end{eqnarray}
where the functions ${\cal A}$ and ${\cal B}$ depend on the
momentum transfer $q_2 \equiv | \vec q_2 |$ of the second
nucleon. These functions govern the long-distance behavior of the $3N$F
and are to be determined by means of the chiral expansion 
\begin{eqnarray}
{\cal A} (q_2) &=&  {\cal A}_{[Q^3]} (q_2)+ {\cal A}_{[Q^4]} (q_2)+ {\cal
                   A}_{[Q^5]} (q_2)+ \ldots , \nonumber \\
{\cal B} (q_2) &=&  {\cal B}_{[Q^3]} (q_2)+ {\cal B}_{[Q^4]} (q_2)+ {\cal
                   B}_{[Q^5]} (q_2)+ \ldots .
\end{eqnarray}  
The dominant contributions to ${\cal A} (q_2)$ and ${\cal B} (q_2)$ at N$^2$LO,
stemming
from the tree diagram in the second column of
Fig.~\ref{fig:3NFchiral}, have the form  \cite{Epelbaum:2002vt,kolk1994}
\begin{eqnarray}
  \label{ABN2LO}
{\cal A}_{[Q^3]} &=& \frac{g_A^2}{8 F_\pi^4} \Big[(2 c_3 - 4 c_1) M_\pi^2 +  c_3 
q_2^2  \Big], \nonumber \\
{\cal B}_{[Q^3]} &=& \frac{g_A^2 c_4}{8 F_\pi^4} , 
\end{eqnarray}  
where $c_i$ denote the $\pi N$ LECs from the subleading Lagrangian
$\mathcal{L}_{\pi N}^{(2)}$.

The leading corrections to Eq.~(\ref{ABN2LO}) are generated at N$^3$LO by
 one-loop diagrams shown in the first line and third column of
 Fig.~\ref{fig:3NFchiral}. Using DimReg, one obtains \cite{Bernard:2007sp,Ishikawa:2007zz}
 \begin{eqnarray}
    \label{ABN3LO}
{\cal A}_{[Q^4]} &=& \frac{g_A^4}{256 \pi  F_\pi^6} \Big[\left(4 g_A^2+1\right) M_\pi^3+2 \left(g_A^2+1\right) M_\pi 
                     q_2^2\nonumber \\
  &+&
                     A(q_2)
\left(2 M_\pi^4+5 M_\pi^2 q_2^2+2 q_2^4 \right) \Big], \\
{\cal B}_{[Q^4]} &=& -\frac{g_A^4 }{256 \pi  F_\pi^6} \big[A(q_2) \left(4 M_\pi^2+q_2^2\right)+(2 g_A^2 
                   +1)M_\pi\big],
                   \nonumber 
\end{eqnarray}
where we have introduced the loop function 
\begin{equation}
A(q_2) = \frac{1}{2 q_2} \arctan \frac{q_2}{2 M_\pi}.
\end{equation}  
The loop function $A(q_2)$ possesses a left-hand cut with the branch point at
$q_2^2 = - (2M_\pi)^2$, which corresponds to the kinematics when both
pions inside the loop of, e.g., the first N$^3$LO diagram can become
on-shell. On the other hand, contributions from diagrams like the second N$^3$LO graph
do not have left-hand cuts and are polynomial in $q_2^2$. Notice
further that loop integrals at N$^3$LO involve only linear
divergences, which vanish in DimReg (and
would have been absorbed into the LEC $c_D$ when using
momentum-dependent regularization schemes). This is consistent with
the already mentioned absence of tree-level $3N$F contributions
involving a single insertion from $\mathcal{L}_{\pi N}^{(3)}$. 

Finally, the N$^4$LO result for the functions ${\cal A} (q_2)$ and ${\cal
  B} (q_2)$, obtained using DimReg, has the form
\cite{Krebs:2012yv}
\begin{eqnarray}
  \label{ABN4LO}
  {\cal A}_{[Q^5]} &=& \frac{g_A^2 \bar e_{14}}{2 F_\pi^4}  \big(2 M_\pi^2 + q_2^2
     \big)^2 +  \frac{g_A^2 \big(M_\pi^2 + 2 q_2^2 \big)}{4608 \pi^2
                       F_\pi^6}
                       \nonumber \\
  &\times&     \Big\{ \big[6 c_1 - 2 c_2 - 3 c_3
   - 2 (6c_1 - c_2 -
  3c_3)  L(q_2) \big] \nonumber \\
  &\times & 12 M_\pi^2  -q_2^2 \big[ 5 c_2 + 18 c_3 - 6    L(q_2) (c_2 + 6 c_3)  \big]\Big\} ,
                          \nonumber \\
  {\cal B}_{[Q^5]} &=& \frac{g_A^2 \bar e_{17}}{2 F_\pi^4} \big( 2
                           M_\pi^2 + q_2^2 \big)                 \;
                           -\;   \frac{g_A^2 c_4}{2304 \pi^2 F_\pi^6} \\
&\times&                           \Big\{q_2^2 \big[5  - 6 L(q_2) \big]+
                           12 M_\pi^2 \big[2 + 9
         g_A^2 - 2 L(q_2) \big] \Big\},
         \nonumber 
\end{eqnarray}
where $\bar e_{14}$ and $\bar e_{17}$ are renormalized LECs from $\mathcal{L}_{\pi
  N}^{(4)}$, evaluated in the $\overline{\rm MS}$ scheme with the
renormalization scale $\mu$ set to $\mu =
M_\pi$. Further, the loop function $L(q_2)$ is given by 
\begin{equation}
  L(q_2)  =  \frac{\sqrt{q_2^2 + 4 M_\pi^2}}{q_2} \log \frac{\sqrt{q_2^2 + 4
    M_\pi^2} + q_2}{2 M_\pi},
\end{equation}  
and it also possesses a left-hand cut that starts at $q_2^2 = - (2M_\pi)^2$. 
Notice that in Eq.~(\ref{ABN4LO}), we
have applied the shifts of the LECs $c_i$ specified in Eq.~(16) of Ref.~\cite{Siemens:2016hdi} to
eliminate certain redundant linear combinations of LECs. 

Given that all LECs entering Eqs.~(\ref{ABN2LO})-(\ref{ABN4LO}) are known from
$\pi N$ scattering, which serves as a sub-process for the $3N$F topology
(a), the above expressions for ${\cal A} (q_2)$ and
${\cal B} (q_2)$ are to be regarded as parameter-free 
predictions of ChEFT. To assess the convergence of the chiral
expansion for the longest-range $3N$F, we employ the numerical values
for the pion mass and decay constant of $M_\pi = 138.03$~MeV and $F_\pi =
92.2$~MeV. The nucleon axial coupling $g_A$ is set to the effective value of
$g_A=1.289$ that accounts for the Goldberger-Treiman
discrepancy. The most reliable values of the higher-order $\pi N$ LECs
are obtained from matching ChPT with the solution of the dispersive Roy-Steiner
equations for  $\pi N$ scattering at the subthreshold kinematical point, see
Refs.~\cite{Siemens:2016jwj,Hoferichter:2015tha} for details. In the following, we employ the central values from the
order-$Q^4$ heavy-baryon-NN fit of Ref.~\cite{Siemens:2016jwj}: $c_1 = -1.11$, $c_2 = 3.61$,
$c_3 = -5.60$, $c_4 = 4.26$, $\bar e_{14}= 1.16$ and $\bar e_{17} =
-0.17$. Here, the values of $c_i$ and $\bar e_{i}$ are given in units
of GeV$^{-1}$ and GeV$^{-3}$, respectively.

\begin{figure}[tbp]
\centering
  \includegraphics[width=0.49\textwidth]{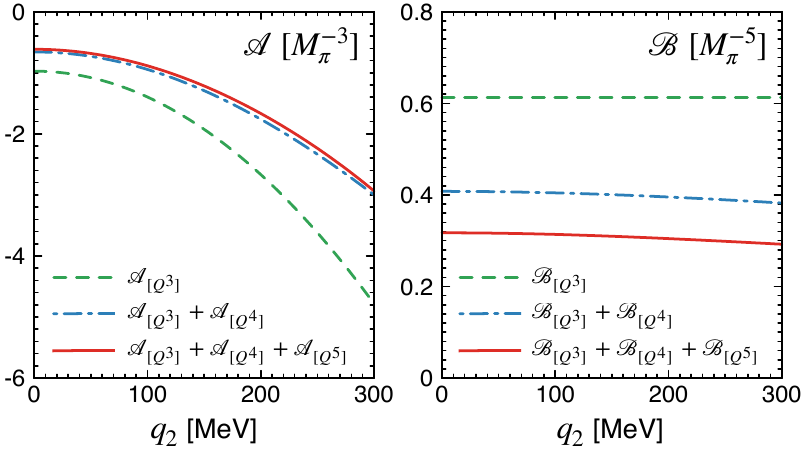}
\caption{\label{fig:AB}
 ChEFT predictions for the functions ${\cal{A}} (q_2)$ and ${\cal{B}} (q_2)$
 that parametrize the longest-range behavior of the $3N$F according to
 Eq.~(\ref{2pi_general}). Green dashed, blue dashed-dotted and red
 solid lines show the results at N$^2$LO, N$^3$LO and N$^4$LO, respectively. }
\end{figure}

In Fig.~\ref{fig:AB}, we
show the predicted behavior of the functions ${\cal A} (q_2)$ and
${\cal B} (q_2)$ at different orders in ChEFT. 
In both cases, the order-$Q^4$ corrections to the dominant order-$Q^3$
contributions amount to about $30\%$. The order-$Q^5$ correction is very
small for ${\cal{A}} (q_2)$ and amounts to less than $15\%$ of the N$^2$LO
result for the function ${\cal{B}} (q_2)$. The observed convergence
pattern for ${\cal{A}} (q_2)$ and ${\cal{B}} (q_2)$ fits well with
expectations based on the power counting with the expansion
parameter $Q \sim {\rm max} (M_\pi, q_2)/\Lambda_b$, see also the
discussion in Sec.~\ref{sec:NN},  and shows that the
low-momentum structure of the $3N$F can be described in ChEFT in a
controlled and systematically improvable fashion. Notice that in
addition to the static contributions considered above, the $3N$F of type
(a) at N$^3$LO receives non-local $1/m_N$ corrections of
relativistic origin \cite{Bernard:2011zr}. These have a
much richer operator structure than the static terms and also do not
involve unknown LECs.

Parameter-free predictions for
the one-pion-two-pion exchange and ring $3N$F topologies corresponding
to diagrams
(b) and (c) in Fig.~\ref{fig:3NFchiral}  can be found in
Refs.~\cite{Bernard:2007sp,Krebs:2013kha} and follow a qualitatively similar pattern.  
We also emphasize that while the static two-pion exchange $3N$F has a rather
restricted form described by just two functions, the long-and
intermediate-range topologies (a), (b) and (c) contribute to all $20$
operators $\hat O_\alpha$ that appear in the parametrization of the $3N$F according to
Eq.~(\ref{20structures}). Interestingly, the results for the
corresponding functions $f_\alpha$ appear to be qualitatively in line with 
estimations based on the large-$N_c$ expansion in QCD \cite{Phillips:2013rsa,Epelbaum:2014sea}. 

As pointed out in the introduction, an intermediate excitation of the
nucleon into the $\Delta (1232)$ resonance was historically recognized as one of the
most important $3N$F mechanisms and is at the heart of the celebrated
Fujita-Miyazawa $3N$F model \cite{10.1143/PTP.17.360} as depicted in
Fig.~\ref{fig:FM-3NF}. How can this important phenomenological
insight be reconciled with the framework of ChEFT? All results 
discussed in this section are obtained using the ChEFT formulation with
pions and nucleons as the only explicit degrees of freedom in the
effective Lagrangian. In such a framework, the information about the
$\Delta$ resonance is included implicitly through its contributions to
various LECs. In particular, the LECs $c_i$ are largely governed by
the $\Delta$ resonance \cite{Bernard:1996gq}:
\begin{equation}
  \label{saturation}
c_2^\Delta = - c_3^\Delta =  2 c_4^\Delta = \frac{4 h_A^2}{9 (m_\Delta
  - m_N)} \simeq 2.7 \; \text{GeV}^{-1} \, ,
\end{equation}  
where $m_\Delta$ refers to  the mass of the $\Delta$ resonance while
$h_A \sim 1.34$ is the $N\Delta$ axial coupling constant. Thus, the
$\Delta$ largely saturates the LECs $c_2$ and $c_4$, and provides
about a half of the $c_3$-value, thereby offering an explanation of the somewhat large
numerical values of these LECs as compared with their expected size
$|c_i | \sim \Lambda_b^{-1}$. This confirms that the intermediate $\Delta$ excitation
indeed provides the dominant mechanism of the two-pion
exchange $3N$F through Eqs.~(\ref{ABN2LO}) and (\ref{saturation}).

Given the strong coupling of the $\Delta$ isobar to the $\pi N$ system and
the smallness of the mass difference $m_\Delta - m_N$, which is numerically of the
order of $2 M_\pi$, it may be advantageous to include the $\Delta$
isobar as an explicit degree of freedom in the effective Lagrangian instead of integrating
it out assuming $m_\Delta - m_N \sim \Lambda_b$, as
done in the standard formulation of ChEFT. In the $\Delta$-full
formulation of ChPT of Ref.~\cite{Hemmert:1997ye}, the mass difference $m_\Delta -
m_N$ is treated as an additional soft scale $m_\Delta - m_N 
\sim M_\pi$\footnote{For an alternative counting of $m_\Delta - m_N$  see
  Ref.~\cite{Pascalutsa:2002pi}.} (in spite of the fact that this quantity does not
vanish in the chiral limit). Accordingly, the $\Delta$ isobar is treated
explicitly in the effective Lagrangian, and the new expansion parameter
is denoted as $\epsilon \in \{p/\Lambda_b, M_\pi/\Lambda_b,
(m_\Delta - m_N)/\Lambda_b \}$. Extensive studies in the
single-nucleon sector using manifestly covariant formulations of ChPT
have revealed that the explicit treatment of the $\Delta$ isobar
indeed often leads to an improved convergence of the EFT expansion, see e.g.~Refs.~\cite{Siemens:2016hdi,Yao:2016vbz,HillerBlin:2014diw,Thurmann:2020mog},
albeit at the cost of more involved calculations and a larger
number of LECs. 
Similar conclusions have been reached for the $NN$ force using the heavy-baryon formulation of $\Delta$-full ChEFT, for which the expressions are presently
available up through N$^2$LO $\epsilon^3$
\cite{Ordonez:1995rz,Kaiser:1998wa,Krebs:2007rh}.
Isospin-breaking contributions to the $NN$ potential have
also been considered within the $\epsilon$-expansion scheme \cite{Epelbaum:2008td}. 

The explicit treatment of the $\Delta$ isobar also has implications
for the $3N$F. In particular, the dominant
Fu\-ji\-ta-Miyazawa-type contribution is shifted from N$^2$LO to NLO in
the $\Delta$-full scheme, since
the diagram in Fig.~\ref{fig:FM-3NF} counts as of order
$\epsilon^2$. Interestingly, this is the only contribution of the
$\Delta$ to the $3N$F up-to-and-including the order N$^2$LO 
(i.e., $\epsilon^3$) \cite{Epelbaum:2007sq}. Recently, the
longest-range $3N$F of type (a) in Fig.~\ref{fig:3NFchiral} has been
worked out at order $\epsilon^4$ \cite{Krebs:2018jkc}. The resulting
parameter-free predictions were found to agree well with the
order-$Q^5$ results of Ref.~\cite{Krebs:2012yv}, which shows
that the effects of the $\Delta$ are well captured by resonance
saturation of the $\pi N$ LECs in the $\Delta$-less ChEFT framework
and suggests that convergence might have been reached for this topology.  
The corresponding results for the intermediate-range $3N$F of types (b)
and (c) within the $\epsilon$-expansion are not yet available. Notice
that in the
$\Delta$-less scheme, the first information about the $\Delta$ isobar
for these topologies appears only at N$^4$LO, since the N$^3$LO
expressions depend solely on the LO LECs, see e.g.~Eq.~(\ref{ABN3LO}),
which contain no information about the $\Delta$ isobar.  The explicit
inclusion of the $\Delta$ might, therefore, be advantageous to achieve
convergence for these topologies.

So far we have left open the question of regularization of the
$3N$F. As already pointed out in Sec.~\ref{sec:ChEFT}, ChEFT allows
one to derive low-momentum approximations of the nuclear forces, but
the method loses its validity 
for large momenta $|\vec p \, | \gtrsim \Lambda_b$. For example, the
ChEFT predictions for the function ${\cal{A}} (q_2)$ in
Eqs.~(\ref{ABN2LO}), (\ref{ABN3LO}) and (\ref{ABN4LO}) behave at large
$q_2$-values as ${\cal{A}}_{[Q^3]} (q_2) \sim q_2^2$, ${\cal{A}}_{[Q^4]}
(q_2)\sim q_2^3$ and ${\cal{A}}_{[Q^5]} (q_2) \sim q_2^4 \, \log (q_2)$,
respectively.
While there is an indication that the expansion for ${\cal{A}}(q_2)$ converges well within the applicability range of ChEFT, with the
order-$Q^5$ result providing a tiny correction as shown in the left panel
of Fig.~\ref{fig:AB}, the N$^4$LO contribution starts exceeding the
N$^3$LO correction at $q_2 \sim 780$~MeV, signalling the breakdown of the
expansion at large momenta. As explained in Sec.~\ref{sec:ChEFT}, the
expressions for nuclear forces derived in ChEFT have to be
regularized by introducing a finite cutoff $\Lambda \sim \Lambda_b$
to remove their uncontrolled behavior at large momenta and 
make the $A$-body Schr\"odinger equation well defined.  
In the $NN$ sector, this can be achieved, e.g., by simply multiplying the
unregularized potentials with some regulator functions. Unfortunately,
following this same strategy for the $3N$F was found in Refs.~\cite{Epelbaum:2019kcf,Epelbaum:2019jbv,Epelbaum:2022cyo}
to violate the chiral symmetry. The problem can be best illustrated
with the $3N$F of type (b) stemming from the first of the two
diagrams shown in the second line and third column of
Fig.~\ref{fig:3NFchiral}. When calculating the scattering amplitude,
the irreducible $3N$F is supplemented with the corresponding reducible
contributions stemming from the ite\-ration of the Faddeev equation. This
reducible contribution involves a linearly divergent piece $\propto
\Lambda$, whose structure violates the chiral symmetry. Accordingly,
the amplitude {\it cannot} be
renormalized by re-adjusting the LEC $c_D$, and the whole ChEFT expansion
for $3N$ scattering breaks down. As explained in Refs.~\cite{Epelbaum:2019kcf,Epelbaum:2019jbv,Epelbaum:2022cyo},
the problem is caused by mixing of two different regularization
schemes, namely DimReg when deriving the $3N$F
and cutoff regularization in the Schr\"odinger equation. The same
issue affects loop contributions to the nuclear current
operators \cite{Krebs:2019ddp,Krebs:2020pii}. This shows that a
{\it consistent} regularization of the $3N$F and 
nuclear currents cannot be achieved by simply multiplying
the corresponding potentials derived using DimReg with some cutoff
functions. Rather, $3N$Fs, 4NFs and nuclear currents starting from N$^3$LO need
to be re-derived using a cutoff instead of DimReg.  

The above conceptual issues have been the main obstacle for the
applications of ChEFT to three- and more-nucleon systems beyond
N$^2$LO. In contrast to DimReg, maintaining the chiral and gauge
symmetries in cutoff regularization represents a non-trivial
task, and the calculations become considerably more involved due to
the appearance of an additional mass scale $\Lambda$. Recently, it was shown
that the symmetry-preserving gradient flow method, which has been successfully applied to
Yang-Mills theories \cite{Luscher:2011bx,Luscher:2013cpa} and is nowadays widely employed in lattice-QCD
simulations \cite{Luscher:2013vga}, can be mer\-ged with ChEFT \cite{Krebs:2023gge}
and applied to obtain consistently
regularized nuclear forces and currents \cite{Krebs:2023ljo}. This is achieved by
generalizing the pion fields to the (artificial) fifth dimension,
usually referred to as the flow ``time'' $\tau$. The flow-time evolution of the
pion fields is governed by the chirally covariant version of the
gradient flow equation and amounts to smothening of the pion field, i.e., a
non-zero flow time $\tau$ acts as a regulator. Moreover, when applied
to the OPEP, the gradient flow method reduces to the SMS regulator of
Ref.~\cite{Reinert:2017usi}
specified in Eq.~(\ref{reg_opep}). The new developments  in
Refs.~\cite{Krebs:2023gge,Krebs:2023ljo} lay down the
foundation for deriving consistently regularized $3N$F and nuclear
currents beyond N$^2$LO. Work along these lines is in progress.

\subsubsection{Selected applications of the lading $3N$F}

We now turn to applications of the chiral nuclear forces to the 3N
and heavier systems, focusing especially on the role of
$3N$Fs. Given the lack of consistently regularized $3N$Fs at N$^3$LO and
N$^4$LO as described in the previous section, the accuracy level of
the applications reviewed below is limited to N$^2$LO.

Nucleon-deuteron ($Nd$) elastic scattering and brea\-kup observables can be
calculated, starting from a given nuclear Hamiltonian, by solving the
Faddeev equations in the partial wave basis as described in detail in
the review article \cite{Gloeckle:1995jg}. 
$Nd$ scattering calculations described below are carried out without
explicit inclusion of the Coulomb interaction and neglecting
relativistic effects, which are known to be small al low and
intermediate energies \cite{Witala:2004pv,Witala:2011yq}.   
We also restrict ourselves to studies based on the
second-generation of chiral $NN$ potentials
introduced in  Refs.~\cite{Epelbaum:2014efa,Epelbaum:2014sza,Reinert:2017usi},
see
Refs.~\cite{epelbaum2009,Epelbaum:2005pn,Kalantar-Nayestanaki_2012,Epelbaum:2012vx,Hammer:2012id} 
and references therein for earlier studies along this line.   
The novel semi-local regularization method employed in these NN
potentials allows one to avoid the appearance of noticeable artifacts in
elastic $Nd$ scattering at large energies reported in Ref.~\cite{Witala:2013ioa} for the first-generation
of N$^3$LO chiral $NN$ potentials of
Refs.~\cite{Epelbaum:2004fk,Entem:2003ft}, which can be traced back to
the
artifacts in the deuteron wave function.  The novel
semi-local potentials thus provide a very good starting ground for
systematic studies of $Nd$ scattering in ChEFT. 

In a series of papers by the LENPIC Collaboration, the semi-local
coordinate-space regularized (SCS) $NN$ potentials of
Refs.~\cite{Epelbaum:2014efa,Epelbaum:2014sza} at all available orders
from LO to N$^4$LO  were employed to
calculate $Nd$ scattering observables and
properties of selected nuclei up to $^{48}$Ca 
\cite{LENPIC:2015qsz,Maris:2016wrd,LENPIC:2018lzt}.  
While these calculations did not include the $3N$F and thus should be
regarded as incomplete starting from N$^2$LO, they have yielded a number
of interesting observations. In particular, calculations beyond the
last complete order NLO were found to differ from experimental data
well outside the estimated truncation uncertainties at the corresponding expansion orders, thus providing  
unambiguous evidence for missing $3N$Fs. Moreover, the amount of
deviations between theory and experimental data was found to be
consistent with estimations based on the power counting. 

In Ref.~\cite{LENPIC:2018ewt}, these studies have been extended to
include the N$^2$LO $3N$F, regularized in coordinate space consistently
with the SCS $NN$ potentials.
The need to perform regularization of
the $3N$F in coordinate space was found to introduce significant
computational overhead for its numerical implementation, which was one
of the motivations to reformulate the SCS regularization scheme to
momentum space \cite{Reinert:2017usi}. Notice that partial wave
decomposition of a general $3N$F can be carried out in an automated way by
numerically performing the required angular integrations as described
in Refs.~\cite{Golak:2009ri,Hebeler:2015wxa}, see also Ref.~\cite{PhysRep21Hebeler}.   

As detailed in Sec.~\ref{sec:3NFTheory}, the $3N$F at N$^2$LO depends on the LECs
$c_D$ and $c_E$ that need to be determined from few-nucleon data. It
is customary to fix the linear combination of these LECs to reproduce
the $^3$H binding energy, which determines $c_E$ as a function of
$c_D$. To fix the second LECs, different observables have been proposed
in the literature including the Nd doublet scattering length \cite{Epelbaum:2002vt,Piarulli:2017dwd},
$^3$H beta decay \cite{Gazit:2008ma}, $^4$He binding energy \cite{Nogga:2005hp}, charge radii
of the $A=3, 4$ nuclei and properties of few- and many-nucleon
systems \cite{Navratil:2007we,Ekstrom:2015rta,Lynn:2017fxg}. 
Clearly, to allow for the most stringent test of the 
nuclear Hamiltonian, the LECs should ideally be fixed from $A
\leq 3$ observables. In Ref.~\cite{LENPIC:2018ewt}, a variety of observables
including the Nd doublet scattering length as well as the Nd total and
differential cross sections at the energies of $E_{\rm lab} = 70$,
$108$ and $135$~MeV have been considered. Taking into account both the
experimental errors and the EFT truncation uncertainty, the strongest
constraint on $c_D$ was found to result from the requirement to
reproduce the proton-deuteron (pd) differential cross section minimum
using the data from Ref.~\cite{sekiguchi2002}.  The resulting
Hamiltonian was then used to calculate $Nd$ elastic scattering
observables, ground state energies and selected excitation energies
of $p$-shell nuclei up to $^{12}$C. For almost all considered nuclei, adding
the $3N$F was found to significantly improve the description of experimental
data. A detailed analysis of elastic $Nd$ scattering and breakup using
the same Hamiltonian is presented in Ref.~\cite{Witala:2019ffj}.  

These studies were further refined in Ref.~\cite{Maris:2020qne} by
employing the high-precision SMS $NN$ interactions of
Ref.~\cite{Reinert:2017usi} along with the consistently regularized
N$^2$LO $3N$Fs, utilizing Bayesian
methods for quantifying EFT truncation errors and extending the range
of considered observables.
\begin{figure}[tbp]
\centering
  \includegraphics[width=0.5\textwidth]{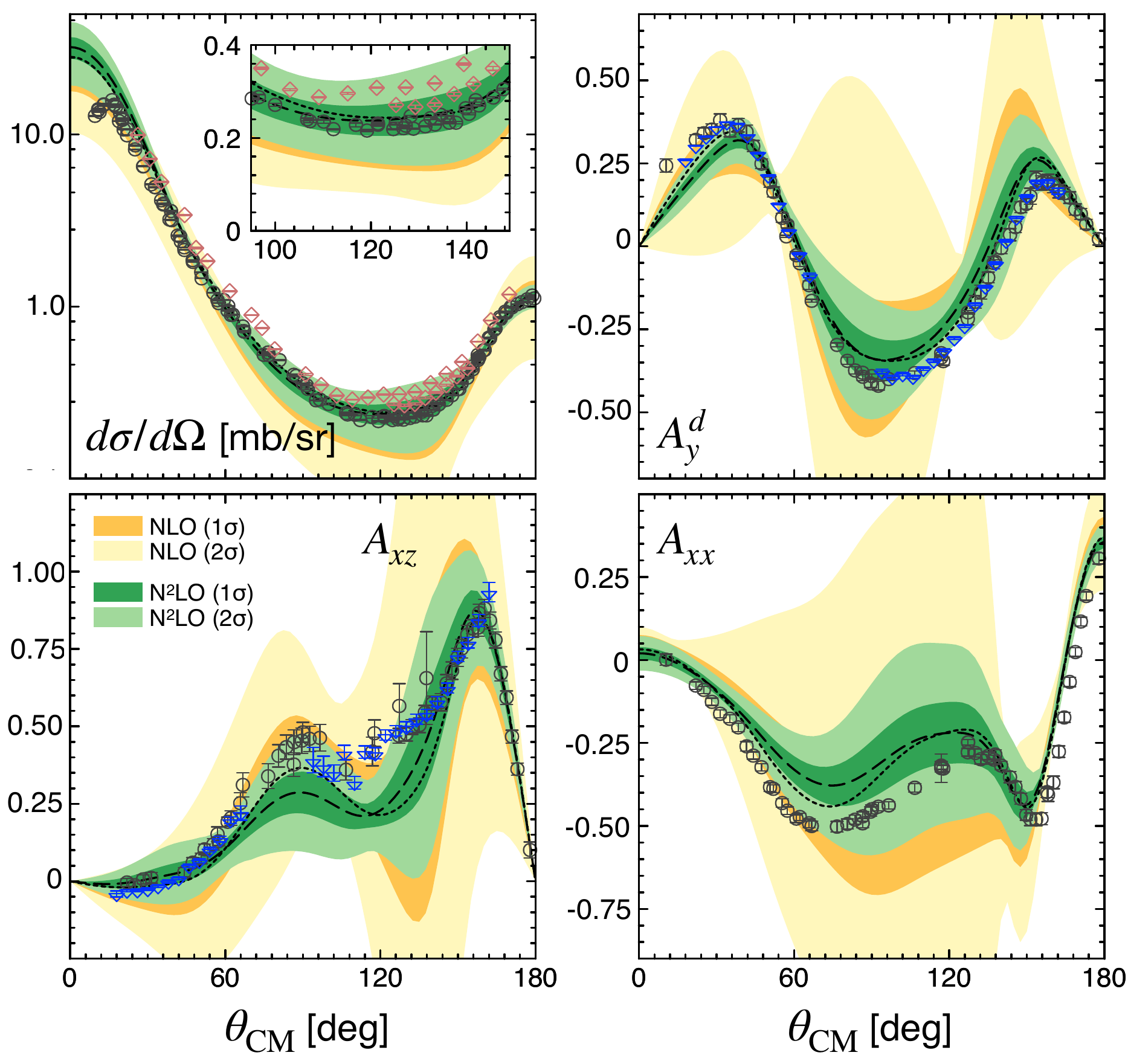}
\caption{\label{fig:Nd135}
 ChEFT predictions for the differential cross section, deuteron vector
analyzing power $A_y^d$ and deuteron tensor analyzing powers $A_{xz}$
and $A_{xx}$ in elastic neutron-deuteron scattering at $E_{\rm lab} =
135$~MeV. Dark-shaded orange and green bands show the NLO and N$^2$LO
results at the $1\sigma$ confidence level, respectively, while the
corresponding light-shaded bands show the $2\sigma$-intervals. 
The data points are experimental $pd$ elastic scattering data taken at RIKEN and RCNP (grey circles)~\cite{sekiguchi2002,nsakamot96,PhysRevLett.95.162301}, KVI (brown diamonds)~\cite{PhysRevC.78.014006,PhysRevLett.86.5862}, and IUCF (blue triangles)~\cite{PhysRevC.74.064003}. Dashed lines in the middle of green bands are the
actual N$^2$LO predictions. Dotted lines are obtained using the NN
interaction at the highest available order N$^4$LO$^+$, supplemented
with the N$^2$LO $3N$F (with the appropriately re-adjusted LECs $c_D$ and
$c_E$). In all calculations, the cutoff is chosen to be $\Lambda = 450$~MeV. }
\end{figure}

 In Fig.~\ref{fig:Nd135}, we show selected results for $Nd$ elastic
 scattering observables at $E_{\rm lab} = 135$~MeV, which may serve as
 representative examples. Given that the LECs $c_D$ and $c_E$ are
 fixed from the $^3$H binding energy and the differential cross
 section minimum at $E_{\rm lab} =70$~MeV, the shown results are to be
 regarded as predictions. The experimental data from
 Refs.~\cite{sekiguchi2002,nsakamot96,PhysRevLett.95.162301,PhysRevC.78.014006,PhysRevLett.86.5862,PhysRevC.74.064003} are mostly in agreement with the
 calculations (within errors), but the N$^2$LO truncation uncertainty
 at this moderate energy appears to be rather large. The description
 of $Nd$ data at N$^2$LO is qualitatively similar to the one for
 proton-proton scattering as a comparable energy, shown in
 Fig.~\ref{fig:pp}. Based on the results in the $NN$ system, it is
 expected that taking into account the $3N$F up through N$^4$LO would
 allow one to achieve a precise description of $Nd$ scattering data,
 comparable to that of the neutron-proton and proton-proton data
 reported in Refs.~\cite{Reinert:2017usi,Reinert:2020mcu}.   

It is interesting to explore the impact of corrections to the $NN$ force
beyond N$^2$LO. To this aim, a set of calculations based on the SMS NN
potentials up through N$^4$LO$^+$, supplemented with the N$^2$LO $3N$F,
has been performed in Ref.~\cite{LENPIC:2022cyu}. In all cases, the
LECs $c_D$ and $c_E$ have been fixed following the
standard LENPIC fitting protocol described above. For the
considered $Nd$ scattering observables, the inclusion of corrections to the
NN force beyond N$^2$LO changes the central N$^2$LO predictions, shown
by the dashed lines in Fig.~\ref{fig:Nd135}, to the dotted lines.
The results visualized by the dashed and dotted lines differ by
N$^3$LO terms, and it is comforting to see that the differences
between these lines are within the estimated N$^2$LO truncation
errors.  While the
higher-order corrections to the $NN$ force do appear to noticeably improve the
description of the tensor analyzing powers $A_{xz}$ and $A_{xx}$ in the angular range of $\theta_{\rm CM}
\in [45^\circ, 100^\circ]$, they still leave room for improvement that
should come from higher-order contributions to the $3N$F.

Further, we mention a comprehensive study of the symmetric space-star
deuteron breakup configuration in Ref.~\cite{Witala:2021zmb}. This particular
configuration is known to exhibit large discrepancies between theory
and data at energies below $E_{\rm lab} \sim 25$~MeV that could so far
not be resolved. Moreover, the calculated cross section appears to be
largely insensitive to the types of $3N$F considered so far. It would be
interesting to study the impact of $3N$F contributions
beyond N$^2$LO on this observable. Finally, a detailed investigation of the
deuteron breakup reaction at $E_{\rm lab} \sim 130$ and $200$~MeV
using the chiral SMS $NN$ and $3N$-forces 
and covering  the whole kinematically allowed phase space has been
carried out in Ref.~\cite{Skibinski:2023nnn}. 

\begin{figure}[tbp]
\centering
  \includegraphics[width=0.48\textwidth]{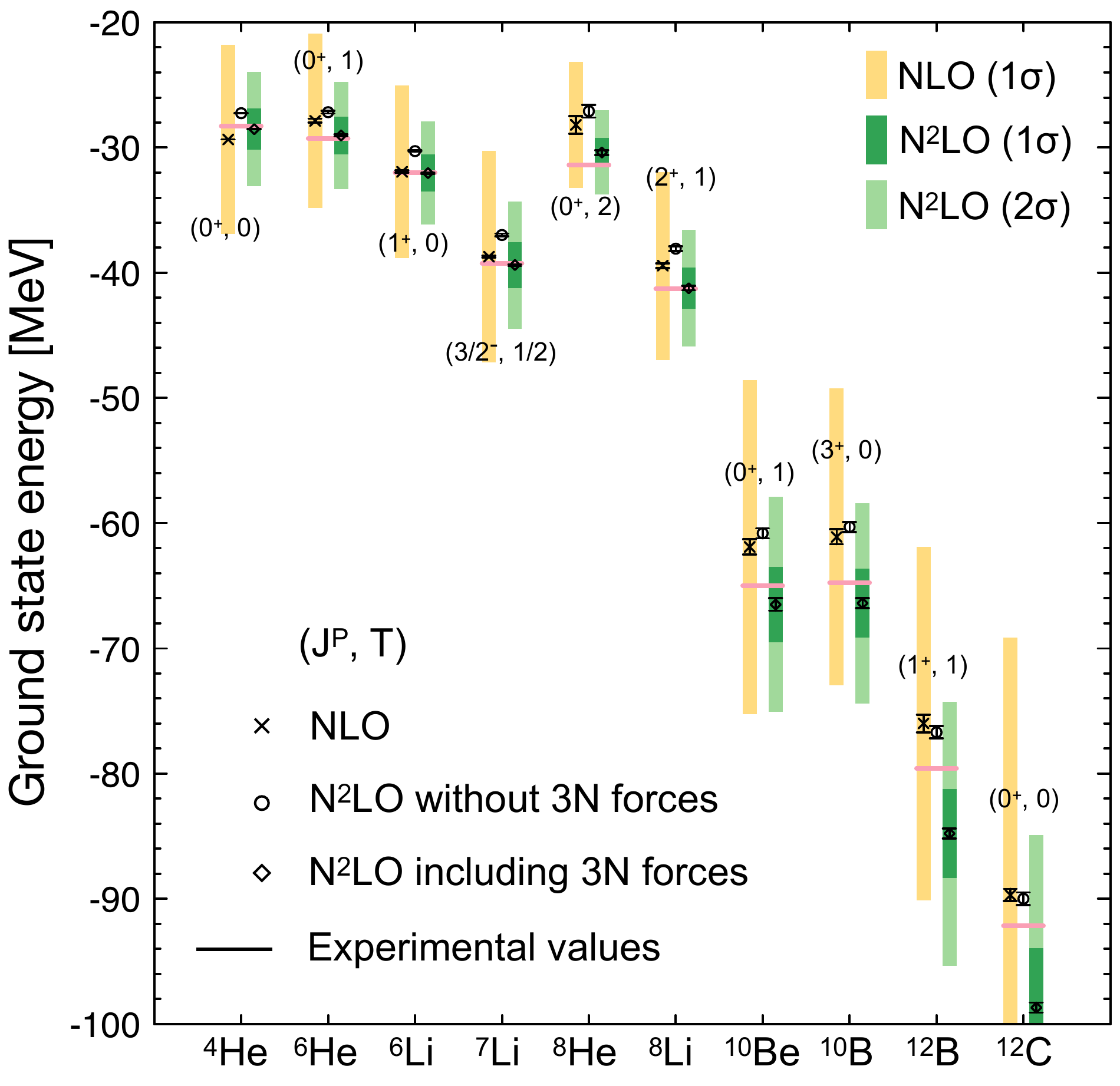}
\caption{\label{fig:Spectrum}
ChEFT predictions for the ground state energies of light nuclei
calculated using the No-Core Configuration Interaction (NCCI) method
\cite{Barrett:2013nh}. Crosses, circles and diamonds show the NLO results,
incomplete N$^2$LO predictions using the $NN$ interactions only and the
complete results at N$^2$LO using both the $NN$ force and $3N$F.
Orange, green dark-shaded and green light-shaded error bars
show the NLO $(1\sigma )$, N$^2$LO $(1\sigma )$ and N$^2$LO $(2\sigma )$ ChEFT
truncation errors, respectively, while black error bars give the numerical
uncertainty of the NCCI method. All theoretical predictions are
obtained using the cutoff value of $\Lambda = 450$~MeV. Horizontal lines depict the
experimental values for the binding energies. See Ref.~\cite{Maris:2020qne} for
details. }
\end{figure}

In Fig.~\ref{fig:Spectrum}, we show the NLO and N$^2$LO predictions
for the ground-state energies of light $p$-shell nuclei from
Ref.~\cite{Maris:2020qne}. For all nuclei, the predicted binding
energies at both NLO and N$^2$LO agree with experimental values within
truncation errors. To facilitate the quantification of $3N$F
effects, we also show the results based on the N$^2$LO $NN$ potential without
inclusion of the $3N$F. For light nuclei up to $^{10}$B, the inclusion
of the $3N$F leads to a significant improvement. However, for both
considered $A=12$-nuclei, the $3N$F effects appear to be too 
large leading to overbinding. This overbinding was shown in 
Ref.~\cite{LENPIC:2022cyu} to be resolved by taking into account the
corrections to the $NN$ force beyond N$^2$LO.  

Last but not least, N$^4$LO short-range contributions to the $3N$F have
been considered in the exploratory studies of $Nd$ scattering reported
in Refs.~\cite{Girlanda:2018xrw,Epelbaum:2019zqc,Witala:2022rzl}. 
While incomplete, these studies demonstrate that the N$^4$LO
contact interactions of natural size have the potential to both
resolve the long-standing $A_y$-puzzle in low-energy elastic Nd
scattering and strongly improve the description of scattering
observables at high energies.


\subsection{\label{sec:nuc3BF_exp}Experimental studies of three-nucleon forces}

To uncover the structure of $3N$Fs
one must utilize systems with more than two nucleons ($A>3$).
Few-nucleon scattering offers a unique opportunity 
to probe dynamical aspects of $3N$Fs, which are momentum, 
spin and isospin dependent, since it provides 
not only the cross sections but also a variety of spin observables 
at different incident nucleon energies. 
A direct comparison between experimental data 
and rigorous numerical calculations using the Faddeev theory 
and based on the realistic
nuclear potentials provides detailed information on the structure of $3N$Fs.

The importance of $3N$Fs in the continuum spectrum was 
shown, for the first time, in nucleon--deuteron ($Nd$) elastic scattering 
at the end of the 1990s \cite{wit98}.
$3N$Fs were found to lead to pronounced effects 
around the cross-section 
minimum occurring at the values of the c.m.~scattering angle of $\theta_{c.m.} \approx 120^{\circ}$  
for incident energies above $60~\rm MeV/nucleon$.
Since then, experiments of proton-deuteron($pd$) and neutron-deuteron($nd$) scattering 
as well as deuteron breakup reactions
at 60--300 MeV/nucleon
have been performed at the facilities, e.g.~RIKEN, RCNP, KVI, IUCF,
TSL, and LANSCE. These experiments have 
provided precise cross-section data as well as
various types of spin observables~\cite{Kalantar-Nayestanaki_2012}.
In this section, we present some representative results of 
nucleon--deuteron elastic scattering and proton--$^3\rm He$ elastic scattering. 
(For the deuteron breakup reactions, see the review paper of Ref.~\cite{Kalantar-Nayestanaki_2012} and 
recent reports such as Refs.~\cite{PhysRevC.102.054002,Ramazani-Sharifabadi2020}).


\begin{figure}[t!]
\centering
  \includegraphics[width=0.45\textwidth]{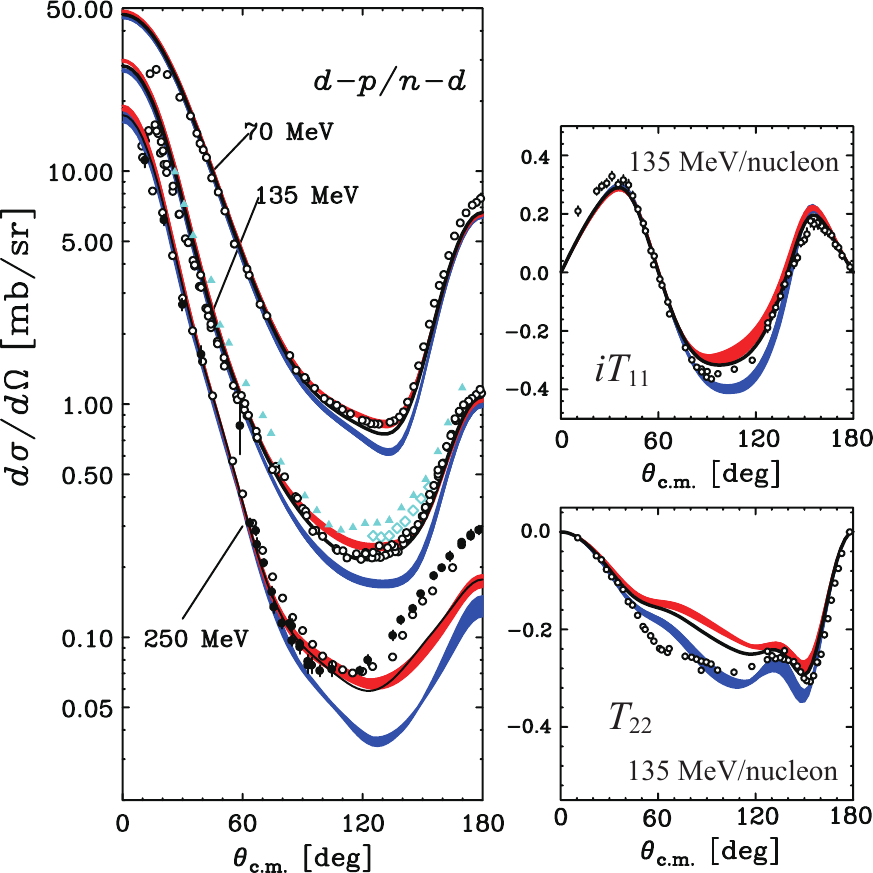}
\caption{\label{fig:dp_exp}
Differential cross section and deuteron analyzing powers $iT_{11}$, 
$T_{22}$ for elastic nucleon-deuteron scattering.
See text for descriptions for the theoretical calculations.}
\end{figure}

Figure~\ref{fig:dp_exp} shows 
experimental results reported in 
Refs.~\cite{sekiguchi2002,PhysRevLett.95.162301,PhysRevC.76.014004,PhysRevC.89.064007}
($dp$ or $pd$ in black open circles, $nd$ in black solid circles).
Data shown in blue are from Refs.~\cite{PhysRevC.68.051001,PhysRevC.78.014006}. 
The experimental data are compared with the Faddeev calculations 
with and w/o $3N$Fs. 
The red (blue) bands are 
the calculations with (without) Tucson-Melbourne 99 
$3N$F based on the modern $NN$ potentials, 
{\it i.e.}\ CD Bonn, AV18, Nijmegen I and II.
The solid lines are the calculations based on the
AV18 potential with including the Urbana IX $3N$F. 
The $3N$Fs considered here are 2$\pi$--exchange types.
For most of the observables shown in the figure, 
large differences are found at the backward angles between the data 
and the calculations based on $NN$ forces only.  
These discrepancies become larger 
with an increasing incident energy.
For the cross section, the $3N$Fs remove 
the discrepancies at lower energies. 
At higher energies, however,
the differences still remain even including the $3N$F potentials 
at the angles $\theta_{\rm c.m.} \gtrsim 120^\circ$,
which extent to the very backward angles $\theta_{\rm c.m.}\sim 180^\circ$
at 250 MeV/nucleon.
For the vector analyzing power $iT_{11}$, the description of the
experimental data by the theoretical calculations is similar 
to that of the cross section.
However, for the tensor analyzing power $T_{22}$ 
a different pattern is observed as the calculations 
including the $3N$Fs do not explain the data at the lower two energies. 

A direct comparison between the data and the Faddeev calculations
in elastic $Nd$ scattering led to the following conclusions so far: 
(1) the $3N$F is definitely needed in elastic $dp$ scattering;
(2) the $3N$F effects are clearly seen at the angles 
where the cross section takes its minimum, and 
their effects become larger with an increasing incident energy;
(3) spin dependent parts of the current $3N$F models are deficient;
(4) the short-range components of  
the $3N$F are probably required for high-momentum transfer region
(at the very backward angles).
These results of comparison between the data and the calculations based 
on the above phenomenological nuclear potentials have been pushing  
into more detailed study of three-nucleon scattering based on the $\chi$EFT 
nuclear potential as described in Sec.~\ref{sec:nuc3BF_EFT}.

\begin{figure}[t]
\centering
  \includegraphics[width=0.45\textwidth]{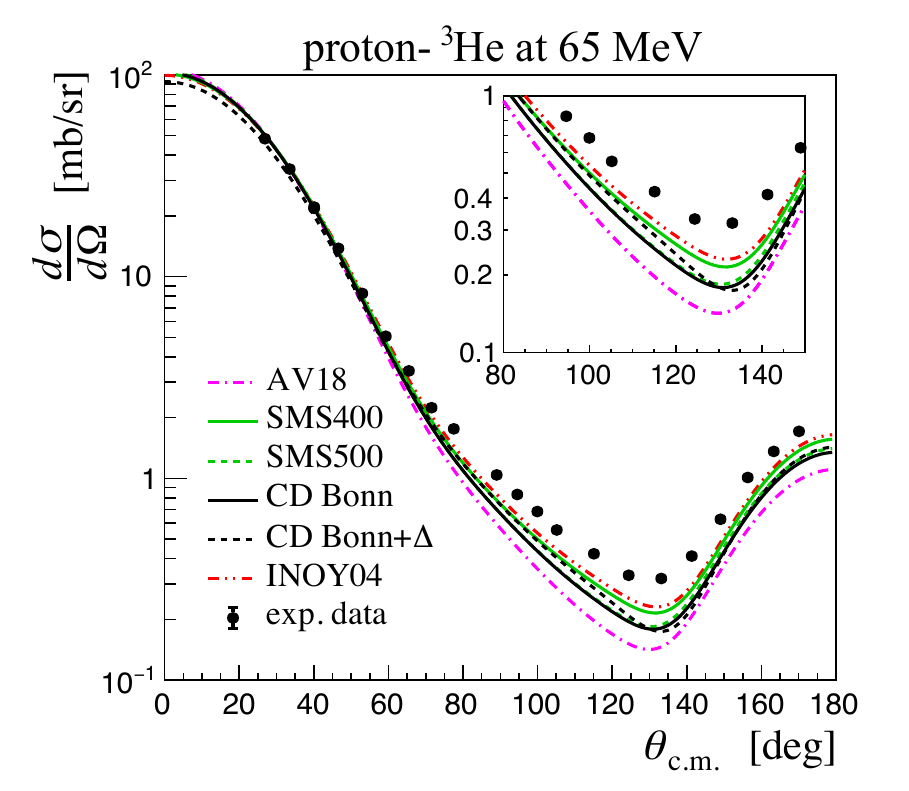}
\caption{\label{fig:p3He_dcs}
Experimental data (solid circles) and the calculations 
from the solutions of exact AGS equations
of the differential cross section $d\sigma/d\Omega$ at 65 MeV 
for $p$-$^3\rm He$ elastic scattering~\cite{PhysRevC.103.044001}. 
Calculations based on the $NN$ potentials are shown with magenta dash-dotted (AV18), 
black solid (CD Bonn), red dot-dot-dashed (INOY04), 
green solid (SMS400), and green dashed (SMS500) lines. 
Black dashed lines are calculations based on the CD Bonn+potential.}
\end{figure}

It is important to keep in mind that the $Nd$ system is dominated by the isospin $T = 1/2$ states. 
Thus, one needs other probes to constrain the properties of the $3N$Fs with total isospin $T = 3/2$, 
whose importance is strongly suggested in the description of asymmetric nuclear matter, 
e.g.~neutron-rich nuclei and pure neutron matter. 
Such aspects could be studied in four-nucleon systems like proton-$\rm ^3He$.
In recent years, remarkable theoretical progress in solving 
the 4N scattering problem using realistic Hamiltonians has been reported
\cite{Deltuva_PRC76,lazauskas2009,viviani_PRL111} even above the 4N breakup threshold
\cite{deltuva_prc87,fonseca_FBS2017}, 
opening new possibilities for nuclear force study in the 4N system at intermediate energies.
In Fig.~\ref{fig:p3He_dcs} the recent results of $p$-$^3\rm He$ scattering at intermediate 
energy are presented~\cite{PhysRevC.103.044001}.
The cross section data at 65 MeV/nucleon are compared 
with the calculations from the solutions of the exact AGS equations as given 
in Refs.~\cite{deltuva_prc87,fonseca_FBS2017} 
using a variety of $NN$ potentials: the AV18, the CD Bonn, and the INOY04~\cite{INOY04},
and the chiral N4LO $NN$ potentials 
with the cutoff parameters $\Lambda$ = 400 ${\rm MeV}/c$ (SMS400) 
and $\Lambda = 500 {\rm MeV}/c$ (SMS500) ~\cite{SMS}.
The calculations based on the CD Bonn+$\Delta$ model 
\cite{CDBD},
which allows an excitation of a nucleon
to an isobar, thereby providing effective $3N$Fs and $4N$Fs, are also presented.
Large contributions of the effective $3N$ and $4N$ forces have been
found to be largely canceled 
by the dispersive isobar effect,
leading to rather small total contributions from the $\Delta$-isobar. 
The results are in contrast to those in $dp$ scattering, where the cancellation is less pronounced \cite{Nemoto1999}. 
The obtained results indicate
that $p$-$^3\rm He$ elastic scattering at intermediate energies is an excellent tool to explore 
nuclear interactions including $3N$Fs, which cannot be accessed in $3N$ scattering. 
It would be interesting to see how the predictions with such $3N$Fs explain the data for the $p$-$^3\rm He$ 
elastic scattering, which will enable us to perform detailed discussions of the effects of $3N$Fs including the $T = 3/2$ isospin channels.

\section{\label{sec:atom3BF}Three-body forces in atoms}

Cold atoms are systems of dilute atomic gases cooled down to nano-Kelvin temperatures. While the interactions among atoms are dominated by two-body forces and their intrinsic three-body forces are negligible, one can engineer and realize cold atoms with an effective three-body force as strong as or even stronger than the two-body force, capitalizing on the high controllability of cold-atom experiments \cite{gross2017quantum,schafer2020tools}. In this section, we overview recent experimental and theoretical attempts to observe, control, and utilize three-body forces in cold atoms.

\subsection{Experimental observations of three-body forces for cold atoms in an optical lattice}
A system of cold atoms such as a Bose-Einstein condensate (BEC) or a Fermi-degenerate gas loaded into an optical lattice is known to be an ideal experimental platform for the quantum simulation of strongly-correlated quantum many-body systems \cite{bloch_many-body_2008}, such as the Hubbard model, owing to the high controllability of its parameters like hopping energy and on-site interaction, and so on.
When cold atoms are trapped in a sufficiently deep optical lattice potential, the hopping between neighboring lattice sites is negligible.
This gives us a novel possibility to simultaneously realize various well-defined few-body systems with definite atom numbers. 
Under these conditions, the trapping potential for the atoms in each lattice site is well approximated by a harmonic potential.
Therefore, a system of cold atoms in an optical lattice is also a useful platform for studying few-body physics in a trap.

Various spectroscopic techniques have been developed in cold atoms, which are quite useful to probe the energy of these few-body systems. 
The first occupancy-resolved high-resolution spectroscopy was reported for a radio-frequency spectroscopy of the ground hyperfine states of rubidium atoms \cite{Campbell2006}.
The observed almost equi-distance between the neighboring resonance frequencies is explained by the pairwise interactions alone.
However, slight deviation of the equi-distance between the neighboring resonance frequencies was also observed, indicating that the simple pairwise interaction is insufficient.
The qualitative explanation in the microscopic description of the system was given as the broadening of the Wannier function due to the two-body interaction.
Similar observations of the slight deviation from the prediction based on the pairwise interaction were reported in the experiments using various methods like matter-wave collapse and revival measurement \cite{Will2010}, resonant lattice modulation \cite{PhysRevLett.107.175301}, and laser spectroscopy \cite{Franchi_2017,Goban2018}.
This deviation is successfully explained by introducing an effective three-body force between the trapped atoms within perturbative treatments \cite{Johnson_2009,Johnson_2012}.
Interestingly, the microscopic origin of the effective three-body force, where one of the three atoms in the lowest vibrational state is excited to the higher vibrational state due to the inter-atomic interaction with the second atom and then is returned to the lowest state via the inter-atomic interaction with the third atom, has close analogy with the Fujita-Miyazawa type nuclear three-body force discussed in \secref{sec:intro} \cite{10.1143/PTP.17.360} (compare Fig.~\ref{fig:FM-3NF} with Fig.~\ref{fig:3-body} in the next subsection).

 Here, we stress that the three-body forces in the nuclear and trapped-atom systems share the same grounds in that they are the effective forces which are considered in low-energy effective descriptions of the systems. The cold-atom system can be a useful testbed to explore the three-body forces, firstly because of its high controllability, and secondly because the intrinsic three-body force between atoms is so small that we can directly investigate the physical mechanism for the emergence of the effective three-body forces.

Following this line of research direction, an occupancy-resolved high-resolution laser spectroscopy has been performed in recent experiments to investigate a new regime of three-body force \cite{honda2024evidence}.
In particular, by working with the ultra-narrow optical transition between the ground ${}^{1}S_0$ and metastable ${}^{3}P_2$ states of ytterbium (Yb) atoms \cite{PhysRevLett.110.173201}, one can utilize both the high resolution in the determination of the binding energy of the few-body atomic system and the high controllability of the two-body interaction through an inter-orbital anisotropy-induced Feshbach resonance \cite{Chin2010}. This enables us to study a new regime of three-body force beyond the perturbative treatment in the weakly-interacting regime.
While the data at small scattering lengths far from the Feshbach resonance are well explained by perturbative calculations, as in the previous works, the results obtained around the Feshbach resonance show significantly different behaviors from the equi-distance between the neighboring resonance frequencies, indicating a strongly-interacting regime of three-body forces, owing to the resonant control of the two-body interaction \cite{honda2024evidence}.
These results obtained by a cold-atom quantum simulator with tunable interactions can be a useful benchmark for developing the theory of the three-body forces beyond the perturbative regime, and will give insights into the nuclear three-body forces where the non-pertubative treatments are generally difficult.

\subsection{Three-body forces in low dimensions}
As demonstrated experimentally, three-body forces naturally appear in physics of cold atoms when they are confined into low dimensions in spite of their interaction being purely pairwise in free space.
While three-body forces are discussed in the previous subsection for quasi-zero-dimensional systems created by three-dimensional optical lattices, they are also possible under two- or one-dimensional optical lattices where atoms have freedom to move in one or two directions.
Here we provide some theoretical accounts of three-body forces in such low dimensions and their physical consequences.
We set $\hbar=k_B=1$ in this and next subsections.

As an illustrative example, let us consider weakly-interacting bosons subjected to a two-dimensional optical lattice that confines them into quasi-one-dimension.
Such a system is described by
\begin{align}\label{eq:H_3D}
\hat{H}_\mathrm{3D}
&= \int\!d^3\bm{r}\,\biggl[\hat\Phi^\dagger(\bm{r})\left\{-\frac{\nabla^2}{2m}
+ \frac{m\omega_\perp^2(y^2+z^2)}{2}\right\}\hat\Phi(\bm{r})\notag \\
&\quad{} + \frac{g_\mathrm{3D}}{2}\hat\Phi^\dagger(\bm{r})\hat\Phi^\dagger(\bm{r})
\hat\Phi(\bm{r})\hat\Phi(\bm{r})\biggr],
\end{align}
where $\hat\Phi(\bm{r})$ is the annihilation operator of bosons and the two-body coupling is related to the scattering length via $g_\mathrm{3D}=4\pi a_\mathrm{3D}/m$.
Due to the transverse confinement, the motion of bosons in $y$ and $z$ directions is quantized by the excitation energy of $\omega_\perp$.
Therefore, as far as low-energy physics relative to $\omega_\perp$ is concerned, the transverse motion cannot be excited so that the motion of bosons is restricted to the $x$ direction only.
Accordingly, such low-energy physics of the system should be described by an effective one-dimensional Hamiltonian in the form of
\begin{align}\label{eq:H_1D}
\hat{H}_\mathrm{1D}
&= \int\!dx\,\biggl[-\hat\phi^\dagger(x)\frac{\partial_x^2}{2m}\hat\phi(x)
+ \frac{g_2}{2}\hat\phi^\dagger(x)\hat\phi^\dagger(x)\hat\phi(x)\hat\phi(x)\notag \\
&\quad{} + \frac{g_3}{6}\hat\phi^\dagger(x)\hat\phi^\dagger(x)\hat\phi^\dagger(x)
\hat\phi(x)\hat\phi(x)\hat\phi(x) + \cdots\biggr].
\end{align}
Here $\hat\phi(x)$ is the annihilation operator of bosons in the transverse ground state and is related to $\hat\Phi(\bm{r})$ via its expansion of
\begin{align}
\hat\Phi(\bm{r}) = \sum_{\ell\in\mathbb{Z}}\sum_{n\geq0}\hat\phi_{\ell n}(x)\varphi_{\ell n}(y,z),
\end{align}
where $\hat\phi(x)=\hat\phi_{00}(x)$ and $\varphi_{\ell n}(y,z)$ is the normalized eigenfunction of a two-dimensional harmonic potential with energy $E_{\ell n}=(1+|\ell|+2n)\,\omega_\perp$.

\begin{figure}
\centering
\includegraphics[width=0.5\columnwidth]{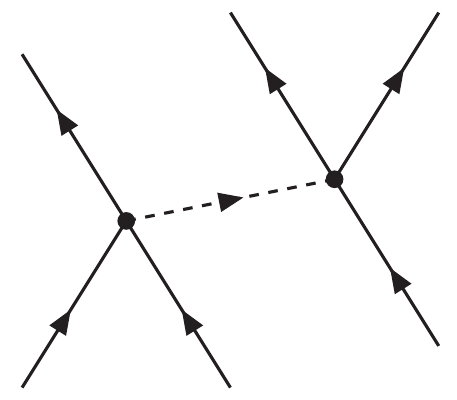}
\caption{\label{fig:3-body}
Three-body scattering process inducing an effective three-body coupling in low dimensions.
Solid and dashed lines represent bosons in transverse ground and excited states, respectively.}
\end{figure}

Because the original three-dimensional Hamiltonian in Eq.~(\ref{eq:H_3D}) has a two-body coupling, the resulting Eq.~(\ref{eq:H_1D}) also has a two-body coupling provided by
\begin{align}\label{eq:g_2}
g_2 = g_\mathrm{3D}\int\!dydz\,[\varphi_{00}(y,z)]^4
= \frac{2a_\mathrm{3D}}{ml_\perp^2}
\end{align}
to the lowest order in $a_\mathrm{3D}$ with $l_\perp=1/\sqrt{m\omega_\perp}$.
Furthermore, Eq.~(\ref{eq:H_1D}) has effective three-body and higher-body couplings induced by virtual excitation of bosons to transverse excited states \cite{PhysRevLett.89.110401,PhysRevLett.96.030406,PhysRevLett.100.210403}.
In particular, the three-body coupling to the lowest order in $a_\mathrm{3D}$ is induced by the three-body scattering process depicted in Fig.~\ref{fig:3-body} and provided by
\begin{align}
g_3 &= 6\sum_{n=1}^\infty\frac{[g_\mathrm{3D}\int\!dydz\,
\{\varphi_{00}(y,z)\}^3\varphi_{0n}(y,z)]^2}{E_{00}-E_{0n}} \notag\\\label{eq:g_3}
&= -12\ln\!\left(\frac43\right)\frac{a_\mathrm{3D}^2}{ml_\perp^2},
\end{align}
where $\varphi_{0n}(y,z)=e^{-(\rho/l_\perp)^2/2}L_n[(\rho/l_\perp)^2]/(\sqrt\pi\,l_\perp)$ with $\rho=\sqrt{y^2+z^2}$ is employed.
We note that $g_3$ presented in Refs.~\cite{PhysRevLett.96.030406,PhysRevLett.100.210403} was four times larger than Eq.~(\ref{eq:g_3}) but was later corrected in Refs.~\cite{PhysRevLett.105.090404,Mazets_2010}.
Dots in Eq.~(\ref{eq:H_1D}) include higher-body couplings as well as couplings involving derivatives such as effective-range corrections.

The two-body coupling in Eq.~(\ref{eq:g_2}) is linear in $a_\mathrm{3D}$ and can be repulsive or attractive depending on the positive or negative sign of $a_\mathrm{3D}$.
On the other hand, the three-body coupling in Eq.~(\ref{eq:g_3}) appears at the quadratic order in perturbation, so that it is always attractive.
Because the former dominates over the latter for weakly-interacting bosons, it is generally expected that physics is essentially determined by the two-body coupling and the three-body coupling only provides quantitative corrections that may be negligible without spoiling essential physics.
However, this is not the case in one dimension because the two-body and three-body couplings have distinct characters: while the two-body coupling preserves the integrability, it is broken by the three-body coupling \cite{PhysRevLett.89.110401,PhysRevLett.96.030406,PhysRevLett.100.210403}.
Therefore, even if the three-body coupling is quantitatively small, it is the leading perturbation to break the integrability and may have some qualitatively significant consequences in one dimension.

In particular, because two-body scatterings in one dimension do not change the momentum distribution, it is three-body scatterings that cause thermalization of a quasi-one-dimensional Bose gas.
The thermalization rate due to the effective three-body coupling was estimated in Refs.~\cite{PhysRevLett.100.210403,Mazets_2010} and was found to be consistent with the time needed for evaporative cooling of a $^{87}$Rb gas \cite{Hofferberth2007,Hofferberth2008}.
Similarly, the thermal conductivity of a weakly-interacting Bose gas in quasi-one-dimension was shown to be dominated by the three-body coupling rather than the two-body coupling \cite{PhysRevE.106.064104}.
Its expression was obtained as
\begin{align}
\kappa = \frac{\mathcal{N}}{m^3g_3^2}
\tilde\kappa\!\left(\frac{mT}{\mathcal{N}^2}\right),
\end{align}
where $T$ is the temperature, $\mathcal{N}$ is the number density, and $\tilde\kappa(*)$ is a dimensionless function determined numerically in Ref.~\cite{PhysRevE.106.064104} (see Fig.~6 therein).

The three-body coupling also has significant consequences on few-body physics in one dimension.
When the two-body coupling is attractive, $N$ bosons in the absence of three-body coupling are known to form a single bound state \cite{10.1063/1.1704156}, whose binding energy is provided by
\begin{align}
E_N^\mathrm{(MG)} = -\frac{N(N^2-1)}{24}mg_2^2.
\end{align}
Such an $N$-body cluster has no interaction (reflection probability) with an extra boson, being another manifestation of the integrability \cite{PhysRevLett.96.163201,PhysRevA.85.062711}.
Therefore, their interaction in quasi-one-dimension is dominated by the effective three-body coupling as the leading perturbation to break the integrability.
The scattering length between one boson and the $(N{-}1)$-body cluster now in the presence of three-body coupling was computed in Ref.~\cite{PhysRevA.97.061603} and was found to be repulsive for $4\leq N\leq38$ but interestingly turn attractive for $N=3$ (see also Refs.~\cite{PhysRevA.97.061604,PhysRevA.97.061605}) and $N\geq39$.
Because infinitesimal pairwise attraction immediately leads to a bound state in one dimension, the latter case exibits new $N$-body cluster formation induced by none other than the three-body coupling.
Its binding energy measured from the dissociation threshold was predicted as
\begin{align}
\Delta E_N^* = -\frac{N\beta_{1,N-1}^2}{8(N-1)}m^3g_2^2g_3^2,
\end{align}
where $\beta_{1,N-1}$ is an $N$-dependent number associated with the boson-cluster scattering length \cite{PhysRevA.97.061603} (see Fig.~2 and Table I therein).

\subsection{Artificial control of three-body forces}
Three-body forces not only naturally appear in low dimensions, but they can also be controlled artificially with cold atoms.
Accordingly, it is even possible to make three-body forces dominate over two-body forces, requiring to go beyond perturbations.
While several such schemes have been proposed theoretically \cite{PhysRevA.89.053619,PhysRevLett.112.103201,PhysRevA.90.021601,PhysRevA.93.043616}, we here introduce the simple and versatile one proposed in Ref.~\cite{PhysRevA.90.021601}, which employs two hyperfine spin components of bosons in an optical lattice.
When the two components are coupled by a nearly resonant field, the system in the tight-binding approximation is described by
\begin{align}\label{eq:H}
\hat{H} = -t\sum_{\sigma=\uparrow,\downarrow}\sum_{\langle i,j\rangle}
\hat{b}_{\sigma i}^\dagger\hat{b}_{\sigma j} + \sum_i\hat{\mathcal{H}}_i
\end{align}
with
\begin{align}\label{eq:H_i}
\hat{\mathcal{H}}_i &= -\frac\Omega2(\hat{b}_{\uparrow i}^\dagger\hat{b}_{\downarrow i}
+ \hat{b}_{\downarrow i}^\dagger\hat{b}_{\uparrow i})
- \frac\Delta2(\hat{b}_{\uparrow i}^\dagger\hat{b}_{\uparrow i}
- \hat{b}_{\downarrow i}^\dagger\hat{b}_{\downarrow i}) \notag\\
&\quad + \sum_{\sigma,\sigma'}\frac{u_{\sigma\sigma'}}{2}
\hat{b}_{\sigma i}^\dagger\hat{b}_{\sigma' i}^\dagger\hat{b}_{\sigma' i}\hat{b}_{\sigma i}.
\end{align}
Here $t$ is the inter-site tunneling amplitude, $\Omega$ is the Rabi frequency, $\Delta$ is the detuning, and $u_{\sigma \sigma'}$ are the on-site interaction energies.
Due to the Rabi coupling, the eigenstates of the spin part of Hamiltonian (first line) in Eq.~(\ref{eq:H_i}) are not $\uparrow$ and $\downarrow$ bosons but their superpositions and their energies are separated by the spin gap of $\sqrt{\Omega^2+\Delta^2}$.
Therefore, as far as low-energy physics relative to the spin gap is concerned, the higher-energy state cannot be excited so that bosons only occupy the lower-energy state.
Accordingly, such low-energy physics of the system should be described by an effective single-component Hamiltonian in the form of
\begin{align}\label{eq:H_eff}
\hat{H}_\mathrm{eff} = -t\sum_{\langle i,j\rangle}\hat{b}_i^\dagger\hat{b}_j
+ \sum_i\sum_{n=1}^\infty\frac{U_n}{n!}\hat{b}_i^{\dagger n}\hat{b}_i^n,
\end{align}
where $U_n$ is an effective $n$-body interaction energy induced by virtual excitation of bosons to the higher-energy state.

When one site is occupied by $N$ bosons, their on-site energy resulting from the second term of Eq.~(\ref{eq:H_eff}) reads
\begin{align}\label{eq:on-site}
\sum_{n=1}^N\frac{N!}{n!(N-n)!}U_n.
\end{align}
In order for the effective Hamiltonian in Eq.~(\ref{eq:H_eff}) to be the correct low-energy description of the original Hamiltonian in Eq.~(\ref{eq:H}), Eq.~(\ref{eq:on-site}) should match the lowest on-site energy of $N$ bosons resulting from Eq.~(\ref{eq:H_i}), whose matrix elements are provided by
\begin{align}
\langle n|\hat{\mathcal{H}}_i|n'\rangle =
\begin{cases}
\,\frac{N-2n}{2}\Delta + \frac{n(n-1)}{2}u_{\uparrow\uparrow}
+ \frac{(N-n)(N-n-1)}{2}u_{\downarrow\downarrow} \\
\quad{} + n(N-n)u_{\uparrow\downarrow} \qquad (n'=n) \\[4pt]
\,-\frac{\sqrt{n(N-n+1)}}{2}\Omega \qquad (n'=n-1) \\[4pt]
\,-\frac{\sqrt{(n+1)(N-n)}}{2}\Omega \qquad (n'=n+1) \\[4pt]
\ 0 \qquad (\mathrm{otherwise})
\end{cases}
\end{align}
with $|n\rangle=[1/\sqrt{n!(N-n)!\,}]\,\hat{b}_{\uparrow i}^{\dagger n}\,\hat{b}_{\downarrow i}^{\dagger N-n}|\mathrm{vac}\rangle$ and $n=0,1,\dots,N$.
Equating its lowest eigenvalue with Eq.~(\ref{eq:on-site}) determines $U_N$ for $N=1,2,\dots$ non-perturbatively as a function of $\Omega$, $\Delta$, and $u_{\sigma\sigma'}$.
Because these parameters are tunable in cold-atom experiments, independent control of $U_N$ is possible at least for several lowest $N$.

\begin{figure}
\centering
\includegraphics[width=\columnwidth]{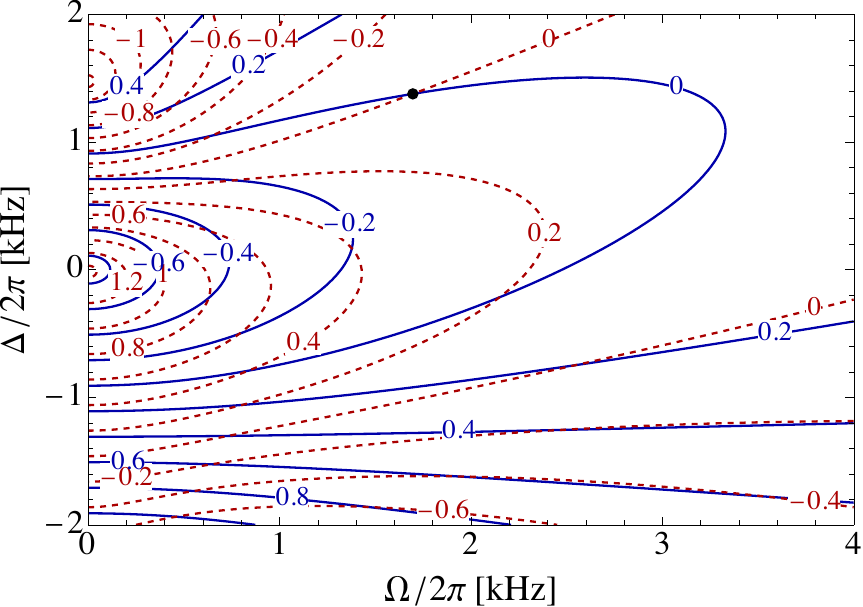}
\caption{\label{fig:U_N}
Contour plots of $U_2/2\pi$ [kHz] (blue solid curves) and $U_3/2\pi$ [kHz] (red dashed curves) in the $\Omega$-$\Delta$ plane.
The black dot marks a point for $U_2=U_3=0$.}
\end{figure}

As a concrete application, Ref.~\cite{PhysRevA.90.021601} considered $^{39}$K atoms subjected to a magnetic field $\approx58$ G in an optical lattice, where the on-site interaction energies in the harmonic approximation were estimated as $u_{\uparrow\uparrow}\approx2\pi\times0.55$ kHz, $u_{\downarrow\downarrow}\approx2\pi\times3.05$ kHz, and $u_{\uparrow\downarrow}\approx-2\pi\times0.91$ kHz for hyperfine spin components of $F=1$, $m_F=-1$ ($\sigma=\ \uparrow$) and $F=1$, $m_F=0$ ($\sigma=\ \downarrow$) \cite{D'Errico_2007,PhysRevA.81.032702}.
The resulting $U_2$ and $U_3$ as functions of $\Omega$ and $\Delta$ are presented in Fig.~\ref{fig:U_N}, where $U_3$ is found to be tunable in both magnitude and sign along the curve corresponding to $U_2=0$.
In particular, $U_2$ and $U_3$ simultaneously vanish at $(\Omega,\Delta)\approx2\pi\times(1.70,1.38)$ kHz so that $U_4\approx2\pi\times0.18$ kHz becomes the leading interaction energy \cite{PhysRevA.90.021601}.
Accordingly, it is possible to realize exotic systems without two-body coupling but with a tunable three-body coupling, and even more exotic systems without two-body nor three-body couplings but with a four-body coupling in any dimensions.

Such exotic systems are expected to exhibit unique physics.
Let us first consider identical bosons in two dimensions without two-body but with a three-body coupling.
When the three-body attraction is tuned to a resonance where a three-body bound state just appears, four such bosons were shown to form an infinite tower of bound states with the universal scaling law of
\begin{align}
E_n \propto e^{-2(\pi n)^2/27} \qquad (n\gg1)
\end{align}
in the binding energy of $n$-th excited state \cite{PhysRevLett.118.230601}.
This newly discovered few-body phenomenon is a unique consequence of the three-body coupling, which was termed the \textit{semisuper Efimov effect} by analogy with the \textit{Efimov effect} \cite{efimov1970energy,efimov1973energy} and the \textit{super Efimov effect} \cite{PhysRevLett.110.235301,PhysRevA.90.063631}.
This trio of effects constitutes universal classes of quantum halos whose spatial extensions can be arbitrarily large compared to the range of interaction potentials (see \secref{sec:Efimov} for more details on the Efimov effect).

Turning to one dimension again, we note that a three-body coupling therein is special not only because it breaks the integrability but also because it is marginally relevant in the sense of the renormalization group if it is attractive \cite{PhysRevA.97.011602,PhysRevLett.120.243002}.
Because the latter character is analogous to that of a two-body coupling in two dimension, similar physics is expected to emerge.
In particular, Ref.~\cite{PhysRevA.97.011602} showed that one-dimensional bosons without two-body but with weak three-body attraction form a many-body cluster stabilized by the quantum mechanical effect, resembling that of two-dimensional bosons \cite{PhysRevLett.93.250408,Bazak_2018}.
Its ground-state energy normalized by that of three bosons is universal, as long as the system remains dilute, and was predicted to grow exponentially as
\begin{align}
\frac{E_N}{E_3} \to \exp\!\left(\frac{8N^2}{\sqrt3\,\pi}\right)
\end{align}
with increasing number of bosons $N\gg1$ \cite{PhysRevA.97.011602}.

Furthermore, a four-body coupling in one dimension is analogous to a two-body coupling in three dimensions in the sense that both of them are irrelevant but have a fixed point at finite attraction corresponding to a resonance where a bound state just appears \cite{PhysRevLett.101.170401,nishida2011liberating}.
As discussed in Sec.~\ref{sec:Efimov}, three-dimensional bosons at a two-body resonance are known to exhibit the Efimov effect \cite{efimov1970energy,efimov1973energy}.
Similarly, one-dimensional bosons without two-body and three-body couplings but at a four-body resonance were shown to exhibit the Efimov effect, where five such bosons form an infinite tower of bound states with the universal scaling law of
\begin{align}
E_n \propto (12.4)^{-2n} \qquad (n\gg1)
\end{align}
in the binding energy of $n$-th excited state \cite{PhysRevA.82.043606}.
The resulting Efimov effect in one dimension is a unique consequence of the four-body coupling since the Efimov effect induced by the two-body coupling is possible only in three dimensions \cite{nielsen2001three}.

\begin{table}
\centering
\caption{\label{tab:cluster}
Ground-state energies of universal $N$-boson clusters for $N\gg1$ with few-body attraction (columns) in various dimensions (rows).
The (semisuper) Efimov effect indicates the universality not in the ground state but only in higher excited states.
The other systems marked by dashes are considered to be non-universal.}\smallskip
\begin{tabular}{cccc}
Dim.\textbackslash Att. & Two-body & Three-body & Four-body \\[2pt]\hline\\[-10pt]
1D & $N^3$ \cite{10.1063/1.1704156} & $e^{8N^2/\sqrt3\pi}$ \cite{PhysRevA.97.011602} & Efimov \cite{PhysRevA.82.043606} \\
2D & $e^{2.15N}$ \cite{PhysRevLett.93.250408} & semisuper \cite{PhysRevLett.118.230601} & --- \\
3D & Efimov \cite{efimov1970energy} & --- & --- \\\hline
\end{tabular}
\end{table}

Our perspective developed so far on the fates of bosons with two-body, three-body, or four-body attraction in various dimensions is summarized in Table~\ref{tab:cluster},
which may be useful to develop further insight into the universality in quantum few-body and many-body physics.

\section{\label{sec:Efimov}Efimov physics in nuclei and atoms}
\subsection{Overview of the Efimov effect}

Among the quantum clusters of few particles, a certain class is remarkable:
those which are very close to dissociation into smaller clusters or
even all the constituent particles. In quantum systems with short-range
interactions, there is indeed a minimum strength of the particles'
attraction that is required for the particles to remain bound to each
other. These clusters are thus realised when the attraction strength
is just above such critical point, and are therefore relatively weakly
bound. What is remarkable about these loosely bound clusters is that
they can be very large, much larger than the range of
the interactions, thanks to the ability of quantum systems to explore
classically forbidden regions. One often speaks of ``quantum halos''~\cite{Jensen2004}
when referring to these states, to emphasise their large extent and
diluteness. Since a dominant part of their wave function is delocalised
outside the region of interaction, it depends upon the interactions
only through a few effective parameters. As a result, these states
are said to be universal in the sense that only these few parameters
are enough to characterise the wave function to good level of approximation,
as well as many other properties such as their energy. In other words,
different systems with very different interactions can nonetheless
lead to the same universal states if their effective interaction parameters
are the same. These universal parameters correspond to the low-energy constants in the ChEFT discussed in Sec.~\ref{sec:nuc3BF_EFT}; The leading order constant arising from the low-energy two-body interaction,  the scattering length $a$, is the prominent one.

\begin{figure}
  \centering
  \includegraphics[width=0.47\textwidth]{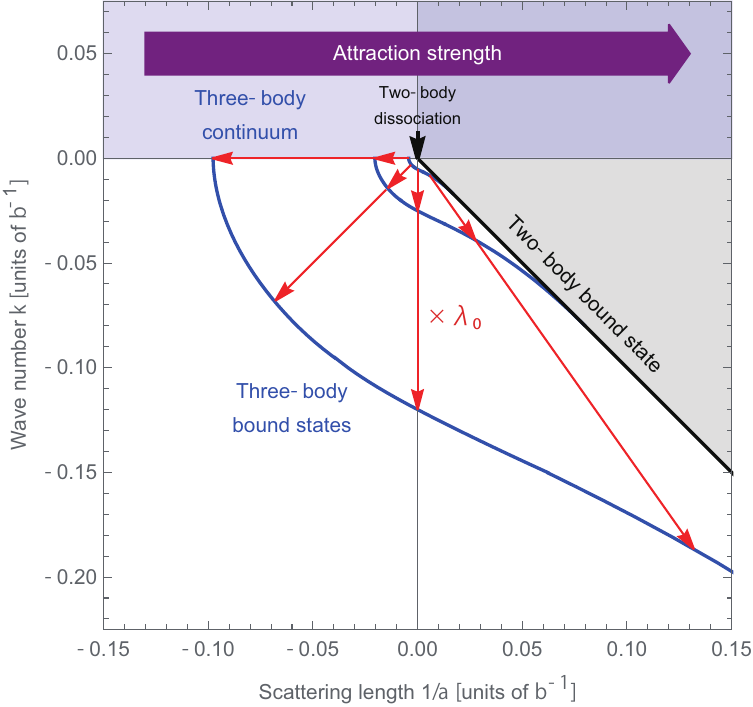}
\caption{\label{fig:Efimov-plot}
Efimov plot: schematic plot of three-particle
energy $E$ as a function of the inverse of the scattering length
$a$ between particles, which is a measure of the particles' attraction
strength indicated by the wide purple arrow. The energy $E$ is rescaled
into a wave number $k=\text{sign}(E)\sqrt{m\vert E\vert}/\hbar$ (where
$m$ is the mass of the particles and $\hbar$ the reduced Planck
constant) so that both coordinates are homogeneous to an inverse length,
and can be expressed in units of the inverse of the range $b$ of
the particles' interactions. As the attraction strength is increased,
the cluster of two particles (black line) appears from the point $1/a=0$
indicated by the black arrow. Conversely, this point can also be regarded
as the two-body dissociation point when the attraction strength is
decreased. There is an infinite number of three-particle clusters
appearing before that point. Their energies form a discrete-scale
invariant pattern, whereby any point on these energy curves can be
mapped to another point by scaling the coordinates ($k,1/a$) by a
factor $\lambda_{0}$, as indicated by the red arrows.}
\end{figure}

Among these universal quantum halo states, a more specific class is
particularly remarkable and has been a centre of attention of physicists
for many years: the so-called ``Efimov states'', discovered by V.
Efimov in the 1970s~\cite{Efimov1970a}. These three-body clusters occur for inter-particle
interactions that are close to the dissociation point of two particles.
This means that they can exist either when their two-body subsystems
themselves form loosely bound two-body halos ($1/a>0$ side of Fig.~\ref{fig:Efimov-plot}), or even when any of
their two-body subsystems is unbound ($1/a<0$ side of Fig.~\ref{fig:Efimov-plot}). In the latter case, these three-body
clusters are said to be ``Borromean'', a fairly common property
of quantum halos. But the unique feature of Efimov states is that
they are bound by an effective {\it long-range three-body} force
called the ``Efimov attraction'' arising from the {\it short-range
two-body} forces between the particles. As a result of this long-range
attraction, an infinite number of three-body bound states exist before
any pair of particles can bind. Moreover, the Efimov attraction is
scale invariant, since it decays as $1/R^{2}$, where $R$ is the
size of the three-particle system. As a result, the infinity of three-body
bound states near the two-body dissociation point is invariant by
scaling transformations with scaling factors that are multiples of
a certain number $\lambda_{0}$. This property is called ``discrete
scale invariance'' and is depicted in Fig.~\ref{fig:Efimov-plot}.
The figure shows in particular that the energy $E_{n}$ of the $n$-th
excited state at the two-body dissociation point is related to that
of the next one by $E_{n}=\lambda_{0}^{2}E_{n+1}$, which gives the scaling law for the excited states:
\begin{equation}
E_n \propto \lambda_0^{-2n} \ \ (n \gg 1).
\end{equation}
Since each three-body
bound state is related to the next one by a fixed scaling transformation,
it is enough to specify the energy of one particular three-body bound
state to determine the energies of all other bound states. This single
energy scale fixing the whole spectrum is referred to as the ``three-body
parameter'' and is conventionally defined to be the limit of $E_{n}\lambda_{0}^{2n}$
for large $n$. The three-body parameter corresponds to the dominant low-energy constant involving three particles (see N$^2$LO in Fig.~\ref{fig:3NFchiral}). Along with the scattering length, these are the two universal parameters that characterise the Efimov states.

The occurrence of discrete-scale invariance in systems of particles
with short-range interactions is called the ``Efimov effect''. There
are certain conditions for the Efimov effect to occur. First of all,
the quantum statistics and spin of the particles play an important
role, because the Pauli exclusion directly competes with the Efimov
attraction~\cite{efimov1973energy,PhysRevA.67.010703,kartavtsev2007low,PhysRevLett.103.153202,PhysRevA.86.062703,endo2011universal}. 
While identical bosons are always subject to the Efimov
effect whenever they are close to their two-body dissociation point,
identical fermions with spin 1/2 or less (polarised fermions) cannot
exhibit the Efimov effect due to the Pauli exclusion. Generally, for
the Efimov effect to occur, at least two pairs of particles must be
able to interact in the $s$ wave, close to their dissociation point.
This means that their corresponding $s$-wave scattering lengths must
be, in absolute value, much larger (typically 10 times~\cite{PhysRevLett.107.120401,PhysRevLett.108.263001,pascaleno3BP1,pascaleno3BP2,johansen2017testing,PhysRevLett.111.053202}) than the range
of their interactions. This is a rather stringent requirement, since
most systems have scattering lengths of the order of the interaction
range, and therefore do not exhibit the Efimov effect.

In systems where the Efimov effect occurs, the value of the scaling
factor $\lambda_{0}$ depends on the quantum statistics and masses
of the particles. For identical bosons, its value is $\lambda_{0}\approx22.7$,
which makes successive states very different in size and energy. For
systems of particles with mass imbalance, the scaling factor can differ
significantly and even approach 1 in the case of a very light particle
interacting with two heavy particles~\cite{efimov1973energy,nielsen2001three,naidon2017efimov}. In this case, the energy spectrum
is denser than that of identical bosons and more easily observable~\cite{Pires2014,Tung2014,PhysRevLett.115.043201}.

The term ``Efimov physics''~\cite{naidon2017efimov} has been coined to loosely designate
the study of any physical situation where the Efimov effect plays
a role, or an Efimov-like effect occurs. For instance, the energy
spectrum of a larger number of particles, such as four bosons, may
also exhibit a discrete-scale invariant pattern with the same scale
factor $\lambda_{0}$ as in the three-particle spectrum, due to the
influence of the 3-body Efimov effect~\cite{PhysRevD.7.2517,hammer2007universalEPJ,von2009signatures,PhysRevA.70.052101,yamashita2006four,PhysRevLett.107.135304,deltuva2011shallow,deltuva2013properties,PhysRevLett.108.073201,PhysRevLett.113.213201}. 
Generalisations of the Efimov effect for systems in mixed dimensions also exhibit discrete-scale
invariance~\cite{PhysRevLett.101.170401,nishida2011liberating,PhysRevA.79.060701,PhysRevA.82.011605,PhysRevA.84.052727}. 
Other systems, such as particles close to dissociation
in the $p$-wave, either in 3D~\cite{PhysRevLett.97.023201,PhysRevA.86.012711,PhysRevA.86.012710,PhysRevA.77.043611,PhysRevLett.96.050401,PhysRevLett.99.210402,PhysRevA.106.023304,PhysRevA.107.033329} or 2D~\cite{PhysRevLett.110.235301,Volosniev_2014,PhysRevA.92.020504,PhysRevA.78.063616,Gridnev_2014,PhysRevA.90.063631}, are not scale invariant but exhibit
an effective long-range attraction similar to the Efimov attraction.

\subsection{Geometry of Efimov states\label{subsec:geometry}}

Although the main feature of the Efimov states is the discrete scaling invariance around the two-body dissociation point, they are also characterised by universal geometric properties. For instance, at the two-body dissociation point, the three-body bound states close to zero energy (a.k.a Efimov trimers) have the same probabilistic distribution of triangular configurations, up to a global scaling by the universal factor $\lambda_{0}$. This distribution favours elongated configurations where one particle remains away from the other two. Quite counterintuitively, these typical configurations get more and more elongated as the attraction between particles gets stronger, until the three-body bound state dissociates into a two-body bound state and a free particle, as shown by the merging of the blue curves with the black curve in Fig.~\ref{fig:Efimov-plot}.
On the opposite side (Borromean region), where the attraction between particles is weaker, the configurations become more equilateral. Interestingly, near the three-body dissociation, they conform to another universal pattern known as ``halo universality''~\cite{Fedorov1994a,Jensen2004,naidon2023universal}. Halo universality is a generic feature of few-body systems close to their full dissociation threshold and is independent of the Efimov effect itself. Unlike Efimov universality, which is characterised by a scaling factor between consecutive states
(and thus difficult to demonstrate), halo universality is characterised by a universal geometry: all distances in the three-body system diverge when the binding energy approaches zero, thereby turning the system into a halo, but their ratios have well defined values. Remarkably, this universal halo geometry generally applies to states close to their three-body dissociation point, including the ground state, in sharp contrast with the Efimov universality which is accurate only for excited states.

In the case of three identical bosons, the universal halo geometry close to the three-body threshold is characterised by an equiprobability of all the triangular configurations of the three bosons. In the case of two identical particles and another particle, the universal halo geometry depends on the scattering length $a\gg b$ between the two identical particles and the binding energy $\kappa=\sqrt{m\vert E\vert}/\hbar\ll b$ of the system: when $\kappa a\ll1$, it is the same universal halo geometry as that of three bosons, but when $\kappa a\gg1$ it goes to a different geometry. These universal geometric properties can be derived analytically as a function of $\kappa a$~\cite{PhysRevLett.128.212501,naidon2023universal}.

\subsection{Efimov states in nuclear physics}

The major requirement for the conventional Efimov effect to occur is having pairs
of particles with scattering lengths much larger than the range of
their interactions. It turns out that the scattering lengths for nucleons
are relatively large, with both the triplet scattering length $a_{t}\approx5$~fm
and the singlet scattering length $a_{s}\approx-20$~fm being larger
than the range of nuclear forces, which is of the order of 1 fm. However,
the Coulomb repulsion between protons, which is a long-range force
and introduces an additional scale, can easily spoil the Efimov effect;
for example, the Hoyle state~\cite{hoyle1954nuclear,efimov1970energy},
the excited state of $^{12}$C nuclei, argued to be analogous to the
Efimov state of three $\alpha$ particles, is significantly affected
by the Coulomb interaction and cannot be described by the universal
Efimov theory~\cite{hammer2008model,PhysRevC.91.014004,otsuka2022alpha}.
Therefore, the Efimov states investigated so far in nuclear physics
mostly involve two neutrons interacting with a nucleus.

\paragraph{Triton.}
The simplest example is the triton, the system composed of two neutrons
and one proton~\cite{efimov1970energy,naidon2017efimov}. This system
fits qualitatively into the Efimov plot of Fig.~\ref{fig:Efimov-plot},
but for a positive scattering length $a=a_{t}$ that is not so large
compared to the interaction range $b$, thus on the right-hand side
of Fig.~\ref{fig:Efimov-plot} where only one three-body bound state
exists. Indeed, there is no excited bound state of the triton. It
is thus impossible to check the discrete-scale invariance, which is
a hallmark of the Efimov effect. Yet, the universal features of the
Efimov effect can readily explain some theoretical properties of the
triton, such as the Phillips line~\cite{phillips1968consistency,efimov1985explanation,efimov1988correlation},
which is a correlation between the triton energy and the neutron-deuteron
spin-doublet scattering length.

\paragraph{Halo nuclei.}
Systems of two neutrons interacting with a nucleus are more promising
candidates for Efimov states. Although a single neutron tends to be
absorbed by a nucleus due to the strong force, the situation changes
when the number of neutrons becomes large and approaches the neutron
dripline where neutrons are no longer tightly bound around the nucleus.
Around this dripline, it may happen for certain nuclides that they
could almost but not quite bind an extra neutron. If one regards
the nucleus and the neutron as two distinct particles, this situation
corresponds to the two-body dissociation point discussed earlier,
around which the Efimov effect can occur. It is thus expected that,
in the presence of two neutrons, such a nucleus could form a three-body Efimov state,
where the two neutrons would remain at large distance forming a halo
around the nucleus. Indeed, nuclei with two-neutron halo structures
have been studied both experimentally and theoretically since the 1980s, such as $^{6}$He,
$^{11}$Li, $^{12}$Be, $^{17}$B, $^{19}$B, $^{20}$C, $^{22}$C, $^{62}$Ca, and $^{72}$Ca~\cite{tanihata1985measurements,AlKhalili2004,Tanihata2013,Fedorov1994,Amorim1997,AnnRev_HamPlatt,naidon2017efimov,FREDERICO2012939,Canham2008,Hammer_2017,ACHARYA2013196,PhysRevLett.111.132501,PhysRevC.105.024310,PhysRevLett.120.052502}. Compared
to the triton, these systems stand as better candidates for Efimov
states~\cite{Fedorov1994,Amorim1997,Canham2008} because some of them would correspond to the Borromean regime
(left-hand side of Fig.~\ref{fig:Efimov-plot}) where the universal
description in terms of Efimov state is more effective. It is indeed
widely believed that these two-neutron halo nuclei are examples of
Efimov states. Three-body models of some nuclei seem to reproduce
observations and confirm this view.

However, the Efimov scenario as shown in Fig.~\ref{fig:Efimov-plot} is difficult to evidence in two-neutron halo nuclei because of several factors. First,
the $s$-wave scattering length between the core nucleus and a neutron
may not be large enough. Typical values for the neutron-rich nuclei
are 5-10 fm. This limits the number of Efimov states as $\dfrac{1}{\pi}\mathrm{ln}\left(\frac{|a|}{b}\right)\lesssim1$,
which precludes the possibility of observing two adjacent trimers
and testing the discrete scale invariance. As the synthesis of the neutron-rich nuclei is currently limited to light nuclei, the good candidates to see the excited Efimov state with current experimental techniques are likely be $^{19}$B and $^{22}$C. In particular,
a large $s$-wave scattering length is reported in $^{17}$B-n as
$\gtrsim100$ fm~\cite{SPYROU2010129,carbonell2020low,PhysRevLett.121.262502}, which may
currently be the best candidate to see the elusive excited Efimov
state in neutron-rich nuclei~\cite{PhysRevC.100.011603}.
Second, the size of the trimer is not large enough and the core may not be
regarded as a featureless inert point particle. Rather, the size and internal
structure of the core nucleus may play a significant role. In
such cases, the collective excitations of the core may couple with
the halo neutrons. The halo neutrons may also have significant correlations
with the neutrons inside the core nucleus, in which case the two neutrons
are better described as forming a BCS-like pair induced by the Fermi
sea of the nuclei, rather than the three-body Efimov picture. Even
when the three-body picture is valid, contributions from higher partial waves 
may play a significant role, as in $^{6}$He, $^{11}$Li~\cite{PhysRevC.50.R550,PhysRevC.90.044004}.
Finally, the effective range (i.e. finite radius of the nucleus) may not be
negligible, so that finite-range corrections to the zero-range universal Efimov theory must be
taken into account. Further experimental observations and comparison with three-body models are thus necessary to confirm whether two-neutron halo nuclei do indeed conform to the geometry of Efimov states.

Interestingly, the geometry of Efimov states near three-body dissociation tends to the universal halo geometry mentioned in Sec.~\ref{subsec:geometry}. Thus, if two-neutron halo nuclei are indeed Borromean Efimov states, they should approach this halo universality. However, the neutron-neutron scattering length is somewhat too small with respect to $b$ to fully reach the universal halo regime, so that they only constitute approximations of such halos. Among the known two-neutron halo nuclei, $^{19}$B and $^{22}$C are good candidates to exhibit the halo universality~\cite{naidon2023universal}.  It can be tested through enhanced measurements of the mean square radii and binding energies, or more directly by measuring the two-neutron correlation~\cite{PhysRevLett.125.252501}.




\paragraph{Excited nuclei.}
Recently, stable nuclei excited around their neutron
breakup thresholds have been proposed as novel nuclear candidates
exhibiting an enormous $s$-wave scattering length and the Efimov
states~\cite{endotanaka2023}. While the stable nuclei well inside the valley of
stability of the nuclear chart do not show any halo structure in their
ground states, they may form an $s$-wave halo of a core nucleus and
a neutron if they are excited in the vicinity of the one-neutron separation
threshold, around which the neutron may be barely bound by the core
nucleus. From the thermal neutron capture cross section data~\cite{shusterman2019surprisingly},
it has been argued that $^{88}\mathrm{Zr}$ and $^{157}\mathrm{Gd}$
should have very large $s$-wave halo, and hence large $s$-wave scattering
lengths $|a|\gtrsim10^{4}$ fm, which is 1-2 orders of magnitude larger
than those in the neutron-rich nuclei. It is predicted that
there are at least one and possibly two Efimov trimers made of the
core plus two neutrons in $^{90}\mathrm{Zr}$ and $^{159}\mathrm{Gd}$,
which appear as sub-threshold excited state around the two-neutron
separation threshold. It has also been conjectured that the $s$-wave
halo and the Efimov states may universally appear in the stripe regions
of the nuclear chart when the new excitation energy axis is added
above the nuclear chart, presenting a global picture to encompass
the Efimov physics in the neutron-rich nuclei and the Efimov physics
in the excited nuclei around the neutron separation threshold. This new perspective needs to be confirmed by further
theoretical and experimental studies.

\subsection{Efimov states in atomic physics}
Atomic physics has been more favourable for the demonstration of the Efimov effect. Below we review experimental progress in observing the Efimov states in atomic systems.

\subsubsection{Helium-4} 
Soon after its prediction, it was realised that helium-4 atoms are good candidates for the observation of Efimov states, since they are bosonic and their $s$-wave scattering length happens to be about 20 times larger than their interaction range. This allows the existence of two trimer states, a ground state and an excited state~\cite{Lim1977}. There was a long experimental effort to observe these states, which involved diffracting beams of helium clusters through a grating, selecting the deflecting beam corresponding to trimers and analysing their geometry. Although the ground state was observed from the 1990s~\cite{Schoellkopf1994,Bruehl2005}, it does not fully conform to the Efimov zero-range theory. Observation of the excited state was finally reported in 2015~\cite{Kunitski2015}, thanks to the Coulomb explosion imaging technique. It revealed that its geometry excellently conforms to the Efimov theory, although the universal scaling ratio $\lambda_{0}$ could not be directly confirmed.

\subsubsection{Cold atoms} Although helium-4 was for a long time the only candidate for the observation of Efimov states, it turned out that the atomic species used in cold-atom experiments are even better for this purpose. Although their natural scattering lengths are usually not large enough to expect the existence of Efimov states in these atomic systems, it is possible to alter the scattering length at will by applying a magnetic field onto these atoms, thanks to the so-called magnetic Feshbach resonances~\cite{Tiesinga1993,Chin2010}. This allows to experimentally probe the Efimov plot of Fig.~\ref{fig:Efimov-plot}, not only at a single scattering length value as in the case of helium-4 or halo nuclei, but over a whole range of scattering lengths. 

\paragraph{Observing Efimov states via three-body loss.}
The simplest way to reveal the presence of Efimov states is to monitor the losses occurring among atoms through three-body collisions and recombinations into deeply bound dimers~\cite{PhysRevLett.83.1751,PhysRevLett.85.908}. Indeed, nearby the scattering lengths at which an Efimov state appears at the three-body zero-energy threshold, these losses are significantly enhanced by the three-body resonances above threshold associated to these Efimov states~\cite{Kraemer2006,PhysRevLett.103.163202,zaccanti2009observation,pollack2009universality}. 
Since these scattering lengths are related to each other by a scaling with the universal scaling ratio $\lambda_{0}$, it makes possible the experimental determination of this ratio. The discrete scale invariance was thus confirmed in both systems of identical bosons~\cite{Huang2014} and systems of heavy bosons interacting with a lighter boson~\cite{Tung2014,Pires2014}. The measured scaling ratios were found to be consistent with the universal predictions, within the experimental uncertainties and corrections due to the finite-range of interactions. To minimize the effects of finite temperatures, trap size, and other aspects deteriorating Efimov signatures, there is an on-going attempt to observe Efimov states in the microgravity environment aboard the International Space Station (ISS)~\cite{elliott2018nasa,oudrhiri2023nasa,frye2021bose}.

\paragraph{Direct association of Efimov states.} In addition to observing the Efimov states through losses, there has also been efforts to directly probe the Efimov states~\cite{LompeScience11,PhysRevLett.106.143201,PhysRevLett.108.210406}. This can be done by shining a radio-frequency (RF) pulse to a mixture of atoms and Feshbach dimers. The atoms are initially prepared in a hyperfine state $| 0 \rangle $, while the dimers composed of two atoms in the $| 1 \rangle $ and $| 2 \rangle $ hyperfine states. A certain frequency of the RF pulse can drive the atoms in hyperfine state $| 0 \rangle $ to a hyperfine state $| 3 \rangle $. If that state has a simultaneous Feshbach resonant interaction with the atoms remaining in states $| 1 \rangle $ and $| 2 \rangle $, there are Efimov states composed of atoms in $| 1 \rangle $, $| 2 \rangle $, and $| 3 \rangle $ states just below this threshold. It is thus possible to form these Efimov states from the initial mixture by shining a slightly reduced RF frequency. By measuring the frequency difference, one can directly obtain the binding energies of the trimers. 
Such a system can be prepared with $^6$Li atoms, for which there is a magnetic field region where the $s$-wave scattering length between the three hyperfine states can be simultaneously large. In Refs.~\cite{LompeScience11,PhysRevLett.106.143201}, such a direct photo-association of an Efimov trimer has been performed, and its binding energy has been directly measured. This RF association can also be performed without the Feshbach dimer; in Ref.~\cite{PhysRevLett.108.210406}, the RF association of an Efimov trimer has been successfully performed with $^7$Li atoms from an initial state where all atoms are unbound. This method, while not requiring three resonant hyperfine states nor Feshbach dimers, relies on a large spatial overlap between the initial three-atom state and the final Efimov state and therefore can only be performed around the dissociation point of the Efimov trimer. More recently, Efimov trimers have also been created from a low-temperature Bose-Einstein condensate of atoms by rapidly sweeping a magnetic field, hence varying the $s$-wave scattering length from small positive values to large positive values for which Efimov trimers exist~\cite{PhysRevLett.119.143401}.

\paragraph{Coherent association of Efimov states.} While the above methods can directly create and probe Efimov trimers, the associated trimers undergo a three-body recombination and soon get lost from the trap. Recently, the group of Lev Khaykovich has observed coherent signatures of Efimov trimers in a manner insensitive to their recombination loss~\cite{PhysRevLett.122.200402,yudkin2023reshape}; they initially prepare cold $^7$Li atoms on the positive $a>0$ side of the Feshbach resonance, and shine a short RF pulse which is tuned to induce a transition into a superposition of an Efimov trimer state and a Feshbach dimer + atom state. The superposition state undergoes a time evolution with a phase factor which differs by $E_t - E_d$ where $E_t$ and $E_d$ are the energies of the Efimov trimer and Feshbach dimer. After a hold time $T$, the second short RF pulse turns the states back into the initial three-atom states. The final atom number count shows an oscillation $\cos[(E_t -E_d)T]$ from which the binding energy of the Efimov trimer relative to the dimer energy can be obtained~\cite{PhysRevLett.122.200402}. The oscillation is coherent and insensitive to the loss of Efimov states as long as the hold time $T$ is smaller than the lifetime of the Efimov states. With this technique, the binding energy of an Efimov trimer can be measured as a function of the $s$-wave scattering length~\cite{yudkin2023reshape}. While this technique has the advantage of enabling coherent control of the Efimov trimer, currently it can only access the $a>0$ region in the vicinity of the dimer+atom breakup threshold of the Efimov state, because the width of the RF pulse must cover both the Efimov state and the dimer+atom state to create their superposition.

\paragraph{Fermionic Efimov states.} 
While the Efimov states for bosonic atoms have been successfully observed for various atomic species, there have been ongoing efforts to observe the Efimov states with fermions. For the fermionic $^6$Li atoms, the Efimov states have been observed for atoms in three different hyperfine states~\cite{PhysRevLett.103.130404,PhysRevA.80.040702,PhysRevLett.106.143201,PhysRevLett.108.210406}. Three atoms in different internal states can be regarded as distinguishable particles, so that the Pauli exclusion principle does not play any role and the properties of the Efimov states are essentially the same as those for bosons. In order to realize Efimov states affected by the Pauli exclusion principle, one needs to prepare fermionic atoms in the same hyperfine state. Since a resonantly large $s$-wave interaction is necessary, the simplest setup is a system of two fermionic atoms in the same hyperfine state interacting with another distinguishable atom via a large $s$-wave scattering length. According to the theoretical arguments~\cite{efimov1973energy,PhysRevA.67.010703,kartavtsev2007low,PhysRevLett.103.153202,PhysRevA.86.062703,endo2011universal}, such a system can form Efimov trimers if the mass ratio is above the critical value of $13.6$. Notably, the predicted Efimov trimers of fermions are characterized by a total orbital angular momentum of $\ell=1$, which is a new feature compared with the bosonic Efimov states with $\ell=0$. Therefore, they are expected to manifest different universal properties than the bosonic ones~\cite{zhao2023effects,SciPostPhys.12.6.185,OiEndo2024}

To realize such a cold-atom system with a large mass ratio, we need a fermionic atomic species with a large enough mass. A mixture of fermionic Yb and Li atoms has been successfully cooled down to their Fermi degeneracy~\cite{PhysRevLett.106.205304}. While the mass ratio is large enough to support  the Efimov states, Yb-Li mixture is found to have no useful magnetic Feshbach resonances in an easily accessible range of the experiments~\cite{PhysRevLett.108.043201,chen2015anisotropy,PhysRevX.10.031037,PhysRevA.96.032711}, owing to closed-shell electronic structure of Yb atom.

More recently, an ultracold mixture of Er and Li atoms has been realized~\cite{PhysRevA.107.L031306}. The Er and Li atoms are cooled down  by a sympathetic cooling method utilizing the Yb atoms as a coolant, which is so powerful that a new kind of quantum degenerate gases of dual BECs of highly-magnetic atoms of Er and totally-nonmagnetic atoms of Yb can also successfully be created. The Er-Li system (mass ratio $\approx 28$) is deemed as a major candidate to realize the fermionic Efimov states. As both Er and Li atoms have open-shell atomic structures, they may have magnetic Feshbach resonances broad enough to accurately control the scattering length~\cite{PhysRevA.92.022708}. Recently, systematic measurements of Er-Li Feshbach resonances for various isotope mixtures have been performed in a magnetic field range below 1~kG, and succeeded in observing many Feshbach resonances~\cite{schafer_feshbach_2022,167Er6Li2023}. While the observed Feshbach resonances are mostly narrow, possibly induced by the anisotropic electrostatic interactions or dipolar interactions~\cite{chen2015anisotropy}, several broad Feshbach resonances were also found.

The observed broad Feshbach resonances in Er-Li is an important step toward observing the fermionic  Efimov trimers. The Efimov trimers are expected to be observed from loss measurement.  In particular, the loss process involving the Er-Er-Li channel, namely two Er accompanied by one Li atoms  should be predominantly enhanced. In Ref.~\cite{167Er6Li2023}, the atom loss behavior around the Feshbach resonance at $455$~G is observed in detail  for the ${}^{167}$Er and ${}^{6}$Li mixture. The atom loss is found to be predominantly induced through the Er-Er-Li channel, namely two Er accompanied by one Li, which is consistent with the formation of the fermionic  Efimov trimers. It is also shown that the shape of the observed resonance is asymmetric, which  is also consistent with the previous observations of multiple losses originating from the formation of several Efimov trimers~\cite{Pires2014,Tung2014}. More detailed measurements, including the determination of the inter-species scattering lengths,   will reveal the behaviors of these resonant losses induced by the fermionic Efimov trimers in this mass-imbalanced Er-Li mixture. Due to the strong magnetic dipole interaction between Er atoms, an interplay of the van der Waals and dipole interaction is predicted to lead to novel universal behaviors of the Efimov states~\cite{OiEndo2024}. We also note that the Er-Li mixture is also a candidate for realizing a novel $p$-wave superfluid of fermions of Er through the induced attraction via a BEC of light Li atoms~\cite{PhysRevA.61.053601,PhysRevLett.117.245302,gurarie2007resonantly,yamaguchi2017novel}, which is also interesting from the viewpoint of the effect of the many-body background of the novel Efimov trimers.

\section{Summary}
We have reviewed three-body forces and related phenomena from an interdisciplinary perspective across nuclei and atoms. The three-body forces have been a key element in accurate accounts of nuclear properties, and we have reviewed recent theoretical and experimental progress on the nuclear three-body forces. We have also reviewed some recent attempts to observe, control, and quantum-simulate three-body and higher-body forces in cold atoms. We have also shown how the Efimov states appear universally across nuclear and atomic systems, and overviewed the recent developments in their observation and understanding. The three-body forces and three-body phenomena have thus been an active area of research, not only as a specific field of research or methodology, but involving various disciplines of physics and scales. With further developments in experimental technologies and theoretical methods, it is expected that three-body forces will be determined with ever more accuracy, and new quantum phenomena will be uncovered by further exploring the implications of three-body physics.

\begin{acknowledgements}
This work is supported by Grant-In-Aid for Scientific Research on Innovative Areas ‘Clustering as a window on the hierarchical structure of quantum systems’. S.E.\ is supported by JSPS KAKENHI Grant Nos.\ JP21H00116, 22K03492, 23H01174, and Matsuo Foundation. S.E.\ acknowledges support from Institute for Advanced Science, University of Electro-Communications.
E.E.\ is supported by ERC under the EU Horizon 2020 research and innovation programme (ERC AdG Nuclear Theory, No.\ 885150), by DFG and NSFC through funds provided to the Sino-German CRC 110 “Symmetries and the Emergence of Structure in QCD” (DFG Project ID 196253076 - TRR 110, NSFC Grant No.\ 11621131001), by the MKW NRW (funding code NW21-024-A), and by the EU Horizon 2020 research and innovation programme (STRONG-2020, No.\ 824093). E.E.\ acknowledges the work of his collaborators Ashot Gasparyan, Hermann Krebs, Ulf-G.~Mei{\ss}ner, and the members of the LENPIC Collaboration.
P.N.\ is supported by JSPS KAKENHI Grant No.\ JP23K03292.
Y.N.\ is supported by JSPS KAKENHI Grant Nos.\ JP18H05405 and JP21K03384.
K.S.\ is supported by JSPS KAKENHI Grant Nos.\ JP25105502, JP16H02171, JP18H05404, JP20H05636, and JST ERATO Grant No.\ JPMJER2304, Japan.
Y.T.\ is supported by the Grant-in-Aid for Scientific Research of JSPS (Nos.\ JP17H06138, JP18H05405, JP18H05228, JP21H01014, JP21K03384, and JP22K20356), JST CREST (Nos.\ JPMJCR1673 and JPMJCR23I3), MEXT Quantum Leap Flagship Program (MEXT Q-LEAP) Grant No.\ JPMXS0118069021, and JST Moonshot R\&D Grant No.\ JPMJMS2269.
\end{acknowledgements}

\bibliographystyle{spphys}

\begin{thebibliography}{100}
\providecommand{\url}[1]{{#1}}
\providecommand{\urlprefix}{URL }
\expandafter\ifx\csname urlstyle\endcsname\relax
  \providecommand{\doi}[1]{DOI \discretionary{}{}{}#1}\else
  \providecommand{\doi}{DOI \discretionary{}{}{}\begingroup
  \urlstyle{rm}\Url}\fi

\bibitem{gibson1988three}
B.F. Gibson, B.H. McKellar, Few-Body Syst. \textbf{3}(4), 143
\newblock  (1988)

\bibitem{Hammer:2012id}
H.W. Hammer, A.~Nogga, A.~Schwenk, Rev. Mod. Phys. \textbf{85}, 197
\newblock  (2013)

\bibitem{10.1143/PTP.17.360}
J.~Fujita, H.~Miyazawa, Progress of Theoretical Physics \textbf{17}(3), 360
\newblock  (1957)

\bibitem{Yukawa193548}
H.~Yukawa, Proc. Phys. Math. Soc. Japan \textbf{17}, 48
\newblock  (1935)

\bibitem{tm99}
S.~Coon, H.~Han, Few-Body Syst. \textbf{30}, 131
\newblock  (2001)

\bibitem{PhysRevC.56.1720}
B.S. Pudliner, V.R. Pandharipande, J.~Carlson, S.C. Pieper, R.B. Wiringa, Phys.
  Rev. C \textbf{56}, 1720
\newblock  (1997)

\bibitem{Weinberg:1990rz}
S.~Weinberg, Phys. Lett. B \textbf{251}, 288
\newblock  (1990)

\bibitem{kolk1994}
U.~van Kolck, Phys. Rev. C \textbf{49}, 2932
\newblock  (1994)

\bibitem{Epelbaum:2002vt}
E.~Epelbaum, A.~Nogga, W.~Gl\"{o}ckle, H.~Kamada, U.-G. Mei\ss{}ner, H.~Witala,
  Phys. Rev. C \textbf{66}, 064001
\newblock  (2002)

\bibitem{PhysRevC.33.1740}
C.R. Chen, G.L. Payne, J.L. Friar, B.F. Gibson, Phys. Rev. C \textbf{33}, 1740
\newblock  (1986)

\bibitem{FewBodySys.1.3}
T.~Sasakawa, S.~Ishikawa, Few-Body Syst. \textbf{1}(1), 3
\newblock  (1986)

\bibitem{AV18}
R.B. Wiringa, V.G.J. Stoks, R.~Schiavilla, Phys. Rev. C \textbf{51}, 38
\newblock  (1995)

\bibitem{cdb}
R.~Machleidt, Phys. Rev. C \textbf{63}, 024001
\newblock  (2001)

\bibitem{nijm}
V.G.J. Stoks, R.A.M. Klomp, C.P.F. Terheggen, J.J. de~Swart, Phys. Rev. C
  \textbf{49}, 2950
\newblock  (1994)

\bibitem{PhysRevC.65.054003}
A.~Nogga, H.~Kamada, W.~Gl\"{o}ckle, B.R. Barrett, Phys. Rev. C \textbf{65}, 054003
\newblock  (2002)

\bibitem{PhysRevC.66.044310}
S.C. Pieper, K.~Varga, R.B. Wiringa, Phys. Rev. C \textbf{66}, 044310
\newblock  (2002)

\bibitem{Piarulli:2017dwd}
M.~Piarulli, et~al., Phys. Rev. Lett. \textbf{120}(5), 052503
\newblock  (2018)

\bibitem{PhysRevC.68.034305}
P.~Navr\'atil, W.E. Ormand, Phys. Rev. C \textbf{68}, 034305
\newblock  (2003)

\bibitem{Hagen:2012sh}
G.~Hagen, M.~Hjorth-Jensen, G.R. Jansen, R.~Machleidt, T.~Papenbrock, Phys.
  Rev. Lett. \textbf{108}, 242501
\newblock  (2012)

\bibitem{Cipollone:2014hfa}
A.~Cipollone, C.~Barbieri, P.~Navr\'atil, Phys. Rev. C \textbf{92}(1), 014306
\newblock  (2015)

\bibitem{Lahde:2019npb}
T.A. L\"ahde, U.-G. Mei\ss{}ner, \emph{{Nuclear Lattice Effective Field Theory}:
  {An introduction}}
\newblock  (Springer, 2019)

\bibitem{PhysRevC.58.1804}
A.~Akmal, V.R. Pandharipande, D.G. Ravenhall, Phys. Rev. C \textbf{58}, 1804
\newblock  (1998)

\bibitem{Gandolfi:2013baa}
S.~Gandolfi, J.~Carlson, S.~Reddy, A.W. Steiner, R.B. Wiringa, Eur. Phys. J. A
  \textbf{50}, 10
\newblock  (2014)

\bibitem{PhysRevLett.116.062501}
J.E. Lynn, I.~Tews, J.~Carlson, S.~Gandolfi, A.~Gezerlis, K.E. Schmidt,
  A.~Schwenk, Phys. Rev. Lett. \textbf{116}, 062501
\newblock  (2016)

\bibitem{epelbaum2009}
E.~Epelbaum, H.W. Hammer, U.-G. Mei\ss{}ner, Rev. Mod. Phys. \textbf{81}, 1773
\newblock  (2009)

\bibitem{Kalantar-Nayestanaki_2012}
N.~Kalantar-Nayestanaki, E.~Epelbaum, J.G. Messchendorp, A.~Nogga, Rep. Prog.
  Phys. \textbf{75}(1), 016301
\newblock  (2011)

\bibitem{PhysRep21Hebeler}
K.~Hebeler, Phys. Rep. \textbf{890}, 1
\newblock  (2021)

\bibitem{Haidenbauer:2016vfq}
J.~Haidenbauer, U.-G. Mei\ss{}ner, N.~Kaiser, W.~Weise, Eur. Phys. J. A
  \textbf{53}(6), 121
\newblock  (2017)

\bibitem{Weinberg:1978kz}
S.~Weinberg, Physica A \textbf{96}(1-2), 327
\newblock  (1979)

\bibitem{Gasser:1983yg}
J.~Gasser, H.~Leutwyler, Ann. Phys. \textbf{158}, 142
\newblock  (1984)

\bibitem{Weinberg:1968de}
S.~Weinberg, Phys. Rev. \textbf{166}, 1568
\newblock  (1968)

\bibitem{Coleman:1969sm}
S.R. Coleman, J.~Wess, B.~Zumino, Phys. Rev. \textbf{177}, 2239
\newblock  (1969)

\bibitem{Callan:1969sn}
C.G. Callan, Jr., S.R. Coleman, J.~Wess, B.~Zumino, Phys. Rev. \textbf{177},
  2247
\newblock  (1969)

\bibitem{Bernard:1995dp}
V.~Bernard, N.~Kaiser, U.-G. Mei\ss{}ner, Int. J. Mod. Phys. E \textbf{4}, 193
\newblock  (1995)

\bibitem{Fettes:1998ud}
N.~Fettes, U.-G. Mei\ss{}ner, S.~Steininger, Nucl. Phys. A \textbf{640}, 199
\newblock  (1998)

\bibitem{Fettes:2000gb}
N.~Fettes, U.-G. Mei\ss{}ner, M.~Mojzis, S.~Steininger, Ann. Phys. \textbf{283},
  273
\newblock  (2000).
\newblock [Erratum: Annals Phys. 288, 249--250 (2001)]

\bibitem{Jenkins:1990jv}
E.E. Jenkins, A.V. Manohar, Phys. Lett. B \textbf{255}, 558
\newblock  (1991)

\bibitem{Bernard:1992qa}
V.~Bernard, N.~Kaiser, J.~Kambor, U.-G. Mei\ss{}ner, Nucl. Phys. B \textbf{388},
  315
\newblock  (1992)

\bibitem{Becher:1999he}
T.~Becher, H.~Leutwyler, Eur. Phys. J. C \textbf{9}, 643
\newblock  (1999)

\bibitem{Gegelia:1999gf}
J.~Gegelia, G.~Japaridze, Phys. Rev. D \textbf{60}, 114038
\newblock  (1999)

\bibitem{Fuchs:2003qc}
T.~Fuchs, J.~Gegelia, G.~Japaridze, S.~Scherer, Phys. Rev. D \textbf{68},
  056005
\newblock  (2003)

\bibitem{Weinberg:1991um}
S.~Weinberg, Nucl. Phys. B \textbf{363}, 3
\newblock  (1991)

\bibitem{Ordonez:1995rz}
C.~Ordonez, L.~Ray, U.~van Kolck, Phys. Rev. C \textbf{53}, 2086
\newblock  (1996)

\bibitem{Pastore:2008ui}
S.~Pastore, R.~Schiavilla, J.L. Goity, Phys. Rev. C \textbf{78}, 064002
\newblock  (2008)

\bibitem{Pastore:2009is}
S.~Pastore, L.~Girlanda, R.~Schiavilla, M.~Viviani, R.B. Wiringa, Phys. Rev. C
  \textbf{80}, 034004
\newblock  (2009)

\bibitem{Pastore:2011ip}
S.~Pastore, L.~Girlanda, R.~Schiavilla, M.~Viviani, Phys. Rev. C \textbf{84},
  024001
\newblock  (2011)

\bibitem{Baroni:2015uza}
A.~Baroni, L.~Girlanda, S.~Pastore, R.~Schiavilla, M.~Viviani, Phys. Rev. C
  \textbf{93}(1), 015501
\newblock  (2016).
\newblock [Erratum: Phys.Rev.C 93, 049902 (2016), Erratum: Phys.Rev.C 95,
  059901 (2017)]

\bibitem{Epelbaum:1998ka}
E.~Epelbaum, W.~Gl\"{o}ckle, U.-G. Mei\ss{}ner, Nucl. Phys. A \textbf{637}, 107
\newblock  (1998)

\bibitem{Epelbaum:1999dj}
E.~Epelbaum, W.~Gl\"{o}ckle, U.-G. Mei\ss{}ner, Nucl. Phys. A \textbf{671}, 295
\newblock  (2000)

\bibitem{Epelbaum:2002gb}
E.~Epelbaum, U.-G. Mei\ss{}ner, W.~Gl\"{o}ckle, Nucl. Phys. A \textbf{714}, 535
\newblock  (2003)

\bibitem{Epelbaum:2005fd}
E.~Epelbaum, U.-G. Mei\ss{}ner, Phys. Rev. C \textbf{72}, 044001
\newblock  (2005)

\bibitem{Epelbaum:2005bjv}
E.~Epelbaum, Phys. Lett. B \textbf{639}, 456
\newblock  (2006)

\bibitem{Epelbaum:2007us}
E.~Epelbaum, Eur. Phys. J. A \textbf{34}, 197
\newblock  (2007)

\bibitem{Bernard:2007sp}
V.~Bernard, E.~Epelbaum, H.~Krebs, U.-G. Mei\ss{}ner, Phys. Rev. C \textbf{77},
  064004
\newblock  (2008)

\bibitem{Bernard:2011zr}
V.~Bernard, E.~Epelbaum, H.~Krebs, U.-G. Mei\ss{}ner, Phys. Rev. C \textbf{84},
  054001
\newblock  (2011)

\bibitem{Krebs:2012yv}
H.~Krebs, A.~Gasparyan, E.~Epelbaum, Phys. Rev. C \textbf{85}, 054006
\newblock  (2012)

\bibitem{Krebs:2013kha}
H.~Krebs, A.~Gasparyan, E.~Epelbaum, Phys. Rev. C \textbf{87}(5), 054007
\newblock  (2013)

\bibitem{Kolling:2009iq}
S.~Kolling, E.~Epelbaum, H.~Krebs, U.-G. Mei\ss{}ner, Phys. Rev. C \textbf{80},
  045502
\newblock  (2009)

\bibitem{Kolling:2011mt}
{Kolling, S. and Epelbaum, E. and Krebs, H. and Mei\ss{}ner, U.-G.}, Phys. Rev.
  C \textbf{84}, 054008
\newblock  (2011)

\bibitem{Krebs:2016rqz}
H.~Krebs, E.~Epelbaum, U.-G. Mei\ss{}ner, Ann. Phys. \textbf{378}, 317
\newblock  (2017)

\bibitem{Krebs:2019aka}
H.~Krebs, E.~Epelbaum, U.-G. Mei\ss{}ner, Few-Body Syst. \textbf{60}(2), 31
\newblock  (2019)

\bibitem{Krebs:2020plh}
H.~Krebs, E.~Epelbaum, U.-G. Mei\ss{}ner, Eur. Phys. J. A \textbf{56}(9), 240
\newblock  (2020)

\bibitem{Kaiser:1997mw}
N.~Kaiser, R.~Brockmann, W.~Weise, Nucl. Phys. A \textbf{625}, 758
\newblock  (1997)

\bibitem{Kaiser:1998wa}
N.~Kaiser, S.~Gerstendorfer, W.~Weise, Nucl. Phys. A \textbf{637}, 395
\newblock  (1998)

\bibitem{Kaiser:1999ff}
N.~Kaiser, Phys. Rev. C \textbf{61}, 014003
\newblock  (2000)

\bibitem{Kaiser:1999jg}
N.~Kaiser, Phys. Rev. C \textbf{62}, 024001
\newblock  (2000)

\bibitem{Kaiser:2001dm}
N.~Kaiser, Phys. Rev. C \textbf{63}, 044010
\newblock  (2001)

\bibitem{Kaiser:2001pc}
N.~Kaiser, Phys. Rev. C \textbf{64}, 057001
\newblock  (2001)

\bibitem{Kaiser:2001at}
N.~Kaiser, Phys. Rev. C \textbf{65}, 017001
\newblock  (2002)

\bibitem{Entem:2015xwa}
D.R. Entem, N.~Kaiser, R.~Machleidt, Y.~Nosyk, Phys. Rev. C \textbf{92}(6),
  064001
\newblock  (2015)

\bibitem{Epelbaum:2005pn}
E.~Epelbaum, Prog. Part. Nucl. Phys. \textbf{57}, 654
\newblock  (2006)

\bibitem{Machleidt:2011zz}
R.~Machleidt, D.R. Entem, Phys. Rep. \textbf{503}, 1
\newblock  (2011)

\bibitem{Epelbaum:2019kcf}
E.~Epelbaum, H.~Krebs, P.~Reinert, Front. Phys. \textbf{8}, 98
\newblock  (2020)

\bibitem{Krebs:2020pii}
H.~Krebs, Eur. Phys. J. A \textbf{56}(9), 234
\newblock  (2020)

\bibitem{deVries:2020iea}
J.~de~Vries, E.~Epelbaum, L.~Girlanda, A.~Gnech, E.~Mereghetti, M.~Viviani,
  Front. Phys. \textbf{8}, 218
\newblock  (2020)

\bibitem{Lepage:1997cs}
G.P. Lepage, {Lecture note of 8th Jorge Andre Swieca Summer School on Nuclear
  Physics} pp. 135--180
\newblock  (1997)

\bibitem{Gasparyan:2021edy}
A.M. Gasparyan, E.~Epelbaum, Phys. Rev. C \textbf{105}(2), 024001
\newblock  (2022)

\bibitem{Gasparyan:2023rtj}
A.M. Gasparyan, E.~Epelbaum, Phys. Rev. C \textbf{107}(4), 044002
\newblock  (2023)

\bibitem{Gross2022hyw}
F.~Gross, et~al., Eur. Phy. J. C \textbf{83}(12), 1125
\newblock  (2023)

\bibitem{Tews:2022yfb}
I.~Tews, et~al., Few Body Syst. \textbf{63}(4), 67
\newblock  (2022)

\bibitem{Reinert:2017usi}
P.~Reinert, H.~Krebs, E.~Epelbaum, Eur. Phys. J. A \textbf{54}(5), 86
\newblock  (2018)

\bibitem{Reinert:2020mcu}
P.~Reinert, H.~Krebs, E.~Epelbaum, Phys. Rev. Lett. \textbf{126}(9), 092501
\newblock  (2021)

\bibitem{Epelbaum:2004fk}
E.~Epelbaum, W.~Gl\"{o}ckle, U.-G. Mei\ss{}ner, Nucl. Phys. A \textbf{747}, 362
\newblock  (2005)

\bibitem{Entem:2003ft}
D.R. Entem, R.~Machleidt, Phys. Rev. C \textbf{68}, 041001
\newblock  (2003)

\bibitem{Epelbaum:2014efa}
E.~Epelbaum, H.~Krebs, U.-G. Mei\ss{}ner, Eur. Phys. J. A \textbf{51}(5), 53
\newblock  (2015)

\bibitem{Epelbaum:2014sza}
E.~Epelbaum, H.~Krebs, U.-G. Mei\ss{}ner, Phys. Rev. Lett. \textbf{115}(12),
  122301
\newblock  (2015)

\bibitem{NavarroPerez:2013usk}
R.~Navarro~P\'erez, J.E. Amaro, E.~Ruiz~Arriola, Phys. Rev. C \textbf{88},
  024002
\newblock  (2013).
\newblock [Erratum: Phys.Rev.C 88, 069902 (2013)]

\bibitem{Epelbaum:2022cyo}
E.~Epelbaum, H.~Krebs, P.~Reinert, {Handbook of Nuclear Physics} pp. 1--25
\newblock  (2022)

\bibitem{Epelbaum:2019zqc}
E.~Epelbaum, et~al., Eur. Phys. J. A \textbf{56}(3), 92
\newblock  (2020)

\bibitem{Cox:1968jxz}
G.F. Cox, G.H. Eaton, C.P. Van~Zyl, O.N. Jarvis, B.~Rose, Nucl. Phys. B
  \textbf{4}, 353
\newblock  (1968)

\bibitem{Jarvis:1971fla}
O.N. Jarvis, C.~Whitehead, M.~Shah, Phys. Lett. B \textbf{36}(4), 409
\newblock  (1971)

\bibitem{TAYLOR1960320}
A.~Taylor, E.~Wood, L.~Bird, Nuclear Physics \textbf{16}(2), 320
\newblock  (1960)

\bibitem{Stoks:1993tb}
V.G.J. Stoks, R.A.M. Klomp, M.C.M. Rentmeester, J.J. de~Swart, Phys. Rev. C
  \textbf{48}, 792
\newblock  (1993)

\bibitem{Rentmeester:1999vw}
M.C.M. Rentmeester, R.G.E. Timmermans, J.L. Friar, J.J. de~Swart, Phys. Rev.
  Lett. \textbf{82}, 4992
\newblock  (1999)

\bibitem{Birse:2003nz}
M.C. Birse, J.A. McGovern, Phys. Rev. C \textbf{70}, 054002
\newblock  (2004)

\bibitem{Entem:2017gor}
D.R. Entem, R.~Machleidt, Y.~Nosyk, Phys. Rev. C \textbf{96}(2), 024004
\newblock  (2017)

\bibitem{Piarulli:2014bda}
M.~Piarulli, L.~Girlanda, R.~Schiavilla, R.~Navarro~P\'erez, J.E. Amaro,
  E.~Ruiz~Arriola, Phys. Rev. C \textbf{91}(2), 024003
\newblock  (2015)

\bibitem{Somasundaram2023sup}
R.~Somasundaram, J.E. Lynn, L.~Huth, A.~Schwenk, I.~Tews, Phys. Rev. C
  \textbf{109}, 034005
\newblock  (2024)

\bibitem{Saha:2022oep}
S.K. Saha, D.R. Entem, R.~Machleidt, Y.~Nosyk, Phys. Rev. C \textbf{107}(3),
  034002
\newblock  (2023)

\bibitem{Phillips:2013rsa}
D.R. Phillips, C.~Schat, Phys. Rev. C \textbf{88}(3), 034002
\newblock  (2013)

\bibitem{Epelbaum:2014sea}
E.~Epelbaum, A.M. Gasparyan, H.~Krebs, C.~Schat, Eur. Phys. J. A
  \textbf{51}(3), 26
\newblock  (2015)

\bibitem{Topolnicki:2017rnt}
K.~Topolnicki, Eur. Phys. J. A \textbf{53}(9), 181
\newblock  (2017)

\bibitem{Weinberg:1992yk}
S.~Weinberg, Phys. Lett. B \textbf{295}, 114
\newblock  (1992)

\bibitem{Ishikawa:2007zz}
S.~Ishikawa, M.R. Robilotta, Phys. Rev. C \textbf{76}, 014006
\newblock  (2007)

\bibitem{Girlanda:2020pqn}
L.~Girlanda, A.~Kievsky, L.E. Marcucci, M.~Viviani, Phys. Rev. C \textbf{102},
  064003
\newblock  (2020)

\bibitem{Girlanda:2023znc}
L.~Girlanda, E.~Filandri, A.~Kievsky, L.E. Marcucci, M.~Viviani, Phys. Rev. C
  \textbf{107}(6), L061001
\newblock  (2023)

\bibitem{Girlanda:2011fh}
L.~Girlanda, A.~Kievsky, M.~Viviani, Phys. Rev. C \textbf{84}(1), 014001
\newblock  (2011).
\newblock [Erratum: Phys.Rev.C 102, 019903 (2020)]

\bibitem{Epelbaum:2004xf}
E.~Epelbaum, U.-G. Mei\ss{}ner, J.E. Palomar, Phys. Rev. C \textbf{71}, 024001
\newblock  (2005)

\bibitem{Friar:2004ca}
J.L. Friar, U.~van Kolck, M.C.M. Rentmeester, R.G.E. Timmermans, Phys. Rev. C
  \textbf{70}, 044001
\newblock  (2004)

\bibitem{Yang:1979zz}
S.N. Yang, Phys. Rev. C \textbf{19}, 1114
\newblock  (1979)

\bibitem{Yang:1983pd}
S.N. Yang, J. Phys. G \textbf{9}, L115
\newblock  (1983)

\bibitem{Siemens:2016hdi}
D.~Siemens, V.~Bernard, E.~Epelbaum, A.~Gasparyan, H.~Krebs, U.-G. Mei\ss{}ner,
  Phys. Rev. C \textbf{94}(1), 014620
\newblock  (2016)

\bibitem{Siemens:2016jwj}
D.~Siemens, J.~Ruiz~de Elvira, E.~Epelbaum, M.~Hoferichter, H.~Krebs, B.~Kubis,
  U.-G. Mei\ss{}ner, Phys. Lett. B \textbf{770}, 27
\newblock  (2017)

\bibitem{Hoferichter:2015tha}
M.~Hoferichter, J.~Ruiz~de Elvira, B.~Kubis, U.-G. Mei\ss{}ner, Phys. Rev. Lett.
  \textbf{115}(19), 192301
\newblock  (2015)

\bibitem{Bernard:1996gq}
V.~Bernard, N.~Kaiser, U.-G. Mei\ss{}ner, Nucl. Phys. A \textbf{615}, 483
\newblock  (1997)

\bibitem{Hemmert:1997ye}
T.R. Hemmert, B.R. Holstein, J.~Kambor, J. Phys. G \textbf{24}, 1831
\newblock  (1998)

\bibitem{Pascalutsa:2002pi}
V.~Pascalutsa, D.R. Phillips, Phys. Rev. C \textbf{67}, 055202
\newblock  (2003)

\bibitem{Yao:2016vbz}
D.L. Yao, D.~Siemens, V.~Bernard, E.~Epelbaum, A.M. Gasparyan, J.~Gegelia,
  H.~Krebs, U.-G. Mei\ss{}ner, JHEP \textbf{05}, 038
\newblock  (2016)

\bibitem{HillerBlin:2014diw}
A.N. Hiller~Blin, T.~Ledwig, M.J. Vicente~Vacas, Phys. Lett. B \textbf{747},
  217
\newblock  (2015)

\bibitem{Thurmann:2020mog}
M.~Th\"urmann, E.~Epelbaum, A.M. Gasparyan, H.~Krebs, Phys. Rev. C
  \textbf{103}(3), 035201
\newblock  (2021)

\bibitem{Krebs:2007rh}
H.~Krebs, E.~Epelbaum, U.-G. Mei\ss{}ner, Eur. Phys. J. A \textbf{32}, 127
\newblock  (2007)

\bibitem{Epelbaum:2008td}
E.~Epelbaum, H.~Krebs, U.-G. Mei\ss{}ner, Phys. Rev. C \textbf{77}, 034006
\newblock  (2008)

\bibitem{Epelbaum:2007sq}
E.~Epelbaum, H.~Krebs, U.-G. Mei\ss{}ner, Nucl. Phys. A \textbf{806}, 65
\newblock  (2008)

\bibitem{Krebs:2018jkc}
H.~Krebs, A.M. Gasparyan, E.~Epelbaum, Phys. Rev. C \textbf{98}(1), 014003
\newblock  (2018)

\bibitem{Epelbaum:2019jbv}
E.~Epelbaum, {Proceedings of 6th International Conference Nuclear Theory in
  Supercomputing Era}
\newblock  (2019)

\bibitem{Krebs:2019ddp}
H.~Krebs, PoS \textbf{CD2018}, 031
\newblock  (2019)

\bibitem{Luscher:2011bx}
M.~Luscher, P.~Weisz, JHEP \textbf{02}, 051
\newblock  (2011)

\bibitem{Luscher:2013cpa}
M.~Luscher, JHEP \textbf{04}, 123
\newblock  (2013)

\bibitem{Luscher:2013vga}
M.~L\"uscher, PoS \textbf{LATTICE2013}, 016
\newblock  (2014)

\bibitem{Krebs:2023gge}
H.~Krebs, E.~Epelbaum, Phys. Rev. C \textbf{110}, 044004
\newblock  (2024)

\bibitem{Krebs:2023ljo}
H.~Krebs, E.~Epelbaum, Phys. Rev. C \textbf{110}, 044003
\newblock  (2024)

\bibitem{Gloeckle:1995jg}
W.~Gl\"{o}ckle, H.~Wita\l{}a, D.~Huber, H.~Kamada, J.~Golak, Phys. Rep.
  \textbf{274}, 107
\newblock  (1996)

\bibitem{Witala:2004pv}
H.~Wita\l{}a, J.~Golak, W.~Gl\"{o}ckle, H.~Kamada, Phys. Rev. C \textbf{71}, 054001
\newblock  (2005)

\bibitem{Witala:2011yq}
H.~Wita\l{}a, J.~Golak, R.~Skibi\'nski, W.~Gl\"{o}ckle, H.~Kamada, W.N. Polyzou,
  Phys. Rev. C \textbf{83}, 044001
\newblock  (2011).
\newblock [Erratum: Phys.Rev.C 88, 069904 (2013)]

\bibitem{Epelbaum:2012vx}
E.~Epelbaum, U.-G. Mei\ss{}ner, Ann. Rev. Nucl. Part. Sci. \textbf{62}, 159
\newblock  (2012)

\bibitem{Witala:2013ioa}
H.~Wita\l{}a, J.~Golak, R.~Skibi\'nski, K.~Topolnicki, J. Phys. G \textbf{41},
  094011
\newblock  (2014)

\bibitem{LENPIC:2015qsz}
S.~Binder, et~al., Phys. Rev. C \textbf{93}(4), 044002
\newblock  (2016)

\bibitem{Maris:2016wrd}
P.~Maris, et~al., EPJ Web Conf. \textbf{113}, 04015
\newblock  (2016)

\bibitem{LENPIC:2018lzt}
S.~Binder, et~al., Phys. Rev. C \textbf{98}(1), 014002
\newblock  (2018)

\bibitem{LENPIC:2018ewt}
E.~Epelbaum, et~al., Phys. Rev. C \textbf{99}(2), 024313
\newblock  (2019)

\bibitem{Golak:2009ri}
J.~Golak, et~al., Eur. Phys. J. A \textbf{43}, 241
\newblock  (2010)

\bibitem{Hebeler:2015wxa}
K.~Hebeler, H.~Krebs, E.~Epelbaum, J.~Golak, R.~Skibi\'nski, Phys. Rev. C
  \textbf{91}(4), 044001
\newblock  (2015)

\bibitem{Gazit:2008ma}
D.~Gazit, S.~Quaglioni, P.~Navratil, Phys. Rev. Lett. \textbf{103}, 102502
\newblock  (2009).
\newblock [Erratum: Phys.Rev.Lett. 122, 029901 (2019)]

\bibitem{Nogga:2005hp}
A.~Nogga, P.~Navratil, B.R. Barrett, J.P. Vary, Phys. Rev. C \textbf{73},
  064002
\newblock  (2006)

\bibitem{Navratil:2007we}
P.~Navratil, V.G. Gueorguiev, J.P. Vary, W.E. Ormand, A.~Nogga, Phys. Rev.
  Lett. \textbf{99}, 042501
\newblock  (2007)

\bibitem{Ekstrom:2015rta}
A.~Ekstr\"om, G.R. Jansen, K.A. Wendt, G.~Hagen, T.~Papenbrock, B.D. Carlsson,
  C.~Forss\'en, M.~Hjorth-Jensen, P.~Navr\'atil, W.~Nazarewicz, Phys. Rev. C
  \textbf{91}(5), 051301
\newblock  (2015)

\bibitem{Lynn:2017fxg}
J.E. Lynn, I.~Tews, J.~Carlson, S.~Gandolfi, A.~Gezerlis, K.E. Schmidt,
  A.~Schwenk, Phys. Rev. C \textbf{96}(5), 054007
\newblock  (2017)

\bibitem{sekiguchi2002}
K.~Sekiguchi, H.~Sakai, H.~Wita\l{}a, W.~Gl\"{o}ckle, J.~Golak, M.~Hatano,
  H.~Kamada, H.~Kato, Y.~Maeda, J.~Nishikawa, A.~Nogga, T.~Ohnishi, H.~Okamura,
  N.~Sakamoto, S.~Sakoda, Y.~Satou, K.~Suda, A.~Tamii, T.~Uesaka, T.~Wakasa,
  K.~Yako, Phys. Rev. C \textbf{65}, 034003
\newblock  (2002)

\bibitem{Witala:2019ffj}
H.~Wita\l{}a, J.~Golak, R.~Skibi\'nski, K.~Topolnicki, E.~Epelbaum, K.~Hebeler,
  H.~Kamada, H.~Krebs, U.-G. Mei\ss{}ner, A.~Nogga, Few-Body Syst.
  \textbf{60}(1), 19
\newblock  (2019)

\bibitem{Maris:2020qne}
P.~Maris, et~al., Phys. Rev. C \textbf{103}(5), 054001
\newblock  (2021)

\bibitem{nsakamot96}
N.~Sakamoto, H.~Okamura, T.~Uesaka, S.~Ishida, H.~Otsu, T.~Wakasa, Y.~Satou,
  T.~Niizeki, K.~Katoh, T.~Yamashita, K.~Hatanaka, Y.~Koike, H.~Sakai, Phys.
  Lett. B \textbf{367}(1), 60
\newblock  (1996)

\bibitem{PhysRevLett.95.162301}
K.~Sekiguchi, H.~Sakai, H.~Wita\l{}a, W.~Gl\"{o}ckle, J.~Golak, K.~Hatanaka,
  M.~Hatano, K.~Itoh, H.~Kamada, H.~Kuboki, Y.~Maeda, A.~Nogga, H.~Okamura,
  T.~Saito, N.~Sakamoto, Y.~Sakemi, M.~Sasano, Y.~Shimizu, K.~Suda, A.~Tamii,
  T.~Uesaka, T.~Wakasa, K.~Yako, Phys. Rev. Lett. \textbf{95}, 162301
\newblock  (2005)

\bibitem{PhysRevC.78.014006}
A.~Ramazani-Moghaddam-Arani, H.R. Amir-Ahmadi, A.D. Bacher, C.D. Bailey,
  A.~Biegun, M.~Eslami-Kalantari, I.~Ga\ifmmode \check{s}\else
  \v{s}\fi{}pari\ifmmode~\acute{c}\else \'{c}\fi{}, L.~Joulaeizadeh,
  N.~Kalantar-Nayestanaki, S.~Kistryn, A.~Kozela, H.~Mardanpour, J.G.
  Messchendorp, A.M. Micherdzinska, H.~Moeini, S.V. Shende, E.~Stephan, E.J.
  Stephenson, R.~Sworst, Phys. Rev. C \textbf{78}, 014006
\newblock  (2008)

\bibitem{PhysRevLett.86.5862}
K.~Ermisch, A.M. van~den Berg, R.~Bieber, W.~Gl\"ockle, J.~Golak, M.~Hagemann,
  V.M. Hannen, M.N. Harakeh, M.A. de~Huu, N.~Kalantar-Nayestanaki, H.~Kamada,
  M.~Ki\ifmmode~\check{s}\else \v{s}\fi{}, J.~Kuro\ifmmode \acute{s}\else
  \'{s}\fi{}-\ifmmode~\dot{Z}\else \.{Z}\fi{}o\l{}nierczuk, M.~Mahjour-Shafiei,
  A.~Micherdzi\ifmmode~\acute{n}\else \'{n}\fi{}ska, A.~Nogga,
  R.~Skibi\ifmmode~\acute{n}\else \'{n}\fi{}ski, H.~Wita\l{}a, H.J. W\"ortche,
  Phys. Rev. Lett. \textbf{86}, 5862
\newblock  (2001)

\bibitem{PhysRevC.74.064003}
B.~v.~Przewoski, H.O. Meyer, J.T. Balewski, W.W. Daehnick, J.~Doskow,
  W.~Haeberli, R.~Ibald, B.~Lorentz, R.E. Pollock, P.V. Pancella, F.~Rathmann,
  T.~Rinckel, S.K. Saha, B.~Schwartz, P.~Th\"orngren-Engblom, A.~Wellinghausen,
  T.J. Whitaker, T.~Wise, Phys. Rev. C \textbf{74}, 064003
\newblock  (2006)

\bibitem{LENPIC:2022cyu}
P.~Maris, et~al., Phys. Rev. C \textbf{106}(6), 064002
\newblock  (2022)

\bibitem{Witala:2021zmb}
H.~Wita\l{}a, J.~Golak, R.~Skibi\'nski, K.~Topolnicki, E.~Epelbaum, H.~Krebs,
  P.~Reinert, Phys. Rev. C \textbf{104}(1), 014002
\newblock  (2021)

\bibitem{Skibinski:2023nnn}
R.~Skibi\'nski, J.~Golak, H.~Wita\l{}a, V.~Chahar, E.~Epelbaum, A.~Nogga,
  V.~Soloviov, Front. Phys. \textbf{11}, 1084040
\newblock  (2023)

\bibitem{Barrett:2013nh}
B.R. Barrett, P.~Navratil, J.P. Vary, Prog. Part. Nucl. Phys. \textbf{69}, 131
\newblock  (2013)

\bibitem{Girlanda:2018xrw}
L.~Girlanda, A.~Kievsky, M.~Viviani, L.E. Marcucci, Phys. Rev. C
  \textbf{99}(5), 054003
\newblock  (2019)

\bibitem{Witala:2022rzl}
H.~Wita\l{}a, J.~Golak, R.~Skibi\'nski, Phys. Rev. C \textbf{105}(5), 054004
\newblock  (2022)

\bibitem{wit98}
H.~Wita\l{}a, W.~Gl\"{o}ckle, D.~Huber, J.~Golak, H.~Kamada, Phys. Rev. Lett.
  \textbf{81}, 1183
\newblock  (1998)

\bibitem{PhysRevC.102.054002}
W.~Parol, et~al., Phys. Rev. C \textbf{102}, 054002
\newblock  (2020)

\bibitem{Ramazani-Sharifabadi2020}
R.~Ramazani-Sharifabadi, H.R. Amir-Ahmadi, M.T. Bayat, A.~Deltuva,
  M.~Eslami-Kalantari, N.~Kalantar-Nayestanaki, S.~Kistryn, A.~Kozela,
  M.~Mahjour-Shafiei, H.~Mardanpour, J.G. Messchendorp, M.~Mohammadi-Dadkan,
  A.~Ramazani-Moghaddam-Arani, E.~Stephan, H.~Tavakoli-Zaniani, Eur. Phys. J. A
  \textbf{56}(9), 221
\newblock  (2020)

\bibitem{PhysRevC.76.014004}
Y.~Maeda, H.~Sakai, K.~Fujita, M.B. Greenfield, K.~Hatanaka, M.~Hatano,
  J.~Kamiya, T.~Kawabata, H.~Kuboki, H.~Okamura, J.~Rapaport, T.~Saito,
  Y.~Sakemi, M.~Sasano, K.~Sekiguchi, Y.~Shimizu, K.~Suda, Y.~Tameshige,
  A.~Tamii, T.~Wakasa, K.~Yako, J.~Blomgren, P.~Mermod, A.~\"Ohrn,
  M.~\"Osterlund, H.~Wita\l{}a, A.~Deltuva, A.C. Fonseca, P.U. Sauer,
  W.~Gl\"{o}ckle, J.~Golak, H.~Kamada, A.~Nogga, R.~Skibi\ifmmode~\acute{n}\else
  \'{n}\fi{}ski, Phys. Rev. C \textbf{76}, 014004
\newblock  (2007)

\bibitem{PhysRevC.89.064007}
K.~Sekiguchi, Y.~Wada, J.~Miyazaki, H.~Wita\l{}a, M.~Dozono, U.~Gebauer,
  J.~Golak, H.~Kamada, S.~Kawase, Y.~Kubota, C.S. Lee, Y.~Maeda, T.~Mashiko,
  K.~Miki, A.~Nogga, H.~Okamura, T.~Saito, H.~Sakai, S.~Sakaguchi, N.~Sakamoto,
  M.~Sasano, Y.~Shimizu, R.~Skibi\ifmmode~\acute{n}\else \'{n}\fi{}ski,
  H.~Suzuki, T.~Taguchi, K.~Takahashi, T.L. Tang, T.~Uesaka, T.~Wakasa,
  K.~Yako, Phys. Rev. C \textbf{89}, 064007
\newblock  (2014)

\bibitem{PhysRevC.68.051001}
K.~Ermisch, H.R. Amir-Ahmadi, A.M. van~den Berg, R.~Castelijns, B.~Davids,
  E.~Epelbaum, E.~van Garderen, W.~Gl\"{o}ckle, J.~Golak, M.N. Harakeh, M.~Hunyadi,
  M.A. de~Huu, N.~Kalantar-Nayestanaki, H.~Kamada, M.~Kis, M.~Mahjour-Shafiei,
  A.~Nogga, R.~Skibi\ifmmode~\acute{n}\else \'{n}\fi{}ski, H.~Wita\l{}a, H.J.
  W\"ortche, Phys. Rev. C \textbf{68}, 051001
\newblock  (2003)

\bibitem{PhysRevC.103.044001}
A.~Watanabe, S.~Nakai, Y.~Wada, K.~Sekiguchi, A.~Deltuva, T.~Akieda, D.~Etoh,
  M.~Inoue, Y.~Inoue, K.~Kawahara, H.~Kon, K.~Miki, T.~Mukai, D.~Sakai,
  S.~Shibuya, Y.~Shiokawa, T.~Taguchi, H.~Umetsu, Y.~Utsuki, M.~Watanabe,
  S.~Goto, K.~Hatanaka, Y.~Hirai, T.~Ino, D.~Inomoto, A.~Inoue, S.~Ishikawa,
  M.~Itoh, H.~Kanda, H.~Kasahara, N.~Kobayashi, Y.~Maeda, S.~Mitsumoto,
  S.~Nakamura, K.~Nonaka, H.J. Ong, H.~Oshiro, Y.~Otake, H.~Sakai, A.~Taketani,
  A.~Tamii, D.T. Tran, T.~Wakasa, Y.~Wakabayashi, T.~Wakui, Phys. Rev. C
  \textbf{103}, 044001
\newblock  (2021)

\bibitem{Deltuva_PRC76}
A.~Deltuva, A.C. Fonseca, Phys. Rev. C \textbf{76}, 021001
\newblock  (2007)

\bibitem{lazauskas2009}
R.~Lazauskas, Phys. Rev. C \textbf{79}, 054007
\newblock  (2009)

\bibitem{viviani_PRL111}
M.~Viviani, L.~Girlanda, A.~Kievsky, L.E. Marcucci, Phys. Rev. Lett.
  \textbf{111}, 172302
\newblock  (2013)

\bibitem{deltuva_prc87}
A.~Deltuva, A.C. Fonseca, Phys. Rev. C \textbf{87}, 054002
\newblock  (2013)

\bibitem{fonseca_FBS2017}
A.C. Fonseca, A.~Deltuva, Few-Body Syst. \textbf{58}(2), 46
\newblock  (2017)

\bibitem{INOY04}
P.~Doleschall, Phys. Rev. C \textbf{69}, 054001
\newblock  (2004)

\bibitem{SMS}
P.~Reinert, H.~Krebs, E.~Epelbaum, Eur. Phys. J. A \textbf{54}(5), 86
\newblock  (2018)

\bibitem{CDBD}
A.~Deltuva, R.~Machleidt, P.U. Sauer, Phys. Rev. C \textbf{68}, 024005
\newblock  (2003)

\bibitem{Nemoto1999}
S.~Nemoto, Elastic nucleon-deuteron scattering with $\delta$-isobar excitation.
\newblock Ph.D. thesis, Hannover University
\newblock  (1999)

\bibitem{gross2017quantum}
C.~Gross, I.~Bloch, Science \textbf{357}(6355), 995
\newblock  (2017)

\bibitem{schafer2020tools}
F.~Sch\"{a}fer, T.~Fukuhara, S.~Sugawa, Y.~Takasu, Y.~Takahashi, Nature Reviews
  Physics \textbf{2}(8), 411
\newblock  (2020)

\bibitem{bloch_many-body_2008}
I.~Bloch, J.~Dalibard, W.~Zwerger, Rev. Mod. Phys. \textbf{80}(3), 885
\newblock  (2008)

\bibitem{Campbell2006}
G.K. Campbell, J.~Mun, M.~Boyd, P.~Medley, A.E. Leanhardt, L.G. Marcassa, D.E.
  Pritchard, W.~Ketterle, Science \textbf{313}(5787), 649
\newblock  (2006)

\bibitem{Will2010}
S.~Will, T.~Best, U.~Schneider, L.~Hackerm{\"u}ller, D.S. L{\"u}hmann,
  I.~Bloch, Nature \textbf{465}(7295), 197
\newblock  (2010)

\bibitem{PhysRevLett.107.175301}
M.J. Mark, E.~Haller, K.~Lauber, J.G. Danzl, A.J. Daley, H.C. N\"agerl, Phys.
  Rev. Lett. \textbf{107}, 175301
\newblock  (2011)

\bibitem{Franchi_2017}
L.~Franchi, L.F. Livi, G.~Cappellini, G.~Binella, M.~Inguscio, J.~Catani,
  L.~Fallani, New J. Phys. \textbf{19}(10), 103037
\newblock  (2017)

\bibitem{Goban2018}
A.~Goban, R.B. Hutson, G.E. Marti, S.L. Campbell, M.A. Perlin, P.S. Julienne,
  J.P. D'Incao, A.M. Rey, J.~Ye, Nature \textbf{563}(7731), 369
\newblock  (2018)

\bibitem{Johnson_2009}
P.R. Johnson, E.~Tiesinga, J.V. Porto, C.J. Williams, New J. Phys.
  \textbf{11}(9), 093022
\newblock  (2009)

\bibitem{Johnson_2012}
P.R. Johnson, D.~Blume, X.Y. Yin, W.F. Flynn, E.~Tiesinga, New J. Phys.
  \textbf{14}(5), 053037
\newblock  (2012)

\bibitem{honda2024evidence}
K.~Honda, Y.~Takasu, Y.~Haruna, Y.~Nishida, Y.~Takahashi, arXiv preprint
  arXiv:2402.16254
\newblock  (2024)

\bibitem{PhysRevLett.110.173201}
S.~Kato, S.~Sugawa, K.~Shibata, R.~Yamamoto, Y.~Takahashi, Phys. Rev. Lett.
  \textbf{110}, 173201
\newblock  (2013)

\bibitem{Chin2010}
C.~Chin, R.~Grimm, P.~Julienne, E.~Tiesinga, Rev. Mod. Phys. \textbf{82}, 1225
\newblock  (2010)

\bibitem{PhysRevLett.89.110401}
A.~Muryshev, G.V. Shlyapnikov, W.~Ertmer, K.~Sengstock, M.~Lewenstein, Phys.
  Rev. Lett. \textbf{89}, 110401
\newblock  (2002)

\bibitem{PhysRevLett.96.030406}
S.~Sinha, A.Y. Cherny, D.~Kovrizhin, J.~Brand, Phys. Rev. Lett. \textbf{96},
  030406
\newblock  (2006)

\bibitem{PhysRevLett.100.210403}
I.E. Mazets, T.~Schumm, J.~Schmiedmayer, Phys. Rev. Lett. \textbf{100}, 210403
\newblock  (2008)

\bibitem{PhysRevLett.105.090404}
S.~Tan, M.~Pustilnik, L.I. Glazman, Phys. Rev. Lett. \textbf{105}, 090404
\newblock  (2010)

\bibitem{Mazets_2010}
I.E. Mazets, J.~Schmiedmayer, New J. Phys. \textbf{12}(5), 055023
\newblock  (2010)

\bibitem{Hofferberth2007}
S.~Hofferberth, I.~Lesanovsky, B.~Fischer, T.~Schumm, J.~Schmiedmayer, Nature
  \textbf{449}(7160), 324
\newblock  (2007)

\bibitem{Hofferberth2008}
S.~Hofferberth, I.~Lesanovsky, T.~Schumm, A.~Imambekov, V.~Gritsev, E.~Demler,
  J.~Schmiedmayer, Nature Physics \textbf{4}(6), 489
\newblock  (2008)

\bibitem{PhysRevE.106.064104}
T.~Tanaka, Y.~Nishida, Phys. Rev. E \textbf{106}, 064104
\newblock  (2022)

\bibitem{10.1063/1.1704156}
J.B. McGuire, J. Math. Phys. \textbf{5}(5), 622
\newblock  (2004)

\bibitem{PhysRevLett.96.163201}
V.A. Yurovsky, A.~Ben-Reuven, M.~Olshanii, Phys. Rev. Lett. \textbf{96}, 163201
\newblock  (2006)

\bibitem{PhysRevA.85.062711}
D.S. Petrov, V.~Lebedev, J.T.M. Walraven, Phys. Rev. A \textbf{85}, 062711
\newblock  (2012)

\bibitem{PhysRevA.97.061603}
Y.~Nishida, Phys. Rev. A \textbf{97}, 061603
\newblock  (2018)

\bibitem{PhysRevA.97.061604}
L.~Pricoupenko, Phys. Rev. A \textbf{97}, 061604
\newblock  (2018)

\bibitem{PhysRevA.97.061605}
G.~Guijarro, A.~Pricoupenko, G.E. Astrakharchik, J.~Boronat, D.S. Petrov, Phys.
  Rev. A \textbf{97}, 061605
\newblock  (2018)

\bibitem{PhysRevA.89.053619}
A.J. Daley, J.~Simon, Phys. Rev. A \textbf{89}, 053619
\newblock  (2014)

\bibitem{PhysRevLett.112.103201}
D.S. Petrov, Phys. Rev. Lett. \textbf{112}, 103201
\newblock  (2014)

\bibitem{PhysRevA.90.021601}
D.S. Petrov, Phys. Rev. A \textbf{90}, 021601
\newblock  (2014)

\bibitem{PhysRevA.93.043616}
S.~Paul, P.R. Johnson, E.~Tiesinga, Phys. Rev. A \textbf{93}, 043616
\newblock  (2016)

\bibitem{D'Errico_2007}
C.~D'Errico, M.~Zaccanti, M.~Fattori, G.~Roati, M.~Inguscio, G.~Modugno,
  A.~Simoni, New J. Phys. \textbf{9}(7), 223
\newblock  (2007)

\bibitem{PhysRevA.81.032702}
M.~Lysebo, L.~Veseth, Phys. Rev. A \textbf{81}, 032702
\newblock  (2010)

\bibitem{PhysRevLett.118.230601}
Y.~Nishida, Phys. Rev. Lett. \textbf{118}, 230601
\newblock  (2017)

\bibitem{efimov1970energy}
V.~Efimov, Phys. Lett. B \textbf{33}(8), 563
\newblock  (1970)

\bibitem{efimov1973energy}
V.~Efimov, Nucl. Phys. A \textbf{210}(1), 157
\newblock  (1973)

\bibitem{PhysRevLett.110.235301}
Y.~Nishida, S.~Moroz, D.T. Son, Phys. Rev. Lett. \textbf{110}, 235301
\newblock  (2013)

\bibitem{PhysRevA.90.063631}
S.~Moroz, Y.~Nishida, Phys. Rev. A \textbf{90}, 063631
\newblock  (2014)

\bibitem{PhysRevA.97.011602}
Y.~Sekino, Y.~Nishida, Phys. Rev. A \textbf{97}, 011602
\newblock  (2018)

\bibitem{PhysRevLett.120.243002}
J.E. Drut, J.R. McKenney, W.S. Daza, C.L. Lin, C.R. Ord\'o\~nez, Phys. Rev.
  Lett. \textbf{120}, 243002
\newblock  (2018)

\bibitem{PhysRevLett.93.250408}
H.W. Hammer, D.T. Son, Phys. Rev. Lett. \textbf{93}, 250408
\newblock  (2004)

\bibitem{Bazak_2018}
B.~Bazak, D.S. Petrov, New J. Phys. \textbf{20}(2), 023045
\newblock  (2018)

\bibitem{PhysRevLett.101.170401}
Y.~Nishida, S.~Tan, Phys. Rev. Lett. \textbf{101}, 170401
\newblock  (2008)

\bibitem{nishida2011liberating}
Y.~Nishida, S.~Tan, Few-Body Syst. \textbf{51}(2-4), 191
\newblock  (2011)

\bibitem{PhysRevA.82.043606}
Y.~Nishida, D.T. Son, Phys. Rev. A \textbf{82}, 043606
\newblock  (2010)

\bibitem{nielsen2001three}
E.~Nielsen, D.V. Fedorov, A.S. Jensen, E.~Garrido, Phys. Rep. \textbf{347}(5),
  373
\newblock  (2001)

\bibitem{Jensen2004}
A.S. Jensen, K.~Riisager, D.V. Fedorov, E.~Garrido, Rev. Mod. Phys.
  \textbf{76}, 215
\newblock  (2004)

\bibitem{Efimov1970a}
V.~Efimov, Yad. Fiz. \textbf{12}, 1080
\newblock  (1970).
\newblock [Sov. J. Nucl. Phys. 12, 589-595 (1971)]

\bibitem{PhysRevA.67.010703}
D.S. Petrov, Phys. Rev. A \textbf{67}, 010703
\newblock  (2003)

\bibitem{kartavtsev2007low}
O.I. Kartavtsev, A.V. Malykh, J. Phys. B \textbf{40}, 1429
\newblock  (2007)

\bibitem{PhysRevLett.103.153202}
J.~Levinsen, T.G. Tiecke, J.T.M. Walraven, D.S. Petrov, Phys. Rev. Lett.
  \textbf{103}, 153202
\newblock  (2009)

\bibitem{PhysRevA.86.062703}
S.~Endo, P.~Naidon, M.~Ueda, Phys. Rev. A \textbf{86}, 062703
\newblock  (2012)

\bibitem{endo2011universal}
S.~Endo, P.~Naidon, M.~Ueda, Few-Body Syst. \textbf{51}(2-4), 207
\newblock  (2011)

\bibitem{PhysRevLett.107.120401}
M.~Berninger, A.~Zenesini, B.~Huang, W.~Harm, H.C. N\"agerl, F.~Ferlaino,
  R.~Grimm, P.S. Julienne, J.M. Hutson, Phys. Rev. Lett. \textbf{107}, 120401
\newblock  (2011)

\bibitem{PhysRevLett.108.263001}
J.~Wang, J.P. D'Incao, B.D. Esry, C.H. Greene, Phys. Rev. Lett. \textbf{108},
  263001
\newblock  (2012)

\bibitem{pascaleno3BP1}
P.~Naidon, S.~Endo, M.~Ueda, Phys. Rev. A \textbf{90}, 022106
\newblock  (2014)

\bibitem{pascaleno3BP2}
P.~Naidon, S.~Endo, M.~Ueda, Phys. Rev. Lett. \textbf{112}, 105301
\newblock  (2014)

\bibitem{johansen2017testing}
J.~Johansen, B.~DeSalvo, K.~Patel, C.~Chin, Nature Physics \textbf{13}(8), 731
\newblock  (2017)

\bibitem{PhysRevLett.111.053202}
S.~Roy, M.~Landini, A.~Trenkwalder, G.~Semeghini, G.~Spagnolli, A.~Simoni,
  M.~Fattori, M.~Inguscio, G.~Modugno, Phys. Rev. Lett. \textbf{111}, 053202
\newblock  (2013)

\bibitem{naidon2017efimov}
P.~Naidon, S.~Endo, Rep. Prog. Phys. \textbf{80}(5), 056001
\newblock  (2017)

\bibitem{Pires2014}
R.~Pires, J.~Ulmanis, S.~H\"afner, M.~Repp, A.~Arias, E.D. Kuhnle,
  M.~Weidem\"uller, Phys. Rev. Lett. \textbf{112}, 250404
\newblock  (2014)

\bibitem{Tung2014}
S.K. Tung, K.~Jim\'{e}nez-Garc\'{\i}a, J.~Johansen, C.V. Parker, C.~Chin, Phys.
  Rev. Lett. \textbf{113}, 240402
\newblock  (2014)

\bibitem{PhysRevLett.115.043201}
R.A.W. Maier, M.~Eisele, E.~Tiemann, C.~Zimmermann, Phys. Rev. Lett.
  \textbf{115}, 043201
\newblock  (2015)

\bibitem{PhysRevD.7.2517}
R.D. Amado, F.C. Greenwood, Phys. Rev. D \textbf{7}, 2517
\newblock  (1973)

\bibitem{hammer2007universalEPJ}
H.W. Hammer, L.~Platter, EPJ. A \textbf{32}, 113
\newblock  (2007)

\bibitem{von2009signatures}
J.~von Stecher, J.P. D'Incao, C.H. Greene, Nature Physics \textbf{5}(6), 417
\newblock  (2009)

\bibitem{PhysRevA.70.052101}
L.~Platter, H.W. Hammer, U.-G. Mei\ss{}ner, Phys. Rev. A \textbf{70}, 052101
\newblock  (2004)

\bibitem{yamashita2006four}
M.T. Yamashita, L.~Tomio, A.~Delfino, T.~Frederico, EPL. \textbf{75}(4), 555
\newblock  (2006)

\bibitem{PhysRevLett.107.135304}
M.R. Hadizadeh, M.T. Yamashita, L.~Tomio, A.~Delfino, T.~Frederico, Phys. Rev.
  Lett. \textbf{107}, 135304
\newblock  (2011)

\bibitem{deltuva2011shallow}
A.~Deltuva, EPL. \textbf{95}(4), 43002
\newblock  (2011)

\bibitem{deltuva2013properties}
A.~Deltuva, Few-Body Syst. \textbf{54}(5-6), 569
\newblock  (2013)

\bibitem{PhysRevLett.108.073201}
Y.~Wang, W.B. Laing, J.~von Stecher, B.D. Esry, Phys. Rev. Lett. \textbf{108},
  073201
\newblock  (2012)

\bibitem{PhysRevLett.113.213201}
D.~Blume, Y.~Yan, Phys. Rev. Lett. \textbf{113}, 213201
\newblock  (2014)

\bibitem{PhysRevA.79.060701}
Y.~Nishida, S.~Tan, Phys. Rev. A \textbf{79}, 060701
\newblock  (2009)

\bibitem{PhysRevA.82.011605}
Y.~Nishida, Phys. Rev. A \textbf{82}, 011605
\newblock  (2010)

\bibitem{PhysRevA.84.052727}
T.~Yin, P.~Zhang, W.~Zhang, Phys. Rev. A \textbf{84}, 052727
\newblock  (2011)

\bibitem{PhysRevLett.97.023201}
J.H. Macek, J.~Sternberg, Phys. Rev. Lett. \textbf{97}, 023201
\newblock  (2006)

\bibitem{PhysRevA.86.012711}
E.~Braaten, P.~Hagen, H.W. Hammer, L.~Platter, Phys. Rev. A \textbf{86}, 012711
\newblock  (2012)

\bibitem{PhysRevA.86.012710}
Y.~Nishida, Phys. Rev. A \textbf{86}, 012710
\newblock  (2012)

\bibitem{PhysRevA.77.043611}
M.~Jona-Lasinio, L.~Pricoupenko, Y.~Castin, Phys. Rev. A \textbf{77}, 043611
\newblock  (2008)

\bibitem{PhysRevLett.96.050401}
L.~Pricoupenko, Phys. Rev. Lett. \textbf{96}, 050401
\newblock  (2006)

\bibitem{PhysRevLett.99.210402}
J.~Levinsen, N.R. Cooper, V.~Gurarie, Phys. Rev. Lett. \textbf{99}, 210402
\newblock  (2007)

\bibitem{PhysRevA.106.023304}
M.D. Higgins, C.H. Greene, Phys. Rev. A \textbf{106}, 023304
\newblock  (2022)

\bibitem{PhysRevA.107.033329}
Y.H. Chen, C.H. Greene, Phys. Rev. A \textbf{107}, 033329
\newblock  (2023)

\bibitem{Volosniev_2014}
A.G. Volosniev, D.V. Fedorov, A.S. Jensen, N.T. Zinner, J. Phys. B
  \textbf{47}(18), 185302
\newblock  (2014)

\bibitem{PhysRevA.92.020504}
C.~Gao, J.~Wang, Z.~Yu, Phys. Rev. A \textbf{92}, 020504
\newblock  (2015)

\bibitem{PhysRevA.78.063616}
J.~Levinsen, N.R. Cooper, V.~Gurarie, Phys. Rev. A \textbf{78}, 063616
\newblock  (2008)

\bibitem{Gridnev_2014}
D.K. Gridnev, J. Phys. A \textbf{47}(50), 505204
\newblock  (2014)

\bibitem{Fedorov1994a}
D.V. Fedorov, A.S. Jensen, K.~Riisager, Phys. Rev. C \textbf{49}, 201
\newblock  (1994)

\bibitem{naidon2023universal}
P.~Naidon, SciPost Phys. \textbf{15}, 123
\newblock  (2023)

\bibitem{PhysRevLett.128.212501}
M.~Hongo, D.T. Son, Phys. Rev. Lett. \textbf{128}, 212501
\newblock  (2022)

\bibitem{hoyle1954nuclear}
F.~Hoyle, Astrophys. J. Suppl. Ser. \textbf{1}, 121
\newblock  (1954)

\bibitem{hammer2008model}
H.W. Hammer, R.~Higa, EPJ. A \textbf{37}, 193
\newblock  (2008)

\bibitem{PhysRevC.91.014004}
H.~Suno, Y.~Suzuki, P.~Descouvemont, Phys. Rev. C \textbf{91}, 014004
\newblock  (2015)

\bibitem{otsuka2022alpha}
T.~Otsuka, T.~Abe, T.~Yoshida, Y.~Tsunoda, N.~Shimizu, N.~Itagaki, Y.~Utsuno,
  J.~Vary, P.~Maris, H.~Ueno, Nature Communications \textbf{13}(1), 2234
\newblock  (2022)

\bibitem{phillips1968consistency}
A.~Phillips, Nucl. Phys. A \textbf{107}(1), 209
\newblock  (1968)

\bibitem{efimov1985explanation}
V.~Efimov, E.~Tkachenko, Phys. Lett. B \textbf{157}(2-3), 108
\newblock  (1985)

\bibitem{efimov1988correlation}
V.~Efimov, E.~Tkachenko, Few-Body Syst. \textbf{4}(2), 71
\newblock  (1988)

\bibitem{tanihata1985measurements}
I.~Tanihata, H.~Hamagaki, O.~Hashimoto, Y.~Shida, N.~Yoshikawa, K.~Sugimoto,
  O.~Yamakawa, T.~Kobayashi, N.~Takahashi, Phys. Rev. Lett. \textbf{55}(24),
  2676
\newblock  (1985)

\bibitem{AlKhalili2004}
J.~Al-Khalili, Lect. Notes Phys. \textbf{651}, 77
\newblock  (2004)

\bibitem{Tanihata2013}
I.~Tanihata, H.~Savajols, R.~Kanungo, Progr. Part. Nucl. Phys. \textbf{68}, 215
\newblock  (2013)

\bibitem{Fedorov1994}
D.V. Fedorov, A.S. Jensen, K.~Riisager, Phys. Rev. Lett. \textbf{73}, 2817
\newblock  (1994)

\bibitem{Amorim1997}
A.E.A. Amorim, T.~Frederico, L.~Tomio, Phys. Rev. C \textbf{56}, R2378
\newblock  (1997)

\bibitem{AnnRev_HamPlatt}
H.W. Hammer, L.~Platter, Ann. Rev. Nucl. Part. Sci. \textbf{60}, 207
\newblock  (2010)

\bibitem{FREDERICO2012939}
T.~Frederico, A.~Delfino, L.~Tomio, M.~Yamashita, Prog. Part. Nucl. Phys.
  \textbf{67}(4), 939
\newblock  (2012)

\bibitem{Canham2008}
D.L. Canham, H.W. Hammer, Eur. Phys. J. A \textbf{37}(3), 367
\newblock  (2008)

\bibitem{Hammer_2017}
H.W. Hammer, C.~Ji, D.R. Phillips, J. Phys. G \textbf{44}(10), 103002
\newblock  (2017)

\bibitem{ACHARYA2013196}
B.~Acharya, C.~Ji, D.~Phillips, Phys. Lett. B \textbf{723}(1), 196
\newblock  (2013)

\bibitem{PhysRevLett.111.132501}
G.~Hagen, P.~Hagen, H.W. Hammer, L.~Platter, Phys. Rev. Lett. \textbf{111},
  132501
\newblock  (2013)

\bibitem{PhysRevC.105.024310}
W.~Horiuchi, Y.~Suzuki, M.A. Shalchi, L.~Tomio, Phys. Rev. C \textbf{105},
  024310
\newblock  (2022)

\bibitem{PhysRevLett.120.052502}
D.~Hove, E.~Garrido, P.~Sarriguren, D.V. Fedorov, H.O.U. Fynbo, A.S. Jensen,
  N.T. Zinner, Phys. Rev. Lett. \textbf{120}, 052502
\newblock  (2018)

\bibitem{SPYROU2010129}
A.~Spyrou, T.~Baumann, D.~Bazin, G.~Blanchon, A.~Bonaccorso, E.~Breitbach,
  J.~Brown, G.~Christian, A.~DeLine, P.~DeYoung, J.~Finck, N.~Frank, S.~Mosby,
  W.~Peters, A.~Russel, A.~Schiller, M.~Strongman, M.~Thoennessen, Phys. Lett.
  B \textbf{683}(2), 129
\newblock  (2010)

\bibitem{carbonell2020low}
J.~Carbonell, E.~Hiyama, R.~Lazauskas, F.M. Marqu{\'e}s, SciPost Phys. Proc.
  (3), 008
\newblock  (2020)

\bibitem{PhysRevLett.121.262502}
S.~Leblond, F.M. Marqu\'es, J.~Gibelin, N.A. Orr, Y.~Kondo, T.~Nakamura,
  J.~Bonnard, N.~Michel, N.L. Achouri, T.~Aumann, et~al., Phys. Rev. Lett.
  \textbf{121}, 262502
\newblock  (2018)

\bibitem{PhysRevC.100.011603}
E.~Hiyama, R.~Lazauskas, F.M. Marqu\'es, J.~Carbonell, Phys. Rev. C
  \textbf{100}, 011603
\newblock  (2019)

\bibitem{PhysRevC.50.R550}
S.~Dasgupta, I.~Mazumdar, V.S. Bhasin, Phys. Rev. C \textbf{50}, R550
\newblock  (1994)

\bibitem{PhysRevC.90.044004}
C.~Ji, C.~Elster, D.R. Phillips, Phys. Rev. C \textbf{90}, 044004
\newblock  (2014)

\bibitem{PhysRevLett.125.252501}
Y.~Kubota, A.~Corsi, G.~Authelet, H.~Baba, C.~Caesar, D.~Calvet, A.~Delbart,
  M.~Dozono, J.~Feng, F.~Flavigny, J.M. Gheller, J.~Gibelin, A.~Giganon,
  A.~Gillibert, K.~Hasegawa, T.~Isobe, Y.~Kanaya, S.~Kawakami, D.~Kim,
  Y.~Kikuchi, Y.~Kiyokawa, M.~Kobayashi, N.~Kobayashi, T.~Kobayashi, Y.~Kondo,
  Z.~Korkulu, S.~Koyama, V.~Lapoux, Y.~Maeda, F.M. Marqu\'es, T.~Motobayashi,
  T.~Miyazaki, T.~Nakamura, N.~Nakatsuka, Y.~Nishio, A.~Obertelli, K.~Ogata,
  A.~Ohkura, N.A. Orr, S.~Ota, H.~Otsu, T.~Ozaki, V.~Panin, S.~Paschalis, E.C.
  Pollacco, S.~Reichert, J.Y. Rouss\'e, A.T. Saito, S.~Sakaguchi, M.~Sako,
  C.~Santamaria, M.~Sasano, H.~Sato, M.~Shikata, Y.~Shimizu, Y.~Shindo,
  L.~Stuhl, T.~Sumikama, Y.L. Sun, M.~Tabata, Y.~Togano, J.~Tsubota, Z.H. Yang,
  J.~Yasuda, K.~Yoneda, J.~Zenihiro, T.~Uesaka, Phys. Rev. Lett. \textbf{125},
  252501
\newblock  (2020)

\bibitem{endotanaka2023}
S.~Endo, J.~Tanaka, arXiv:2309.04131
\newblock  (2023)

\bibitem{shusterman2019surprisingly}
J.A. Shusterman, N.D. Scielzo, K.J. Thomas, E.B. Norman, S.E. Lapi, C.S.
  Loveless, N.J. Peters, J.D. Robertson, D.A. Shaughnessy, A.P. Tonchev, Nature
  \textbf{565}(7739), 328
\newblock  (2019)

\bibitem{Lim1977}
T.K. Lim, S.K. Duffy, W.C. Damert, Phys. Rev. Lett. \textbf{38}, 341
\newblock  (1977)

\bibitem{Schoellkopf1994}
W.~Sch{\"o}llkopf, J.P. Toennies, Science \textbf{266}(5189), 1345
\newblock  (1994)

\bibitem{Bruehl2005}
R.~Br\"uhl, A.~Kalinin, O.~Kornilov, J.P. Toennies, G.C. Hegerfeldt, M.~Stoll,
  Phys. Rev. Lett. \textbf{95}, 063002
\newblock  (2005)

\bibitem{Kunitski2015}
M.~Kunitski, S.~Zeller, J.~Voigtsberger, A.~Kalinin, L.P.H. Schmidt,
  M.~Sch{\"o}ffler, A.~Czasch, W.~Sch{\"o}llkopf, R.E. Grisenti, T.~Jahnke,
  D.~Blume, R.~D{\"o}rner, Science \textbf{348}(6234), 551
\newblock  (2015)

\bibitem{Tiesinga1993}
E.~Tiesinga, B.J. Verhaar, H.T.C. Stoof, Phys. Rev. A \textbf{47}, 4114
\newblock  (1993)

\bibitem{PhysRevLett.83.1751}
B.D. Esry, C.H. Greene, J.P. Burke, Phys. Rev. Lett. \textbf{83}, 1751
\newblock  (1999)

\bibitem{PhysRevLett.85.908}
P.F. Bedaque, E.~Braaten, H.W. Hammer, Phys. Rev. Lett. \textbf{85}, 908
\newblock  (2000)

\bibitem{Kraemer2006}
T.~Kraemer, M.~Mark, P.~Waldburger, J.G. Danzl, C.~Chin, B.~Engeser, A.D.
  Lange, K.~Pilch, A.~Jaakkola, H.C. N\"{a}gerl, R.~Grimm, Nature \textbf{440},
  315
\newblock  (2006)

\bibitem{PhysRevLett.103.163202}
N.~Gross, Z.~Shotan, S.~Kokkelmans, L.~Khaykovich, Phys. Rev. Lett.
  \textbf{103}, 163202
\newblock  (2009)

\bibitem{zaccanti2009observation}
M.~Zaccanti, B.~Deissler, C.~D'Errico, M.~Fattori, M.~Jona-Lasinio,
  S.~M\"{u}ller, G.~Roati, M.~Inguscio, G.~Modugno, Nature Physics
  \textbf{5}(8), 586
\newblock  (2009)

\bibitem{pollack2009universality}
S.E. Pollack, D.~Dries, R.G. Hulet, Science \textbf{326}(5960), 1683
\newblock  (2009)

\bibitem{Huang2014}
B.~Huang, L.A. Sidorenkov, R.~Grimm, J.M. Hutson, Phys. Rev. Lett.
  \textbf{112}, 190401
\newblock  (2014)

\bibitem{elliott2018nasa}
E.R. Elliott, M.C. Krutzik, J.R. Williams, R.J. Thompson, D.C. Aveline, npj
  Microgravity \textbf{4}(1), 16
\newblock  (2018)

\bibitem{oudrhiri2023nasa}
K.~Oudrhiri, J.M. Kohel, N.~Harvey, J.R. Kellogg, D.C. Aveline, R.L. Butler,
  J.~Bosch-Lluis, J.L. Callas, L.Y. Cheng, A.P. Croonquist, et~al.,
  arXiv:2305.13285
\newblock  (2023)

\bibitem{frye2021bose}
K.~Frye, S.~Abend, W.~Bartosch, A.~Bawamia, D.~Becker, H.~Blume, C.~Braxmaier,
  S.W. Chiow, M.A. Efremov, W.~Ertmer, et~al., EPJ Quantum Technology
  \textbf{8}(1), 1
\newblock  (2021)

\bibitem{LompeScience11}
T.~Lompe, T.B. Ottenstein, F.~Serwane, A.N. Wenz, G.~Z\"{u}rn, S.~Jochim,
  Science \textbf{330}(6006), 940
\newblock  (2010)

\bibitem{PhysRevLett.106.143201}
S.~Nakajima, M.~Horikoshi, T.~Mukaiyama, P.~Naidon, M.~Ueda, Phys. Rev. Lett.
  \textbf{106}, 143201
\newblock  (2011)

\bibitem{PhysRevLett.108.210406}
O.~Machtey, Z.~Shotan, N.~Gross, L.~Khaykovich, Phys. Rev. Lett. \textbf{108},
  210406
\newblock  (2012)

\bibitem{PhysRevLett.119.143401}
C.E. Klauss, X.~Xie, C.~Lopez-Abadia, J.P. D'Incao, Z.~Hadzibabic, D.S. Jin,
  E.A. Cornell, Phys. Rev. Lett. \textbf{119}, 143401
\newblock  (2017)

\bibitem{PhysRevLett.122.200402}
Y.~Yudkin, R.~Elbaz, P.~Giannakeas, C.H. Greene, L.~Khaykovich, Phys. Rev.
  Lett. \textbf{122}, 200402
\newblock  (2019)

\bibitem{yudkin2023reshape}
Y.~Yudkin, R.~Elbaz, J.P. D'Incao, P.S. Julienne, L.~Khaykovich, Nature
  Communications \textbf{15}(1), 2127
\newblock  (2024)

\bibitem{PhysRevLett.103.130404}
J.R. Williams, E.L. Hazlett, J.H. Huckans, R.W. Stites, Y.~Zhang, K.M. O'Hara,
  Phys. Rev. Lett. \textbf{103}, 130404
\newblock  (2009)

\bibitem{PhysRevA.80.040702}
A.N. Wenz, T.~Lompe, T.B. Ottenstein, F.~Serwane, G.~Z\"urn, S.~Jochim, Phys.
  Rev. A \textbf{80}, 040702
\newblock  (2009)

\bibitem{zhao2023effects}
C.Y. Zhao, H.L. Han, T.Y. Shi, J. Phys. B \textbf{57}(13), 135301
\newblock  (2024)

\bibitem{SciPostPhys.12.6.185}
P.~Naidon, L.~Pricoupenko, C.~Schmickler, SciPost Phys. \textbf{12}, 185
\newblock  (2022)

\bibitem{OiEndo2024}
K.~Oi, P.~Naidon, S.~Endo, Phys. Rev. A \textbf{110}, 033305
\newblock  (2024)

\bibitem{PhysRevLett.106.205304}
H.~Hara, Y.~Takasu, Y.~Yamaoka, J.M. Doyle, Y.~Takahashi, Phys. Rev. Lett.
  \textbf{106}, 205304
\newblock  (2011)

\bibitem{PhysRevLett.108.043201}
D.A. Brue, J.M. Hutson, Phys. Rev. Lett. \textbf{108}, 043201
\newblock  (2012)

\bibitem{chen2015anisotropy}
T.~Chen, C.~Zhang, X.~Li, J.~Qian, Y.~Wang, New J. Phys. \textbf{17}(10),
  103036
\newblock  (2015)

\bibitem{PhysRevX.10.031037}
A.~Green, H.~Li, J.H. See~Toh, X.~Tang, K.C. McCormick, M.~Li, E.~Tiesinga,
  S.~Kotochigova, S.~Gupta, Phys. Rev. X \textbf{10}, 031037
\newblock  (2020)

\bibitem{PhysRevA.96.032711}
F.~Sch\"afer, H.~Konishi, A.~Bouscal, T.~Yagami, Y.~Takahashi, Phys. Rev. A
  \textbf{96}, 032711
\newblock  (2017)

\bibitem{PhysRevA.107.L031306}
F.~Sch\"afer, Y.~Haruna, Y.~Takahashi, Phys. Rev. A \textbf{107}, L031306
\newblock  (2023)

\bibitem{PhysRevA.92.022708}
M.L. Gonz\'alez-Mart\'{\i}nez, P.S. \ifmmode~\dot{Z}\else \.{Z}\fi{}uchowski,
  Phys. Rev. A \textbf{92}, 022708
\newblock  (2015)

\bibitem{schafer_feshbach_2022}
F.~Sch\"{a}fer, N.~Mizukami, Y.~Takahashi, Phys. Rev. A \textbf{105}(1), 012816
\newblock  (2022)

\bibitem{167Er6Li2023}
F.~Sch\"{a}fer, Y.~Haruna, Y.~Takahashi, J. Phys. Soc. Jap. \textbf{92}(5),
  054301
\newblock  (2023)

\bibitem{PhysRevA.61.053601}
M.J. Bijlsma, B.A. Heringa, H.T.C. Stoof, Phys. Rev. A \textbf{61}, 053601
\newblock  (2000)

\bibitem{PhysRevLett.117.245302}
Z.~Wu, G.M. Bruun, Phys. Rev. Lett. \textbf{117}, 245302
\newblock  (2016)

\bibitem{gurarie2007resonantly}
V.~Gurarie, L.~Radzihovsky, Ann. Phys. \textbf{322}(1), 2
\newblock  (2007)

\bibitem{yamaguchi2017novel}
T.~Yamaguchi, D.~Inotani, Y.~Ohashi, J. Phys. Soc. Jpn. \textbf{86}(1), 013001
\newblock  (2017)

\end{thebibliography}

\end{document}